\documentclass[a4paper,fleqn,usenatbib]{mnras}

\usepackage{newtxtext,newtxmath}

\usepackage[T1]{fontenc}
\usepackage{ae,aecompl}

\usepackage{graphicx}	
\usepackage{amsmath}	
\usepackage{amssymb}	
\usepackage{hyperref} 
\usepackage{tabularx} 

\newcommand{\mytilde}{\raise.17ex\hbox{$\scriptstyle\mathtt{\sim}$}}
\newcommand{\diffd}{\mathrm {d}}
\defcitealias{Peng2015}{P15}

\title[Starvation and outflows drive galaxy quenching]{Both starvation and outflows drive galaxy quenching}

\author[J. Trussler et al.]{James Trussler,$^{1,2}$\thanks{E-mail: jaat2@cam.ac.uk}
Roberto Maiolino,$^{1,2}$
Claudia Maraston,$^{3}$
Yingjie Peng,$^{4}$
\newauthor
Daniel Thomas,$^{3}$
Daniel Goddard$^{3}$
and Jianhui Lian$^{3}$
\\
$^{1}$Cavendish Laboratory, University of Cambridge, 19 J.J.\@ Thomson Avenue, Cambridge, CB3 0HE, UK\\
$^{2}$Kavli Institute for Cosmology, University of Cambridge, Madingley Road, Cambridge CB3 0HA, UK\\
$^{3}$Institute of Cosmology and Gravitation, University of Portsmouth, Portsmouth PO1 3FX, UK\\
$^{4}$Kavli Institute for Astronomy and Astrophysics, Peking University, Beijing 100871, China
}

\date{Accepted XXX. Received YYY; in original form ZZZ}

\pubyear{2019}

\begin{document}
\label{firstpage}
\pagerange{\pageref{firstpage}--\pageref{lastpage}}
\maketitle

\begin{abstract}
Star-forming galaxies can in principle be transformed into passive systems by a multitude of processes that quench star formation, such as the halting of gas accretion (starvation) or the rapid removal of gas in AGN-driven outflows. However, it remains unclear which processes are the most significant, primary drivers of the SF-passive bimodality. We address this key issue in galaxy evolution by studying the chemical properties of 80,000 local galaxies in SDSS DR7. 
In order to distinguish between different quenching mechanisms, we analyse the stellar metallicities of star-forming, green valley and passive galaxies. We find that the significant difference in stellar metallicity between passive galaxies and their star-forming progenitors implies that for galaxies at all masses, quenching must have involved an extended phase of starvation. However, some form of gas ejection also has to be introduced into our models to best match the observed properties of local passive galaxies, indicating that, while starvation is likely to be the prerequisite for quenching, it is the combination of starvation and outflows that is responsible for quenching the majority of galaxies. Closed-box models indicate that the duration of the quenching phase is 2--3~Gyr, with an $e$-folding time of 2--4~Gyr, after which further star formation is prevented by an ejective/heating mode. Alternatively, leaky-box models find a longer duration for the quenching phase of 5--7~Gyr and an $e$-folding time of $\sim$1~Gyr, with outflows becoming increasingly important with decreasing stellar mass. Finally, our analysis of local green valley galaxies indicates that quenching is slower in the local Universe than at high-redshift.
\end{abstract}

	


\section{Introduction}\label{sec:intro}

Galaxies in the local Universe exhibit a bimodality in many of their key properties, such as colour \citep[e.g.\@][]{Strateva2001,
Blanton2003, Baldry2004}, star formation rate \citep[e.g.\@][]{Noeske2007, McGee2011, Wetzel2012}, stellar age
\citep[e.g.\@][]{Kauffmann2003a, Gallazzi2008} and morphology \citep[e.g.\@][]{Wuyts2011, VanderWel2014}. Galaxies are therefore broadly
divided into two main classes: star-forming and passive. Star-forming galaxies typically have blue colours, young stellar populations and
late-type morphologies. On the other hand, passive galaxies are not actively forming stars and typically have red colours, old stellar
populations and early-type morphologies. Deeper observations have revealed evidence for galaxy bimodality up to $z$ {$\sim$} 4
\citep[e.g.\@][]{Brammer2009, Muzzin2013}, identifying a population of massive, passive galaxies at high redshift. These ancient passive
systems must have assembled their mass very quickly and then rapidly quenched over a short time-scale. Despite this significant evidence
for galaxy bimodality, it is not yet clear which quenching mechanisms are primarily responsible for shutting down star formation and
transforming the blue star-forming spiral galaxies into `red-and-dead' galaxies. A major focus of current research in astrophysics is to determine the nature of the dominant quenching mechanisms, the typical quenching time-scales and how galaxy quenching depends on key parameters such as stellar mass, environment and redshift. \\
\indent 
Large galaxy surveys allow the dependence of galaxy quenching on critical parameters, such as mass (which traces internal processes) and environment (which traces external processes), to be studied. Since
different quenching mechanisms operate over different mass regimes and environments, the mass- and environment-dependence of galaxy
quenching can reveal the relative importance of different quenching mechanisms, which puts valuable constraints on the nature of the
primary quenching mechanism in the Universe. \citet{Peng2010} showed for galaxies in the SDSS ($z$ \textless\ 0.1) and zCOSMOS (0.3 \textless\ $z$ \textless\ 0.6) that the effects of mass and environment are largely separable, implying that there are two distinct quenching processes at work: one that depends on
galaxy mass (`mass quenching', which is mostly independent of environment) and one that depends on environment (`environmental quenching', which is mostly independent of mass).

Mass quenching corresponds to internal processes that quench star formation, with the effectiveness of the quenching mechanisms mainly
depending on total galaxy mass. In the low mass regime, outflows driven by stellar feedback (stellar radiation, stellar winds or
supernova explosions) are thought to be effective at reducing star formation, although they may not be sufficient to completely quench
star formation \citep [e.g.\@][]{Larson1974, Dekel1986}. The shallower potential wells of these low-mass galaxies mean that the gas is
less strongly bound and hence escapes more easily in the form of outflows. On the other hand, AGN feedback is thought to be more
effective at quenching massive galaxies. As massive galaxies tend to host more massive black holes which can reach larger AGN luminosities, these galaxies drive more powerful outflows that can clean the galaxy of its gas content
(ejective feedback) and/or heat the surrounding circumgalactic medium via energy injection through radio jets and
winds, which in turns results into diminished cold gas accretion, hence quenching by starvation
\citep[e.g.\@][]{Fabian2012, Cicone2014, King2015, Fluetsch2019}. In addition, it is thought that infalling gas from the intergalactic medium (IGM) is shock-heated as it accretes on to galaxies with halo masses above $10^{12}$~$\mathrm{M}_{\odot}$, resulting in the heating of the halo gas, hence suppressing cold accretion on to the galaxy, therefore halting the supply of the fuel for star formation in galaxies \citep[`halo quenching', e.g.\@][]{Birnboim2003, Keres2005, Dekel2006}.

Environmental quenching corresponds to external processes that quench star formation, through interactions between a galaxy and its
environment, i.e.\@ other galaxies or the intracluster medium (ICM). These physical processes preferentially operate in dense
environments and hence can be more important for galaxies residing in clusters and groups rather than galaxies in the field. One example
of environmental quenching is ram pressure stripping, which can occur when a satellite galaxy falls into a cluster. Gas in the
interstellar medium (ISM) can be rapidly removed as the galaxy moves through the hot ICM, causing rapid quenching \citep[e.g.\@][]{Gunn1972,
Abadi1999}. 
Additionally, galaxies plunging into the hot ICM are likely prevented from
accreting further gas from the IGM, 
shutting down the fuel supply for star formation and effectively resulting in galaxy `starvation', a process referred to as `strangulation' \citep[e.g.\@][]{Larson1980, VanDenBosch2008}.

Additional insights into galaxy quenching have come from detailed studies of the chemical content of galaxies.  Since the abundance of
metals in the ISM results from stellar nucleosynthesis, and is affected by the accretion of gas from the IGM as well as the ejection of
material through galactic winds, the metallicity of a galaxy is a tracer for both the full star formation history and the flow of baryons
into and out of the galaxy.  Hence measurements of gas-phase and stellar metallicities can serve as a powerful method to constrain galaxy
evolutionary processes and the relative importance of different quenching mechanisms. 
For example, observations have revealed that local
galaxies follow a clear correlation between stellar mass and gas-phase metallicity, with the more massive galaxies being more metal
enriched \citep[e.g.\@][]{Tremonti2004, Lee2006, Kewley2008, Andrews2013}.
Although it is not entirely clear what physical processes drive the mass-metallicity relation
(MZR), it has been thought to primarily arise from supernova-driven galactic outflows of metal-rich gas which are preferentially expelled
from low-mass galaxies \citep[e.g.\@][]{Larson1974, Tremonti2004}. 
Furthermore, observations of distant galaxies have revealed that the
MZR holds at least out to $z \sim 3$ \citep[e.g.\@][]{Savaglio2005, Erb2006, Maiolino2008, Mannucci2009, Zahid2011, Zahid2014a, Onodera2016},
with the normalisation of the MZR
decreasing with redshift, which could potentially be indicating that high-redshift galaxies were more strongly accreting pristine
(low-metallicity) gas from the IGM, or perhaps that these galaxies are less evolved, and so have transformed less of their gas into stars, resulting in a lower amount of metals and hence a smaller metallicity. 

Studies investigating the scatter in the MZR have revealed that the gas metallicity also
has a secondary dependence on SFR. This three dimensional relationship between stellar mass, gas metallicity and SFR is known as the fundamental metallicity
relation \citep[FMR,][]{Mannucci2010}. For galaxies of a given stellar mass, the metallicity decreases with increasing SFR. 
This anti-correlation has been seen in observations
\citep[e.g.\@][]{Ellison2008, Lara-Lopez2010, Mannucci2010, Cresci2012, Stott2013, Hunt2016a, Almeida2019a}, as well as simulations/semi-analytic models \citep[e.g.\@][]{Yates2012, Lagos2016, DeRossi2015, DeRossi2017b, Torrey2018, Torrey2019} and has also been investigated using gas regulator models \citep[e.g.\@][]{Dayal2013, Lilly2013, Forbes2014, Peng2014b, Hunt2016b}.
This anti-correlation is thought to be primarily driven by inflows of
pristine gas, where the accreted gas dilutes the gas metallicity but also boosts the star formation rate by increasing the gas content. The small scatter in the FMR suggests that there is a smooth secular connection between star formation and gas flows.

Stellar metallicities offer a complementary method for studying the chemical enrichment of galaxies. Studies of local galaxies have
revealed a stellar mass--stellar metallicity relation \citep[e.g.\@][]{Gallazzi2005, Thomas2005, Panter2008, Thomas2010}, where more
massive galaxies typically have higher stellar metallicities. Comparisons of gas and stellar metallicities show that the stellar
metallicities are typically 0.25 dex lower than the metallicity of the gas \citep[e.g.\@][]{Finlator2008, Halliday2008, Peng2014b,
Pipino2014}. \citet{Lian2018a, Lian2018b} simultaneously analyse the gas and stellar MZR of local star-forming galaxies and find that,
due to the relatively low stellar metallicity in low-mass galaxies, both MZRs can only be reproduced simultaneously if the
metal-enrichment in low-mass galaxies is suppressed at early times in their evolution. This suppression can be achieved with either a
time-dependent metal outflow with larger metal loading factors in galactic winds at early times (i.e.\@ less metal retention), or through
a time-dependent IMF, with steeper IMF slopes at early times (i.e.\@ less metal production). More recent observations have begun to probe
the stellar metallicities of high-redshift galaxies \citep[e.g.\@][]{Halliday2008, Sommariva2012, Gallazzi2014}. However, these studies at high redshift currently
lack the statistics to confirm the existence of a tight MZR. 
One significant benefit of studying stellar metallicities is that they can
also be reliably measured for passive galaxies \citep[e.g.\@][]{Thomas2005, Thomas2010, Gallazzi2014, Onodera2015, Lonoce2015, Kriek2016, Toft2017} thus making it possible to compare the metallicities of star-forming and passive
galaxies. Such comparisons have not been possible for gas metallicities, as the nebular emission in passive galaxies is often too
weak (due to the lack of gas) to derive reliable gas-phase metallicities and also because proper calibration of metallicity diagnostics
for the gas phase in non star-forming regions have only recently become available \citep{Kumari2019}.

\citet[P15 hereafter]{Peng2015} pioneered the idea that the stellar metallicity difference between star-forming and passive galaxies can be used to determine the nature of the primary quenching mechanism in the Universe.
During the evolution of a star-forming galaxy, gas is converted into stars and metals are continuously released into the ISM. As a result, subsequent generations of stars that form have progressively higher metallicities. However, due to the diluting effect of
the accreting (pristine/low-metallicity) gas, the metallicity increase is modest. If at some point star formation is rapidly
halted because of gas removal (ejective mode), then few metals and few stars are formed in the quenching phase and the
resulting passive galaxy has the same stellar mass and stellar metallicity as its star-forming progenitor. If, instead, gas accretion
on to the galaxy is halted by some mechanism (e.g.\@ halo heating), hence resulting in quenching by starvation, then 
the galaxy keeps forming stars with the gas reservoir still available in the ISM, but the dilution effect from the accreting
gas is no longer present and this results in a much steeper increase of the metallicity (and of the newly formed generations
of stars) during the quenching phase; the result is a passive galaxy with slightly higher stellar mass and much higher stellar
metallicity than the star-forming progenitor.

In their study, \citetalias{Peng2015} analysed the stellar metallicities of $26,000$ galaxies in SDSS DR4 \citep{Adelman-McCarthy2006}.
At $M_* < 10^{11}~\mathrm{M}_{\odot}$ passive galaxies were found to have a systematically larger stellar metallicity than star-forming galaxies of the same stellar mass, a clear signature of quenching by starvation. These observed differences in stellar metallicity between star-forming and passive galaxies were
then compared with the predictions of gas-regulator models in order to put quantitative constraints on the possible quenching mechanisms and
time-scales. Their analysis confirmed that starvation (i.e.\@ the halting of the supply of cold gas) is the primary mechanism responsible
for shutting down star formation in galaxies with stellar masses below $10^{11}$ $\mathrm{M_{\odot}}$. \citetalias{Peng2015} inferred
a typical mass-independent quenching time-scale of 4~Gyr. They further found that their models are unable to reproduce the observed stellar
metallicity differences when outflows are included, suggesting that outflows play a minor role in quenching galaxies. In order to
distinguish between different origins for the starvation mechanism (e.g.\@ halo quenching, strangulation or the shutdown of cosmological
gas accretion), an analysis of the environmental dependence of the stellar metallicity difference was also undertaken. Satellite galaxies
were found to have slightly higher metallicity differences than central galaxies at stellar masses below $10^{10}$~$\mathrm{M_\odot}$,
suggesting an environmental origin for the starvation mechanism in this low-mass regime. However, at stellar masses above
$10^{10}$~$\mathrm{M_\odot}$ no difference was found between satellites and centrals (neither in terms of overdensity), suggesting
that in this higher mass range environmental effects do not contribute significantly to the quenching.

The metallicity difference between passive and star-forming galaxies found by \citetalias{Peng2015} was also investigated by \cite{Spitoni2017} in the context
of their exponentially declining accretion scenario. They interpret the difference in terms of a faster decline of the accretion, which is 
qualitatively very similar to the more abrupt halt of accretion adopted by \citetalias{Peng2015}, hence independently confirming the need of
a starvation phase to explain the metallicity difference between the two galaxy populations.

In this work we build upon the original analysis by \citetalias{Peng2015}, using stellar metallicity differences to determine the primary mechanism responsible for quenching star formation. We focus on the role of mass in galaxy quenching and the similarities and differences between the quenching of galaxies at high-redshift and the quenching of galaxies in the local Universe. We note that an extension of the \citetalias{Peng2015} analysis on the role of environment in galaxy quenching will be undertaken in a future paper. Our analysis in this work improves upon the original study in several ways. Firstly, we undertake a more extensive comparison between models and observations, deriving stronger constraints on the mass-dependent role of outflows in quenching star formation, by simultaneously reproducing the stellar metallicities and the star formation rates observed in local passive galaxies in our models. Secondly, using the much larger spectroscopic sample of galaxies (930,000 cf.\@ 566,000 in the original work) available in SDSS DR7 \citep{Abazajian2009}, we now also study the quenching of star formation in local green valley galaxies, in addition to the quenching of passive galaxies that was studied in the original work. Thirdly, we have introduced a more sophisticated treatment of the progenitor--descendant comparison to allow for a better assessment of the amount of chemical enrichment during the quenching phase. We now compare the stellar metallicity difference between local passive galaxies and their high-$z$ star-forming progenitors, rather than the difference between local passive and local star-forming galaxies. This new comparison results in a widening of the gap in stellar metallicity between star-forming and passive galaxies, which qualitatively strengthens the case for starvation. Fourthly, we use the mass-weighted stellar ages of local galaxies derived in this work to provide a more empirically-motivated estimation of the redshift associated with the onset of quenching in the models. Finally, we adopt a consistent treatment of stellar metallicity in this work, comparing mass-weighted metallicity predictions from models with mass-weighted stellar metallicities from observations, rather than comparing against light-weighted stellar metallicities from observations like in \citetalias{Peng2015}.

The paper is structured as follows. In Section \ref{sec:data}, we describe our galaxy sample and the parameters used in our study. In Section \ref{sec:method}, we outline how the differences in stellar metallicity between star-forming and passive (or green valley) galaxies can be used to distinguish between different quenching mechanisms. We also show the stellar mass--stellar metallicity relations which form the basis for all of our subsequent analysis. In Section \ref{sec:model}, we describe the gas regulator models that are used to interpret the observed stellar metallicity differences. In Section \ref{sec:quenching_mechanism}, we compare the observed stellar metallicity differences with the predictions from gas regulator models to put constraints on the relative role of different quenching mechanisms, as well as the associated quenching time-scale. In Section \ref{sec:caveats}, we discuss assumptions and modelling techniques that need to be taken into account when comparing our results to the literature. Finally, in Section \ref{sec:summary} we summarise our main findings and conclude. Throughout this work, we assume that solar metallicity $Z_\odot = 0.02$.

\section{Data}  \label{sec:data}

\subsection{Sample} \label{subsec:sample}

We use the spectroscopic sample of galaxies in the Sloan Digital Sky Survey Data Release 7 \citep[SDSS DR7,][]{York2000, Abazajian2009}
dataset, obtained using the Sloan 2.5m telescope \citep{Gunn2006}. Briefly, the SDSS DR7 dataset includes five-band photometry \citep[$u,
g, r, i, z, $][]{Gunn1998, Doi2010} for 357 million distinct objects and spectroscopy \citep{Smee2013} for over 1.6 million sources,
including 930,000 galaxies which we study in our analysis. The spectroscopic sample of galaxies is substantially larger than the 566,000
galaxies in DR4 \citep{Adelman-McCarthy2006} that were studied by \citetalias{Peng2015}. The galaxies chosen for spectroscopic follow-up
consist of two main samples. Firstly, a sample complete to a \citet{Petrosian1976} magnitude limit of $r=17.77$ \citep[`Main Galaxy
Sample',][]{Strauss2002}. Secondly, two smaller and deeper samples of luminous red ellipticals up to $r=19.2$, corresponding to an
approximately volume-limited sample to $z=0.38$ and $z=0.55$, respectively \citep[`Luminous Red Galaxy Sample',][]{Eisenstein2001}.
Spectroscopic observations are in the optical/NIR (3800--9200 \AA), have a spectral resolution $R \sim 2000$ and a typical
signal-to-noise ratio (S/N) $\sim$ 10 for galaxies near the main sample flux limit. Since the SDSS sample suffers from incompleteness
at $M_* < 10^{10}~\mathrm{M_\odot}$, we apply the $V_{\mathrm{max}}$ weightings from \citet{Blanton2003} to correct for volume incompleteness, allowing our analysis to be safely extended down to $M_* = 10^{9}~\mathrm{M_\odot}$. However, we do note that these $V_{\mathrm{max}}$ corrections only have a very minor effect on our stellar mass--stellar metallicity and stellar mass--stellar age relations, and so do not affect the results from our study.

Similar to \citetalias{Peng2015}, we restrict our analysis to galaxies with reliable spectroscopic redshifts in the range $0.02 < z < 0.085$. This redshift cut was applied
for several reasons. Firstly, to reduce the effect of cosmological evolution on the analysis ($z_{\mathrm{max}} \sim
0.55$ in the full SDSS sample corresponds to roughly 40 per cent of the age of the Universe). Secondly, to reduce the impact of aperture effects associated with the
projected physical aperture of the SDSS spectroscopic fibre. If a broader redshift range were used, then the more distant galaxies would be studied over larger effective radii than nearby galaxies, which could result in unwanted biases in the analysis. Finally, to ensure that the $V_{\mathrm{max}}$ correction remains reliable for the sample studied.

We also restrict our study to galaxies with reliable stellar metallicities and stellar ages, requiring that the median signal-to-noise
ratio per spectral pixel is higher than 20. Such a high S/N cut could potentially introduce biases into our analysis as low surface brightness galaxies are preferentially removed from the sample. However, we find that the trends seen in our results do not change significantly with the chosen S/N criterion. Hence we have selected a S/N threshold of 20, since this provides a healthy balance between good statistics and reliability of measurements.

\subsection{Derived Parameters}

We use the spectral fitting code {\footnotesize FIREFLY} \citep{Comparat2017, Goddard2017b, Wilkinson2017} to obtain stellar metallicities and stellar ages for each galaxy in the SDSS sample. Briefly, {\footnotesize FIREFLY} is a  $\chi ^2 $ minimisation fitting code that fits input galaxy spectra using an arbitrarily weighted, linear combination of simple stellar populations (SSPs) which can have a range of metallicities and ages. The weighted sum of metallicities and ages of each of the SSPs is used to derive the stellar metallicity and stellar age of the galaxy. The code returns both light-weighted and mass-weighted stellar ages and stellar metallicities. The light-weighted properties are obtained by weighting each SSP by its total luminosity across the fitted wavelength range (3500--7429 \AA). On the other hand, mass-weighted properties are obtained by weighting each SSP by its stellar mass contribution. These two weightings are complementary. The light-weighted quantities primarily trace the properties of the younger stellar populations, as these tend to be brighter and dominate the light in the galaxy spectrum. On the other hand, the mass-weighted quantities trace the cumulative evolution of the galaxy. We study the mass-weighted ages and metallicities in our analysis, as these are directly comparable with our simple gas regulator models. This is in contrast with the light-weighted ages and metallicities, which in the model would have to be computed using detailed stellar population synthesis modelling, that assesses the evolution of the relative light contributions from different stellar populations across the fitted wavelength range.

{\footnotesize FIREFLY} fits the observed galaxy spectra using the stellar population models of \citet{Maraston2011}, together with input stellar spectra from the empirical stellar library MILES \citep{Sanchez-Blazquez2006} and a Kroupa IMF \citep{Kroupa2001}.  We have chosen to use MILES in our analysis as it had the most comprehensive sampling range in stellar metallicity and stellar age out of the empirical libraries that were available. The MILES library has metallicities [Z/H] = -2.3, -1.3, -0.3, 0.0, 0.3 and ages that span from 6.5 Myr to 15~Gyr. The spectral resolution is 2.5 \AA  \ FWHM and the wavelength coverage is 3500--7429 \AA. We have also repeated our analysis using the Salpeter \citep{Salpeter1955} and Chabrier {\citep{Chabrier2003} IMFs and find that our results are essentially unchanged. We have chosen to use the Kroupa IMF in this work.

We make use of the publicly available MPA-JHU DR7 release of spectral measurements\footnote{The MPA-JHU data release is available at
\url{https://wwwmpa.mpa-garching.mpg.de/SDSS/DR7/}.}, which provides derived galaxy parameters for all galaxies in SDSS DR7. Stellar
masses are obtained from fits to the photometry, using the Bayesian methodology of \citet{Kauffmann2003a}. Star formation rates within
the spectroscopic fibre aperture are computed from the $\mathrm{H}~\alpha$ emission \citep{Brinchmann2004}, which are then aperture-corrected using
photometry \citep{Salim2007} to obtain total SFRs which extends beyond the spectroscopic fibre aperture. We use the total SFRs in our analysis. For AGN or galaxies with faint emission lines, such as passive galaxies, SFRs are obtained from photometry. It should be noted that the SFRs derived for passive galaxies are most likely upper limits.

\section{Method and results} \label{sec:method}

\subsection{Stellar metallicity}

\citetalias{Peng2015} pioneered the idea that stellar metallicities can be used to distinguish between different quenching mechanisms. We
build upon this idea in our analysis. Although already briefly mentioned in the introduction, here we discuss the basic idea behind the approach in more detail. Consider a typical star-forming galaxy that is evolving along the main sequence, with its gas
reservoir in near equilibrium, where gas depletion (e.g.\@ through star formation and galactic winds) is balanced by gas replenishment
through accretion. Over time, the stellar mass of the galaxy increases as new stars are formed out of the ISM. Furthermore, both the
gas-phase metallicity and the stellar metallicity increase with time, as the elements produced through stellar nucleosynthesis enrich the
ISM and increase the gas metallicity, which causes successively more metal-rich stars to form out of the gas, resulting in a steady
increase in the stellar metallicity. While a galaxy evolves along the star-forming main sequence, it accretes an appreciable amount of
gas from the IGM to fuel its star formation and maintain its gas reservoir in a rough equilibrium. Since this accreted gas is pristine
(i.e.\@ low metallicity), the metal content within the galaxy is diluted, and so the rate of metal enrichment is slowed down, with both
the gas-phase metallicity and the stellar metallicity growing less steeply (per unit stellar mass formed) than would have been the case
in the absence of any gas accretion. However, star-forming galaxies do not remain on the main sequence indefinitely. Indeed, the
star-forming progenitors of passive galaxies must have been thrown off of the main sequence at some epoch, at which point their gas
reservoir begins to decline, star formation shuts down, and a quiescent system is ultimately produced. We imagine that this quenching
process beings at some epoch $z_\mathrm{q}$, when the galaxy is thrown out of equilibrium and begins quenching. During this quenching phase both
the stellar mass ($\Delta M_*$) and the stellar metallicity ($\Delta Z_*$) of the galaxy grow with time, as new metal-rich stars form out
of the gas. \\
\indent The amount by which the stellar metallicity is enhanced during quenching depends on the quenching mechanism. \\
\indent Galaxies that quench rapidly through powerful outflows driven by AGN-feedback or ram pressure stripping quickly deplete their gas reservoirs and so only a small number of metal-enriched stars are produced during the quenching phase. In this case the stellar metallicity is only enhanced by a small amount during quenching and the stellar mass increase is negligible. Hence $\Delta Z_*$ and $\Delta M_*$ are small.\\
\indent On the other hand, galaxies that deplete their gas reservoirs over long time-scales and quench slowly, e.g.\@ through starvation (in the
absence of outflows), produce a significant amount of metal-enriched stars during the quenching phase. In addition to this, in the starvation scenario the supply of cold gas has halted and there is therefore no longer any dilution of the ISM by pristine gas accreted from the IGM, so the stellar metallicity grows much more steeply (per unit stellar mass formed). Hence galaxies that quench through starvation undergo a significant increase in stellar metallicity during quenching, resulting in a large $\Delta Z_*$. \\
\indent Since small stellar metallicity enhancements correspond to quenching by rapid gas removal and large enhancements correspond to
starvation, it is possible to distinguish between the two quenching mechanisms
by measuring the amount by which the stellar metallicity is increased ($\Delta Z_*$) during the quenching phase. Of course, we cannot
make this measurement directly as it is not possible to track the evolution of the stellar metallicity of an individual galaxy across
cosmic time. Instead, we can measure this enhancement indirectly by statistically
studying the difference in stellar metallicity between star-forming and passive galaxies. Star-forming galaxies represent galaxies prior
to quenching ($t < t_\mathrm{q}$), while passive galaxies represent galaxies after quenching has completed. Hence the stellar metallicity enhancement during quenching can be inferred from the stellar metallicity difference between star-forming and passive galaxies. By measuring this stellar metallicity difference for many galaxies, we aim to put constraints on the nature of the primary mechanism responsible for shutting down star formation in galaxies. Our work therefore focusses on the chemical enrichment and decline in star formation that takes place during the quenching phase. However, as discussed in Section \ref{sec:intro}, there is also a strong connection between star formation activity and morphology, with star-forming galaxies typically having late-type morphologies, while passive galaxies tend to have early-type morphologies, which may indicate that galaxies often undergo a morphological transformation during quenching as well \citep[e.g.\@][]{Wuyts2011}. Despite this important connection between galaxy morphology and galaxy quenching, we acknowledge that our analysis and models do not take the star-formation--morphology correlation nor morphological transformation into account.

\subsection{Classification}

We use the bimodality in the star formation rate--stellar mass (SFR-M$_*$) plane to classify galaxies as either star-forming, green valley or
passive. In a similar fashion to \citet{Renzini2015}, we define the boundary of the star-forming (quenched) region in the SFR-M$_*$ plane by
the locus of points, given by a best straight-line fit, where the number of star-forming (quenched) galaxies per SFR-M$_*$ bin has dropped
below some threshold with respect to the peak value at a given mass. In this way, the SFR-M$_*$ plane is partitioned into three regions,
which correspond to star-forming, green valley and passive galaxies, respectively. The adopted boundary between star-forming and green valley galaxies is given by
\begin{equation}
\log \mathrm{SFR} = 0.70 \log M_* - 7.52
\end{equation}
\noindent and the adopted boundary between green valley and passive galaxies is given by
\begin{equation}
\log \mathrm{SFR} = 0.70 \log M_* - 8.02.
\end{equation}
\noindent This selection criterion is illustrated in Fig.\@ \ref{fig:sfr_m}, where we show the subsample of SDSS DR7 galaxies in the redshift range $0.02 < z < 0.085$. It should be noted that the results from our analysis do not change significantly when the slopes or intercepts of the boundaries between the star-forming, green valley and passive regions are changed. 
\begin{figure}
\centering
\centerline{\includegraphics[width=0.9\linewidth]{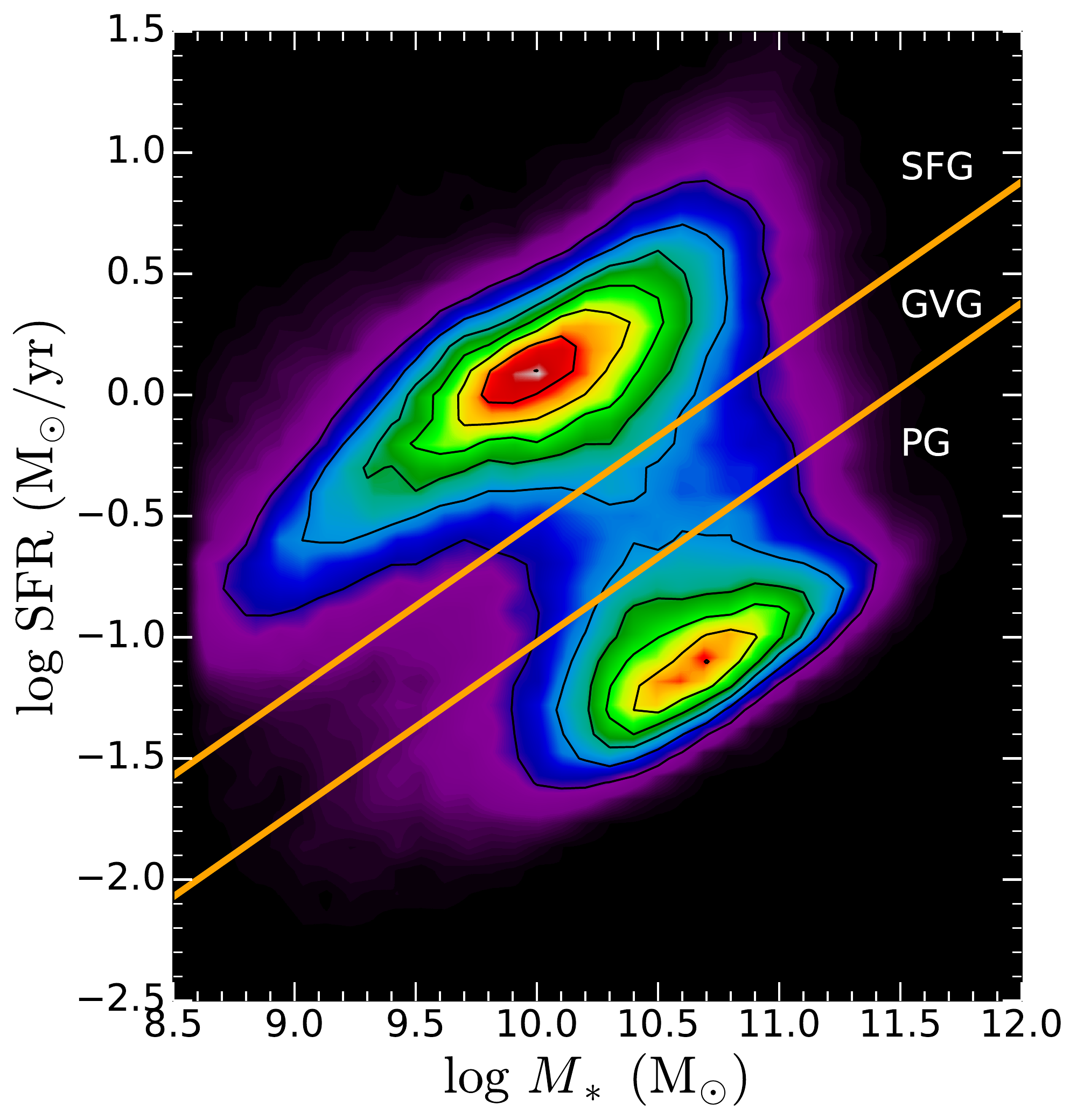}}
\caption{The bimodality of local galaxies in the star formation rate--stellar mass (SFR--M$_*$) plane. We only show the subsample of SDSS DR7 galaxies in the redshift range $0.02 < z < 0.085$. The colour shown reflects the number of galaxies in each SFR-M$_*$ bin, ranging from low counts (purple) to high counts (red). The orange lines mark the boundaries of the star-forming, green valley and passive regions of the plane. Galaxies in the upper left are classified as star-forming (SFG), intermediate galaxies are classified as green valley (GVG), and galaxies in the lower right are classified as passive (PG).}
\label{fig:sfr_m}
\end{figure}

Star-forming galaxies are also required to have their BPT classification set to `star-forming' according to the [NII]-BPT
diagnostic diagram \citep{Brinchmann2004}. This excludes objects hosting an AGN, ensuring that we only analyse true star-forming galaxies
in our study (as the presence of the AGN may affect the estimation of the star formation rate, through its additional contribution
to the nebular line emission, and may also affect the determination of the stellar metallicity through the additional contribution
to the continuum emission). After applying our cuts on redshift, S/N and this selection criterion, our final sample consists of 16,685 star-forming galaxies, 8,445 green valley galaxies and 53,661 passive galaxies. All of these galaxies have reliable stellar masses, star formation rates, stellar metallicities and stellar ages.

Note that the relative fraction of star-forming, green valley and passive galaxies does not reflect the real relative census of these
different populations of galaxies as their relative number is also convolved with our selection criteria (in particular the
requirement of high S/N on the continuum, to reliably measure the stellar metallicities).

\subsection{Scaling relations}

We now apply the previously discussed cuts in redshift and S/N, as well as the SFR-M$_*$ classification criterion to investigate relations between stellar mass and stellar metallicity, as well as stellar mass and stellar age, for star-forming, green valley and passive galaxies. All galaxies of a particular class (e.g.\@ star-forming) are binned in stellar mass bins of 0.02 dex and the median stellar metallicity and stellar age in each mass bin is determined. Error bars on the median stellar metallicities and stellar ages are given by the $1\sigma$ uncertainty on the median (1.253$\sigma / \sqrt N$). We also apply a running average of 0.2 dex to smooth the data. These scaling relations in stellar mass--stellar metallicity and stellar mass--stellar age will form the basis for all of our subsequent analysis. The differences in stellar age between star-forming and passive (green valley) galaxies, will be used to estimate the epoch when the progenitors of local passive (green valley) galaxies began quenching. Furthermore, the differences in stellar metallicity between star-forming and passive (green valley) galaxies will enable us to put constraints on the role that different quenching mechanisms played in quenching the progenitors of local passive (green valley) galaxies.

\subsubsection{Stellar metallicity}

The mass-weighted stellar mass--stellar metallicity relations for star-forming (blue), green valley (green) and passive galaxies (red) are
shown in the top panel of Fig.\@ \ref{fig:combined_mzr_sgq}. We find that the stellar metallicity increases with stellar mass for
star-forming, green valley, and passive galaxies alike. This result is qualitatively similar to what has been seen in previous studies,
where the stellar metallicity tends to increase with stellar mass. For example, \citet{Gallazzi2005} studied the light-weighted stellar
mass--stellar metallicity relation for the total galaxy population (i.e.\@ without distinguishing between star-forming, green valley
and passive galaxies) in SDSS DR4 and found that more massive galaxies typically have larger stellar metallicities. \citetalias{Peng2015}
further divided the \citet{Gallazzi2005} sample into star-forming and passive galaxies, and found that the stellar metallicity tends to
increase with stellar mass for both of these subpopulations. Similar to \citetalias{Peng2015}, but using mass-weighted rather than
light-weighted stellar metallicities in our study, we also find that at a given stellar mass the stellar metallicity of passive galaxies
is systematically larger than that of star-forming galaxies. In addition to this, we find that green valley galaxies typically have metallicities that are intermediate between those of star-forming and passive galaxies. 

\begin{figure}
\centering
\centerline{\includegraphics[width=1.0\linewidth]{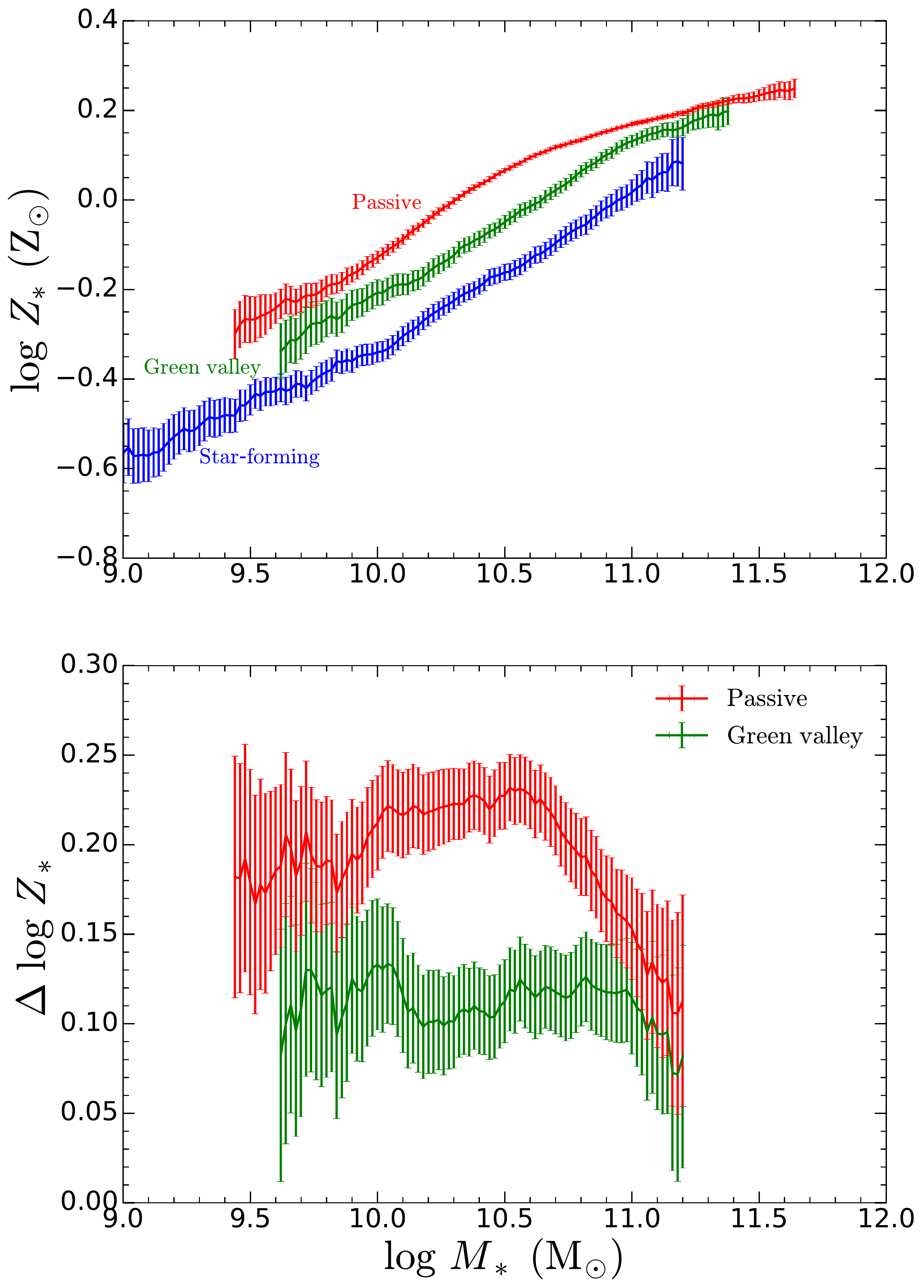}}
\caption{Top panel: The stellar mass--stellar metallicity relation for star-forming (blue), green valley (green) and passive (red) galaxies. Galaxies are binned in 0.02 dex stellar mass bins and the median stellar metallicity in each bin is plotted. Error bars correspond to the $1\sigma$ uncertainty on the median. Bottom panel: The observed difference in stellar metallicity between star-forming and passive galaxies (red), as well as the difference between star-forming and green valley galaxies (green). Error bars represent the 1$\sigma$ error on the stellar metallicity difference.}
\label{fig:combined_mzr_sgq}
\end{figure}

The difference in stellar metallicity between star-forming and passive galaxies (red), as well as the difference between star-forming and green valley galaxies (green) is shown in the bottom panel of Fig.\@ \ref{fig:combined_mzr_sgq}. The stellar metallicity differences are computed by determining the logarithmic difference between the appropriate populations for each stellar mass bin. For example, the stellar metallicity difference between star-forming and passive galaxies is given by  $\log Z_\mathrm{PG} (M_*) -  \log Z_\mathrm{SFG} (M_*)$. 
We find, similar to \citetalias{Peng2015}, that the stellar metallicity difference between star-forming and passive galaxies decreases with increasing stellar mass. As we shall see later on, this metallicity difference is even higher if one compares the metallicity of passive galaxies with the metallicity of their star-forming progenitors at high redshift. Our results therefore suggest that galaxies typically undergo significant chemical enrichment during the quenching phase. This significant difference in stellar metallicity between star-forming and passive galaxies is qualitatively consistent with quenching through starvation, and inconsistent with quenching through simple gas removal (ejective mode). Within the starvation scenario, the decrease in stellar metallicity difference can be attributed to the smaller gas fractions present in more massive galaxies, which therefore have relatively smaller amounts of gas available to convert into metals during starvation.
 
In contrast to the \citetalias{Peng2015} study, we find that there is a non-zero difference in stellar metallicity above $10^{11}~\mathrm{M_\odot}$ (which becomes even larger when comparing passive galaxies with their high-z star-forming progenitors). This enables us to investigate the nature of the primary quenching mechanism for massive galaxies in this study, which is something that was not possible in the original \citetalias{Peng2015} analysis. Furthermore, we find that, at the low-mass end, the stellar metallicity differences between star-forming and passive galaxies found in this work is 0.2~dex smaller than what was found in \citetalias{Peng2015} \citep[who used the metallicity measurements from][]{Gallazzi2005}. We discuss the similarities and differences between the stellar metallicities and ages derived using {\footnotesize FIREFLY} and those derived by \citet{Gallazzi2005}, as well as the implications this has on our results, in more detail in Section \ref{subsec:sp}.

\subsubsection{Stellar age}

We show the mass-weighted stellar mass--stellar age relation for star-forming, green valley and passive galaxies in Fig.\@
\ref{fig:combined_mar_sgq}. The stellar ages of the different galaxy subpopulations tend to increase with increasing stellar mass.
Furthermore, passive galaxies and green valley galaxies are always systematically older than star-forming galaxies of the same stellar
mass. Combined, Figs.\@ \ref{fig:combined_mzr_sgq} and \ref{fig:combined_mar_sgq} indicate that green valley galaxies are intermediate between star-forming and passive galaxies in both metallicity and age. Their intermediate properties are perhaps to be expected for the simple reason that, if we assume one-way evolution (i.e.\@ ignoring rejuvenation), green valley galaxies are currently making the transition from star-forming to passive. We note that the mass-weighted ages for star-forming and passive galaxies in our study are typically 2--4~Gyr higher than the
light-weighted ages in \citetalias{Peng2015}. We discuss this point in more detail in Section \ref{subsec:sp}, but briefly remark that this is mostly due to the fact that mass-weighted ages tend to be larger than light-weighted ages, but we also acknowledge that the different modelling and fitting techniques adopted in the two works likely also play a role.

Fig.\@ \ref{fig:combined_mar_sgq} also shows the stellar age difference $\Delta t$ between star-forming and passive galaxies, as well as the age difference between star-forming and green valley galaxies. We find that the star-forming--passive age difference is typically between 2.5--4~Gyr and seems to be roughly constant within the error bars, with perhaps a weakly declining trend with increasing stellar mass. This is different to the roughly mass-independent $\sim$4~Gyr that was found in \citetalias{Peng2015}, perhaps due to the reasons mentioned earlier. The star-forming--green valley age difference is smaller, typically between 1.5--3~Gyr, with perhaps a weakly increasing trend with increasing stellar mass.

\begin{figure}
\centering
\centerline{\includegraphics[width=1.0\linewidth]{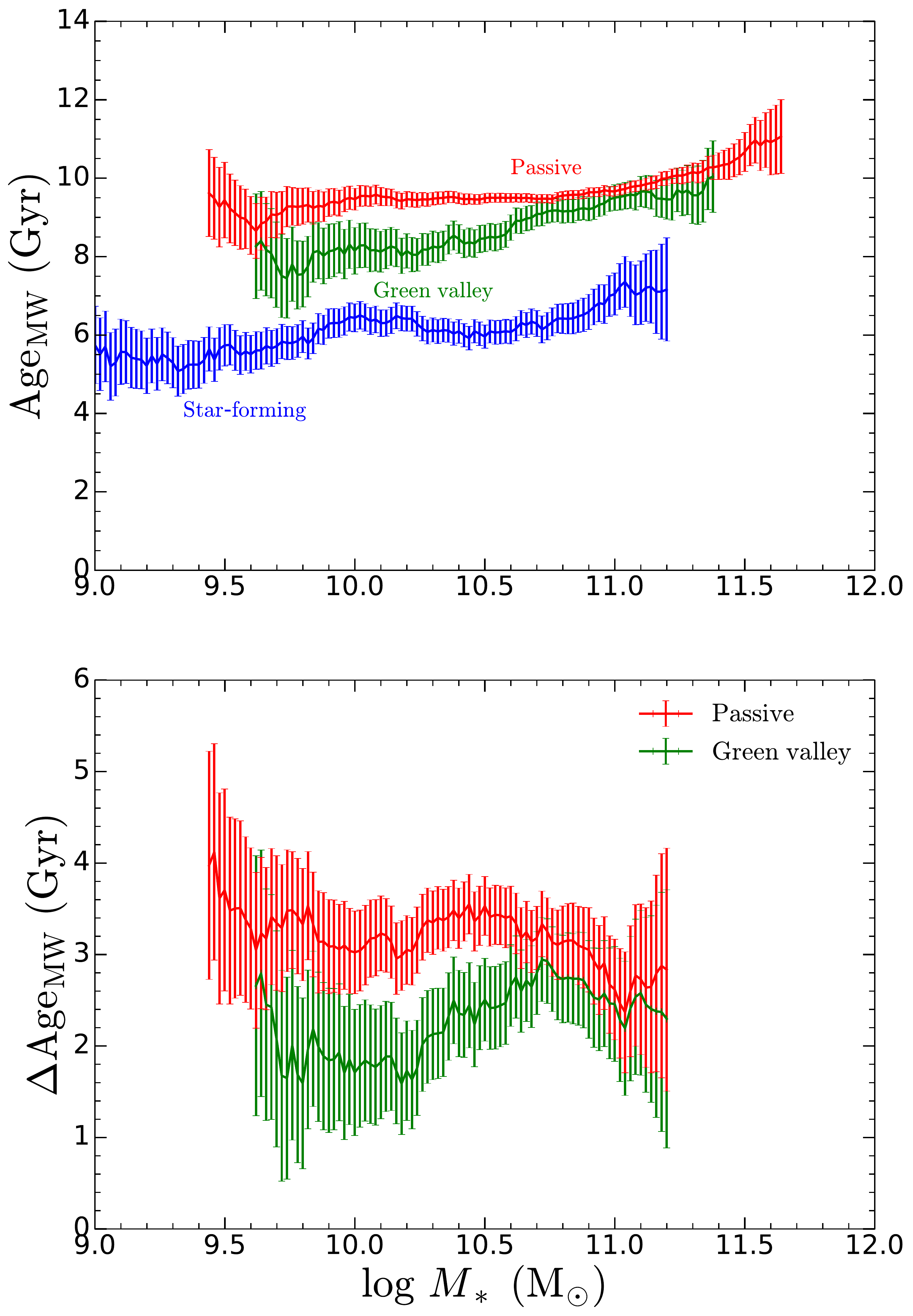}}
\caption{Top panel: The stellar mass--stellar age relation for star-forming (blue), green valley (green) and passive (red) galaxies. Bottom panel: The observed difference in stellar age between star-forming and passive galaxies (red), as well as the difference between star-forming and green valley galaxies (green).}
\label{fig:combined_mar_sgq}
\end{figure}

\section{Model} \label{sec:model}

In order to determine the role that different quenching mechanisms play in shutting down star formation, we compare the observed stellar metallicity differences between star-forming and passive (or green valley) galaxies with the predictions made by analytical models for galaxy evolution. In this section we discuss the models we use to determine how the stellar metallicity of a galaxy changes during quenching. There are three key factors that influence our model predictions. Firstly, our results depend on the differential equations that we solve to determine the evolution of the mass and metallicity of gas and stars in the galaxy. Secondly, we must estimate the cosmic epoch when the progenitors of local passive (or green valley) galaxies began quenching. Finally, we must specify the initial properties these star-forming progenitors had at the onset of quenching. 
 
\subsection{Differential Equations}\label{subsec:diff_eq}

We make use of the analytical framework developed in the gas regulator model of \citet{Peng2014b}, which does not assume any equilibrium
conditions for galaxy evolution. Star formation is regulated, near equilibrium by the mass of the gas reservoir, which itself is affected
by ongoing star formation, gas accretion and gas outflows. Furthermore, the ISM is enriched by metals produced through stellar
nucleosynthesis, it is diluted by accretion of gas from the IGM (assumed to be near pristine)
and metals are removed from the galaxy in galactic winds.

We use the instantaneous recycling approximation (IRA), which assumes that massive stars instantly die upon formation, returning some fraction of their (chemically enriched) gas to the ISM. Furthermore, this enriched material is instantaneously mixed uniformly with the gas in the ISM. On the other hand, low-mass stars are assumed to remain on the main sequence indefinitely, returning zero material to the ISM. Our model only tracks the global evolution of the stars and gas in the galaxy, and does not consider the spatial dependence of gas flows, metal enrichment or star formation. We also assume that the IMF does not change with time.

We parametrize star formation through an integrated, linear ($n=1$) Schmidt-Kennicutt law \citep{Schmidt1959, Kennicutt1998}, with the star formation rate $\Psi$ directly proportional to the gas mass $g$. That is, 
\begin{equation}
\Psi = \epsilon g,
\label{eq:sk_law}
\end{equation}
\noindent where $\epsilon$ is the `star formation efficiency'. Although there is still debate on whether the Schmidt-Kennicutt relation is linear or super-linear, we assume the linear approximation in our model. We use the total gas mass for $g$, which includes both the atomic and molecular components. Furthermore, we use the total star formation efficiency for $\epsilon$. We assume that $\epsilon$ remains constant during quenching. The star formation efficiency is related to the total gas depletion time-scale $t_{\mathrm{depl}}$, through $\epsilon = 1/t_{\mathrm{depl}}$. The depletion time-scale is defined as the time needed to convert the entire reservoir of gas in the galaxy into stars, assuming that the star formation rate remains fixed at the current value. We explore how our model predictions change when only the molecular gas component is considered in Appendix \ref{app:molec_gas}.

We assume that the mass outflow rate $\Lambda$ is directly proportional to the star-formation rate, 
\begin{equation}
\Lambda = \lambda _\mathrm{eff}\Psi,
\end{equation}
\noindent where $\lambda _\mathrm{eff}$ is the so-called outflow `effective' mass-loading factor. The term `effective'
refers to the outflowing mass that effectively escapes the galaxy (i.e. permanently removed from the system)
or which is reaccreted by the galaxy only on very long
time-scales ($\ge$ Hubble time). Outflowing gas which is reaccreted on to the galaxy on short time-scales 
is not accounted into the outflowing budget, as it is effectively
recycled for further star formation. Therefore, galaxies with prominent outflows, and with (`classical') outflow loading factor
$\lambda >0$,
may still have $\lambda _\mathrm{eff}=0$ if the outflow does not escape the galaxy or the halo \citep[as seems to be
the case for many massive galaxies,][]{Fluetsch2019}.

In our subsequent analysis we will be modelling quenching purely through starvation (no inflow and no `effective' outflows, i.e.
with $\lambda _\mathrm{eff}$ = 0), as well as quenching through a combination of starvation and outflows (i.e. no inflow
and $\lambda _\mathrm{eff} > 0$). 

Gas flows and star formation can change the amount of gas contained within a galaxy. Star formation and galactic winds deplete the gas reservoir, while gas accretion replenishes it. For a galaxy forming stars at a rate $\Psi$, ejecting gas at a rate $\Lambda$ and accreting material at a rate $\Phi$, the evolution of the gas mass is given by
\begin{equation}
\frac{\diffd g}{\diffd t} = -(1-R)\Psi - \Lambda + \Phi.
\label{eq:gas_mass}
\end{equation}
\noindent Here $R$ is the return fraction, which is the fraction of the mass of newly formed stars that is quickly returned (IRA) to the
ISM through stellar winds and supernovae. $1-R$ represents the fraction of mass that is locked up in long-lived stars and stellar
remnants. We assume $R=0.425$ \citep{Vincenzo2016a} in our analysis. 

Gas flows and star formation also affect the metal content of a galaxy. The metals produced by stellar nucleosynthesis enrich the ISM, while the accretion of pristine gas dilutes it and galactic winds remove metals from the galaxy. In general, the evolution of the gas metallicity $Z_g$ \citep{Tinsley1980} is given by 
\begin{equation}
g\frac{\diffd Z_g}{\diffd t} = (1-R)y\Psi - (Z_\Lambda - Z_g)\Lambda - (Z_g - Z_\Phi)\Phi,
\label{eq:gas_metallicity}
\end{equation}
\noindent where $y$ is the net yield, representing the amount of newly-forged metals released into the ISM per unit mass locked up in long-lived stars. $Z_\Lambda$ and $Z_\Phi$ are the metallicity of the outflowing and inflowing gas, respectively. 

\citet{Vincenzo2016a} showed that the value of the net yield is quite sensitive to the IMF, the IMF upper mass cutoff, as well as the set of stellar yields (which specify the dependence of the mass of metals released on the initial mass of a star) that are adopted. We assume the $y = 0.054$ value corresponding to the \citet{Kroupa2001} IMF, as this is the same IMF that was used to determine the observed stellar metallicities in {\footnotesize FIREFLY}. We do note that using a smaller yield value will affect our model predictions, as chemical enrichment is slower and so it will take longer to reproduce the observed stellar metallicity differences, resulting in a longer quenching time-scale. Furthermore, models using small $y$ values that incorporate outflows will have more difficulty reproducing the observed stellar metallicity differences. 

We will make the simplifying assumption that the metallicity of the outflowing gas and the metallicity of the ISM are equal ($Z_\Lambda =
Z_g$), i.e.\@ we assume that outflows do not preferentially remove metals from the galaxy. However, this assumption may not be completely
true for low-mass galaxies, as \citet{Lian2018a, Lian2018b} find that some metal-loading in the outflow (relative to the ISM metallicity) is required to simultaneously
match gas-phase and stellar metallicities. \cite{Vincenzo2016b} also find that preferential ejection of oxygen (and other core-collapse SNe
products) provides a better description of the observed chemical abundances and metallicities. This is also observed
in a few galactic outflows \citep[e.g.\@][]{Ranalli2008}. However, this differential effect is not major and our simplified analysis, which does not include this effect, is expected to provide a good description of the overall metallicity evolution, especially, obviously, for the
starvation scenario (in the outflow scenario the preferential ejection of metals makes an even stronger case for explaining the observed
metallicity difference between passive and star-forming galaxies in terms of starvation).

Furthermore, we will be investigating the effect that starvation (the halting of gas accretion) has on the evolution of stellar metallicities. Hence we will assume that there is no gas accretion during the quenching period and we set the inflow rate $\Phi = 0$. 

Under these set of assumptions, equations (\ref{eq:gas_mass}) and (\ref{eq:gas_metallicity}) are simplified. The evolution of the gas mass is now given by
\begin{equation}
\frac{\diffd g}{\diffd t} = -(1-R)\Psi - \Lambda.
\end{equation}
Furthermore, the evolution of the gas metallicity is now given by
\begin{equation}
\frac{\diffd Z_g}{\diffd t} = (1-R)y\epsilon,
\end{equation}
\noindent which continues to grow with time as stars return their metal-enriched gas to the ISM.

The stellar mass of the galaxy continuously increases due to the conversion of gas into stars. The evolution of the stellar mass $s$ is given by
\begin{equation}
\frac{\diffd s}{\diffd t} = (1-R)\Psi,
\end{equation}
\noindent where $(1-R)\Psi$ is the net star formation rate that contributes to the stellar mass increase of the galaxy.

Finally, the mass-weighted stellar metallicity rises with time as increasingly more metal-rich stars form out of the enriched gas. The evolution of the mass-weighted stellar metallicity is given by
\begin{equation}
\frac{\diffd Z_*}{\diffd t} = \frac{\Psi}{s}(1-R)(Z_g - Z_*).
\end{equation}
The stellar metallicity is a weighted average of the metallicities of all the stars in the galaxy. Since galaxies consist of a mixture of old metal-poor stars that formed early on out of pristine gas, as well as young metal-rich stars that recently formed out of more enriched gas, the mass-weighted stellar metallicity traces the cumulative chemical evolution of the galaxy. On the other hand, the gas metallicity only traces the current state of chemical enrichment. As a result, the mass-weighted stellar metallicity of a galaxy tends to lag behind the gas metallicity. 

\subsection{Estimating the onset of quenching}

Prior to the onset of quenching, the star-forming progenitors of local passive
galaxies probably evolved along the star-forming main sequence, where their gas
reservoir was kept relatively fixed due to the balance between the depletion of
gas driven by star formation and galactic winds and the replenishment of gas by accretion. However, these progenitors must have started quenching at
some cosmic epoch in order to form the passive galaxies we see in the local
Universe. In our model we assume that these progenitors began quenching through
starvation at a redshift $z_{\mathrm{q}}$, which is the epoch when the
accretion of gas is halted in the starvation scenario.

We assume that $z_{\mathrm{q}}$ depends on the stellar mass $M_*$ of the local passive
galaxies. Indeed, studies such as \citet{Thomas2005, Thomas2010} have shown how the
typical star formation epoch and time-scale depend on galaxy mass, finding that the more massive
galaxies tended to form the bulk of their stars earlier on in cosmic history than less
massive galaxies. Hence we would expect the progenitors of the most massive passive
galaxies to have started quenching at higher redshift than the progenitors of low-mass
passive galaxies. We use the mass-weighted stellar ages of local passive galaxies
$t_0(M_*)$ that were obtained in this work (see Fig.\@ \ref{fig:combined_mar_sgq}) to
estimate the epoch associated with the onset of quenching  $z_{\mathrm{q}}$. We make
the simplifying assumption that a negligible amount of additional stellar mass is
formed during the quenching phase, i.e.\@ that the gas mass available for additional
star formation is small compared with the mass of stars already assembled. Since
galaxies that quench through starvation form additional stars during the starvation
phase, the lookback times to the onset of quenching that we estimate here through
the mass-weighted ages are actually underestimates. We will revisit this caveat in Section \ref{subsubsec:gvg_pure}. Under this simplifying assumption there is a simple relationship between the stellar age of the local passive galaxy $t_0(M_*)$ and the stellar age of its star-forming progenitor at the onset of quenching $t(z_{\mathrm{q}}, M_*)$. The difference between these two ages is just given by the (mass-dependent) lookback time $t_\mathrm{lb}(M_*)$ to the onset of quenching. We have that 
\begin{equation}
t_0(M_*) = t(z_{\mathrm{q}}, M_*) + t_\mathrm{lb}(M_*).
\label{eq:passive_age_evolution}
\end{equation}
\noindent As discussed earlier, the star-forming progenitors of the most massive passive galaxies will have begun quenching at an earlier epoch in cosmic history, so these galaxies will have the largest lookback times. We therefore need to determine the epoch when these progenitors began quenching, $z_{\mathrm{q}}(M_*)$, which is equivalent to determining the lookback time to the onset of quenching, $t_\mathrm{lb}(M_*)$.

For local passive galaxies, it is reasonable to assume that the ages of their
star-forming progenitors were much smaller than the lookback time to the onset of quenching, i.e.\@
$t(z_{\mathrm{q}}, M_*) \ll t_\mathrm{lb}(M_*)$, as the stellar ages of the progenitors
are usually $<$ 1~Gyr (see e.g.\@ \citet{Reddy2012, Sklias2017} for massive galaxies at $z=1$--$3$, and \citet{Gallazzi2014} for low-mass galaxies at $z\sim 0.7$ if the observed relation is extrapolated down to lower masses), while the lookback times are usually $\sim$10~Gyr. In this case, the lookback times (and therefore $z_{\mathrm{q}})$ are simply given by $t_0(M_*)$, the stellar ages of the local passive galaxies. That is, we set $t_\mathrm{lb}(M_*) = t_0(M_*)$.

For local green valley galaxies, we have to be more careful, as these are still in the quenching phase. In this case the ages of the
progenitors are comparable to the lookback times, i.e.\@$\ $$t(z_{\mathrm{q}}, M_*)$
{$\sim$} $t_\mathrm{lb}(M_*)$. We estimate the lookback time,
$t_\mathrm{lb}(M_*)$, by assuming that the ages of the star-forming progenitors of the same mass, $t(z_{\mathrm{q}}, M_*)$, are given by the stellar ages of local star-forming galaxies, $t_{0, \mathrm{SFG}}(M_*)$. The lookback time to the onset of quenching is then given by the stellar age difference between local green valley galaxies and local star-forming galaxies of mass $M_*$, i.e.\@ $t_\mathrm{lb}(M_*) = t_{0, \mathrm{GVG}}(M_*) - t_{0, \mathrm{SFG}}(M_*)$. 

We show the redshifts $z_\mathrm{q}$ when the star-forming progenitors of local passive and local green valley galaxies began quenching through starvation in our models in Fig.\@ \ref{fig:quenching_epochs}.

\begin{figure}
\centering
\centerline{\includegraphics[width=1.0\linewidth]{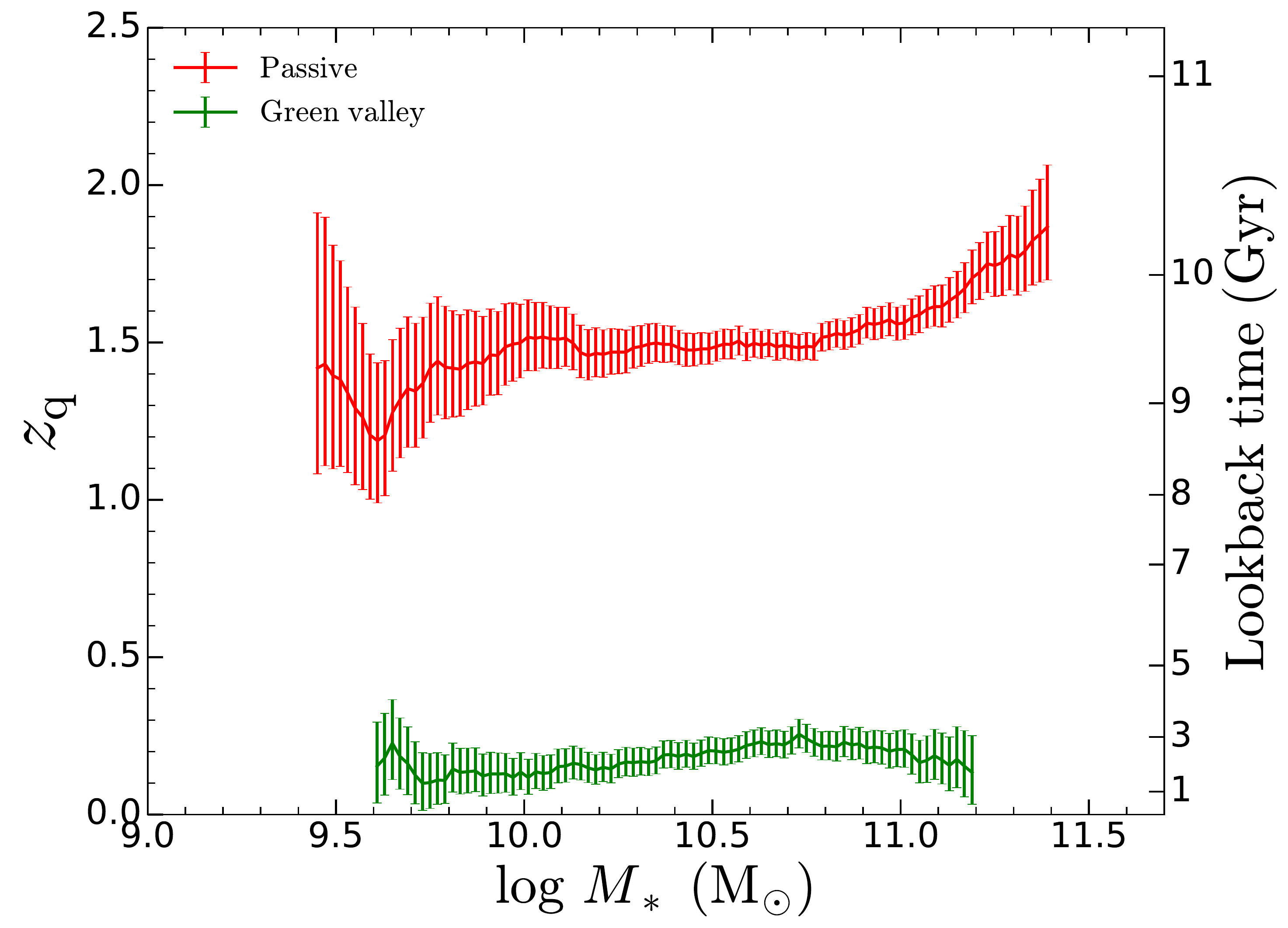}}
\caption{The redshift $z_\mathrm{q}$ when the star-forming progenitors of local passive galaxies (red) and local green valley galaxies (green) begin quenching through starvation, as a function of stellar mass, as inferred by our analysis. The onset of quenching for passive and green valley galaxies were estimated from the mass-weighted stellar ages of local passive galaxies, and the difference in mass-weighted stellar age between local green valley and local star-forming galaxies, respectively. }
\label{fig:quenching_epochs}
\end{figure}

\subsection{Initial conditions} \label{subsec:ic}

In order to model the change in stellar metallicity during quenching, the initial
state of the star-forming progenitor at the onset of quenching has to be specified. The
key initial conditions in our model are the gas mass $g$, the gas metallicity $Z_g$,
the stellar metallicity $Z_*$ and the star formation efficiency $\varepsilon$, i.e.\@
the star formation rate per unit gas mass
($\rm \varepsilon = SFR/M_{gas}$), often identified with the inverse of
the depletion time-scale $t_{\mathrm{depl}}$, i.e.\@ the time required by the star
formation to deplete all available gas into stars, in the ideal case in which
star formation remains
constant and gas is not replenished. We use mass- and redshift-dependent scaling relations for star-forming galaxies, that span the stellar mass range $\log (M_*/{\rm M_\odot}) = 9.4\text{--}11.3$ and redshift range $0 \leq z \leq 2.0$,  to determine the values for these quantities.  We try, whenever possible, to use empirical scaling relations as opposed to theoretical scaling relations in our models. A summary of the initial conditions and other important quantities used in our gas regulator model is provided in Table \ref{tab:grm_ic}.

{
\begin{table*}
\begin{center} 
\caption{Summary of the initial conditions and other key quantities used in the gas regulator model.}
\begin{tabular}{ p{1.5cm}  p{4cm}  p{5cm}  p{5cm} }
\hline
Quantity& Description& Passive galaxies& Green valley galaxies\\
\hline
$g$& Initial total gas mass (molecular + atomic) & Molecular gas mass: \citet{Tacconi2018}& Molecular gas mass: \citet{Boselli2014}\\
& & & \\
&& Atomic gas mass: \citet{Popping2014}& Atomic gas mass: \citet{Boselli2014}\\
& & & \\
& & Redshift-evolution: \citet{Tacconi2018}& Redshift-evolution: \citet{Tacconi2018}\\
\\ \\
$t_\mathrm{depl}$ (= $\epsilon ^{-1}$) & Total gas depletion time (molecular + atomic)& Molecular gas depletion time: \citet{Tacconi2018}& Molecular gas depletion time: \citet{Boselli2014}\\
& & & \\
& & Atomic gas depletion time: \citet{Popping2014}& Atomic gas depletion time: \citet{Boselli2014}\\
& & & \\
& & Redshift-evolution: \citet{Tacconi2018} & Redshift-evolution: \citet{Tacconi2018}\\
\\ \\
$Z_*$& Initial mass-weighted stellar metallicity & Local $Z_*$: $Z_\mathrm{MW}$ for star-forming galaxies from this work& Local $Z_*$: $Z_\mathrm{MW}$ for star-forming galaxies from this work\\
& & & \\
& & Redshift-evolution: \citet{Maiolino2008}& Redshift-evolution: \citet{Maiolino2008}\\
\\ \\
$Z_\mathrm{g}$& Initial gas-phase metallicity & 0.25 dex larger than $Z_*$& 0.25 dex larger than $Z_*$\\
\\ \\
$z_\mathrm{q}$& Redshift when the star-forming progenitor began quenching through starvation (i.e.\@ when the accretion of gas is halted)& Given by the mass-weighted age of local passive galaxies from this work& Given by the mass-weighted age difference between local green valley and star-forming galaxies from this work\\
\\ \\
$t_\mathrm{quench}$& Duration of quenching (i.e.\@ how long a star-forming progenitor must quench before its stellar metallicity is equal to the stellar metallicity of local passive/green valley galaxies) & Given by the time elapsed since the onset of quenching when $\qquad \qquad$ $Z_{*, \mathrm{model}} = Z_{*, \mathrm{passive}}$ &  Given by the time elapsed since the onset of quenching when $\qquad \qquad $ $Z_{*, \mathrm{model}} = Z_{*, \mathrm{green\ valley}}$ \\
& & & \\
& & Represents the time required to complete quenching & Represents the time elapsed since the onset of quenching (since green valley galaxies have not yet completely quenched)\\
\\ \\
$\tau_\mathrm{q}$& $e$-folding time for quenching (i.e.\@ the typical time-scale over which most of the star formation and metal enrichment takes place)& Using equation (\ref{eq:tau_q}), together with the $\lambda _\mathrm{eff}$ required to simultaneously reproduce the observed $Z_*$ and SFR in local passive galaxies & Using equation (\ref{eq:tau_q}), together with the $\lambda _\mathrm{eff}$ required to simultaneously reproduce the observed $Z_*$ and SFR in local green valley galaxies\\
\hline
\label{tab:grm_ic}
\end{tabular}
\end{center}
\end{table*}
}

In order to determine the total gas mass and star formation efficiency, we require
measurements of both the molecular and atomic gas components.
\citet{Tacconi2018} used a combination of CO and dust measurements to determine the
molecular gas properties of star-forming galaxies over a broad range in stellar mass
$\log (M_*/{\rm M_\odot}) = 9.0\text{--}11.9$ and redshift $z = 0\text{--}4.4$. We use their
relations for molecular gas fractions (giving molecular gas masses) and molecular gas
depletion time-scales (giving molecular star formation efficiencies) in our model. Due
to the intrinsic faintness of HI in emission, there are currently no atomic gas
measurements for star-forming galaxies within our required redshift range $0.5 \leq z
\leq 2.0$ \citep[see e.g.\@][]{Rhee2016, Cortese2017a}. We have chosen to use
theoretical relations between the molecular and atomic components of galaxies to
estimate atomic gas masses and atomic gas depletion times in our model, which are supported by indirect measurements. We use the
results from \citet{Popping2014}, who used pressure-based H$_2$ formation recipes to
determine the evolution of the molecular-to-atomic gas mass ratio $R_{\mathrm{mol}}$
across cosmic time, indicating that at high redshift, at least out to $z\sim3$,
most of the gas in a galaxy is in the molecular phase.
These results are in agreement with HI constraints given by the (lack of) evolution of damped $\mathrm{Ly}~\alpha$ absorption systems \citep{Prochaska2009}, indicating that the amount of HI in galaxies remains roughly constant out to $z\sim 3$, while the amount of molecular gas increases substantially \citep{Tacconi2018}. We estimate the atomic gas mass $a$ in our model galaxies from the molecular gas mass $m$ and $R_{\mathrm{mol}}$ by computing  $a = m/R_{\mathrm{mol}}$. In a similar fashion, the atomic depletion time $t_{\mathrm{depl, a}}$ is obtained from the molecular depletion time $t_{\mathrm{depl, m}}$ through $t_{\mathrm{depl, a}} = t_{\mathrm{depl, m}}/R_{\mathrm{mol}}$.

In contrast with local passive galaxies, which are thought to have mostly begun quenching at higher redshift, local green valley galaxies are a transitionary population that have only recently begun quenching. Therefore, in our analysis that investigates the quenching of star formation in local green valley galaxies, we use the local relations \citep{Boselli2014} for the total gas mass and total gas depletion time measured for star-forming galaxies in our models. Since our models estimate that the star-forming progenitors of these green valley galaxies began quenching between $0.15 \leq z_{\mathrm{q}} \leq 0.25$ (based off of the stellar age difference between local star-forming and green valley galaxies), we evolve the $z=0$ relations to this redshift range, using the redshift-dependence of the gas mass and gas depletion time-scale measured by \citet{Tacconi2018}.

Deep observations of galaxies have begun to probe the stellar mass--stellar
metallicity relation at high redshift, such as the works by \citet{Choi2014} at $z\sim0.4$, \citet{Gallazzi2014} at
$z=0.7$, \citet{Lonoce2015} and \citet{Onodera2015} at $z\sim1.5$, \citet{Halliday2008} at $z=2$, \citet{Sommariva2012} at $z=3$ and
\citet{Lian2018c}. However, current studies of the stellar metallicity
relation at high redshift do not yet have the broad coverage in stellar mass and
redshift, nor the statistics (resulting in very large scatter) that is required by our
models. Therefore, we do not use the observed stellar mass--stellar metallicity
relations at high redshift to determine the initial stellar metallicities in our model.
Instead, we make the assumption that the gas MZR and stellar MZR of star-forming
galaxies evolve similarly across cosmic time \citep{Peng2014b}. In that case the evolution of stellar metallicity with redshift is simply given by the evolution of the gas metallicity. We have used the redshift evolution of the gas MZR measured by \citet{Maiolino2008} to evolve the local relation for mass-weighted stellar metallicities to higher redshift, and this is shown in Fig.\@ \ref{fig:mzr_progenitor}. We investigate the validity of our assumed evolution of the stellar MZR in Section \ref{subsec:mzr_evo}. 

Studies of gas metallicities can suffer from calibration issues, where different
diagnostics can yield significantly different metallicities, resulting in
mass--metallicity relations with different shapes and different normalisations
\citep[e.g.\@][]{Kewley2008, Curti2017}. We wish to avoid any issues associated
with metallicity calibrations in our analysis. We could in principle use the stellar
metallicities from {\footnotesize FIREFLY} and the gas metallicities from
\citet{Maiolino2008}, who estimated the evolution of the MZR as a function of redshift by using consistent calibrations within the gas-phase metallicity diagnostics.
However, we would then be using two separate metallicity
calibrations for gas and stars, which are likely to be inconsistent with one another.
We restrict our
analysis to a single calibration scheme and do not directly use the gas metallicities from
\citet{Maiolino2008} in our model. Instead, we use only the metallicity evolution (i.e.\@ $\Delta \log Z_g / \Delta z$)
inferred by \cite{Maiolino2008} and renormalize it locally using the stellar
metallicities obtained by {\footnotesize FIREFLY}.

Observational and theoretical studies have shown that there is typically a 0.2--0.3 dex difference between the gas-phase metallicity and stellar metallicity in star-forming galaxies \citep[e.g.\@][]{Finlator2008, Halliday2008, Yates2012, Yates2014, Peng2014b, Pipino2014, Ma2016, DeRossi2017b, Lian2018c}. This result seems to apply across a broad range in stellar mass and in redshift. Hence we set the initial gas metallicity to be 0.25 dex larger than the initial mass-weighted stellar metallicity in our models. 

\indent 
For local descendants (i.e.\@ passive or green valley galaxies) of stellar mass $M_*$, we make the simplifying assumption that their progenitors also had a stellar mass $M_*$ at the onset of quenching. We investigate how the assumed stellar mass offset (which will affect the initial properties of the star-forming progenitors) between progenitors and descendants can affect our results in Section \ref{subsec:prog_mass_offset}. Note that this is in contrast to our analysis in Section \ref{sec:quenching_mechanism} which focusses on the change in galaxy properties (i.e.\@ stellar mass and stellar metallicity) during quenching and does take the stellar mass evolution into account. The initial state [$g, \epsilon, Z_g, Z_*$] of a star-forming progenitor of stellar mass $M_*$ that begins quenching at $z_{\mathrm{q}}$ is then given by evaluating the scaling relations discussed above at ($M_*, z_{\mathrm{q}}$).

\section{Quenching Mechanisms} \label{sec:quenching_mechanism}

In this section we compare the observed differences in stellar metallicity between star-forming and passive (or green valley) galaxies with the predictions from gas regulator models to put constraints on the relative role of different quenching mechanisms. We will investigate closed-box models, where galaxies quench purely through starvation and with no outflows ($\lambda _\mathrm{eff}= 0$), but will also consider leaky-box models, where galaxies quench through a combination of starvation and outflows ($\lambda _\mathrm{eff}> 0$). This will allow us to study how the relative role of different quenching mechanisms, as well as the associated quenching time-scale depends on stellar mass. On the one hand, we will study how the star-forming progenitors of local passive galaxies quenched, providing insights into the processes responsible for shutting down star formation at high redshift ($z\sim1$--$2$). On the other hand, we will also address the quenching of star formation in the local Universe, by studying local green valley galaxies that are currently in the process of quenching. Combined, these two studies enable us to probe the evolution of galaxy quenching across cosmic time. 

\subsection{Evolution of SFR and $Z_*$ during the quenching phase} 

In this subsection, we briefly introduce and highlight the differences between the closed-box and leaky-box models that will be explored in the remainder of this section. 

We will consider two different leaky-box models. In Sections \ref{subsubsec:qg_of} and \ref{subsubsec:gvg_of}, we assume that $\lambda_{\mathrm{eff}}=1$ for galaxies of all stellar masses. In Section \ref{subsubsec:qg_sfr} (Section \ref{subsubsec:gvg_sfr}), we instead find the values of $\lambda_\mathrm{eff}$ for which our model simultaneously reproduces the $Z_*$ and SFR seen in passive (green valley) galaxies. 

We schematically show the evolution of the star formation rate and the evolution of the logarithmic stellar metallicity during the quenching phase in our models in Fig.\@ \ref{fig:sfr_and_z_evolution}. The main aspects of these two quenching models is summarised in the following. Prior to the onset of quenching, the galaxy evolves along the star-forming main sequence, where its gas reservoir is in quasistatic equilibrium, so the star formation rate increases slowly, and the stellar metallicity increase is modest due to dilution by accretion of pristine gas. At a time $t(z_\mathrm{q})$, the accretion of cold gas is halted and the galaxy begins quenching through starvation. The blue curve shows the evolution in the case of pure starvation ($\lambda _\mathrm{eff} = 0$) that we will explore in Section \ref{subsubsec:qg_pure}, while the red curve shows the evolution in the case of starvation with outflows ($\lambda _\mathrm{eff} > 0$) which we will explore in Sections \ref{subsubsec:qg_of} and \ref{subsubsec:qg_sfr}. In the absence of gas accretion, the star formation rate declines exponentially according to 
\begin{equation}
\Psi (t) = \Psi _0e^{- \frac{t}{\tau _\mathrm{q}}},
\end{equation}
as gas is converted into stars and removed in galactic winds. Therefore, $\tau _\mathrm{q}$ represents the $e$-folding time-scale over which the star formation rate decreases by a factor of $e$, and is given by
\begin{equation}
\tau _ \mathrm{q} = \frac{1}{\epsilon(1-R+\lambda _\mathrm{eff})}.
\label{eq:tau_q}
\end{equation}
The $e$-folding time-scale for the model incorporating outflows $\tau _\mathrm{q}^\prime$ is shorter than for the model including only starvation $\tau _\mathrm{q}$, as the additional gas that is lost through galactic winds causes the star formation rate to decline more quickly. In the pure starvation scenario, the galaxy quenches for a time-scale $t_\mathrm{quench}$ before it has a stellar metallicity that matches the stellar metallicity observed for a local passive galaxy of the same stellar mass. However, as we shall see in Section \ref{subsubsec:qg_pure}, these galaxies still harbour a relatively large gas reservoir, and so continued star formation beyond $t = t(z_\mathrm{q}) + t_\mathrm{quench}$ in our model would result in a galaxy that is too metal-rich (as shown by the dashed blue curves). In order to prevent further star formation and chemical enrichment, an ejective or heating mode is required, which completely quenches the galaxy, reducing the star formation rate to zero and keeping the stellar metallicity constant at a fixed value. In the starvation with outflows scenario, the quenching phase has a duration $t_\mathrm{quench}^{\prime}$, at which point, for the model explored in Section \ref{subsubsec:qg_sfr}, both the stellar metallicity and the star formation rate of the galaxy are similar to those seen in local passive galaxies of the same stellar mass.
 
\begin{figure}
\centering
\centerline{\includegraphics[width=1.0\linewidth]{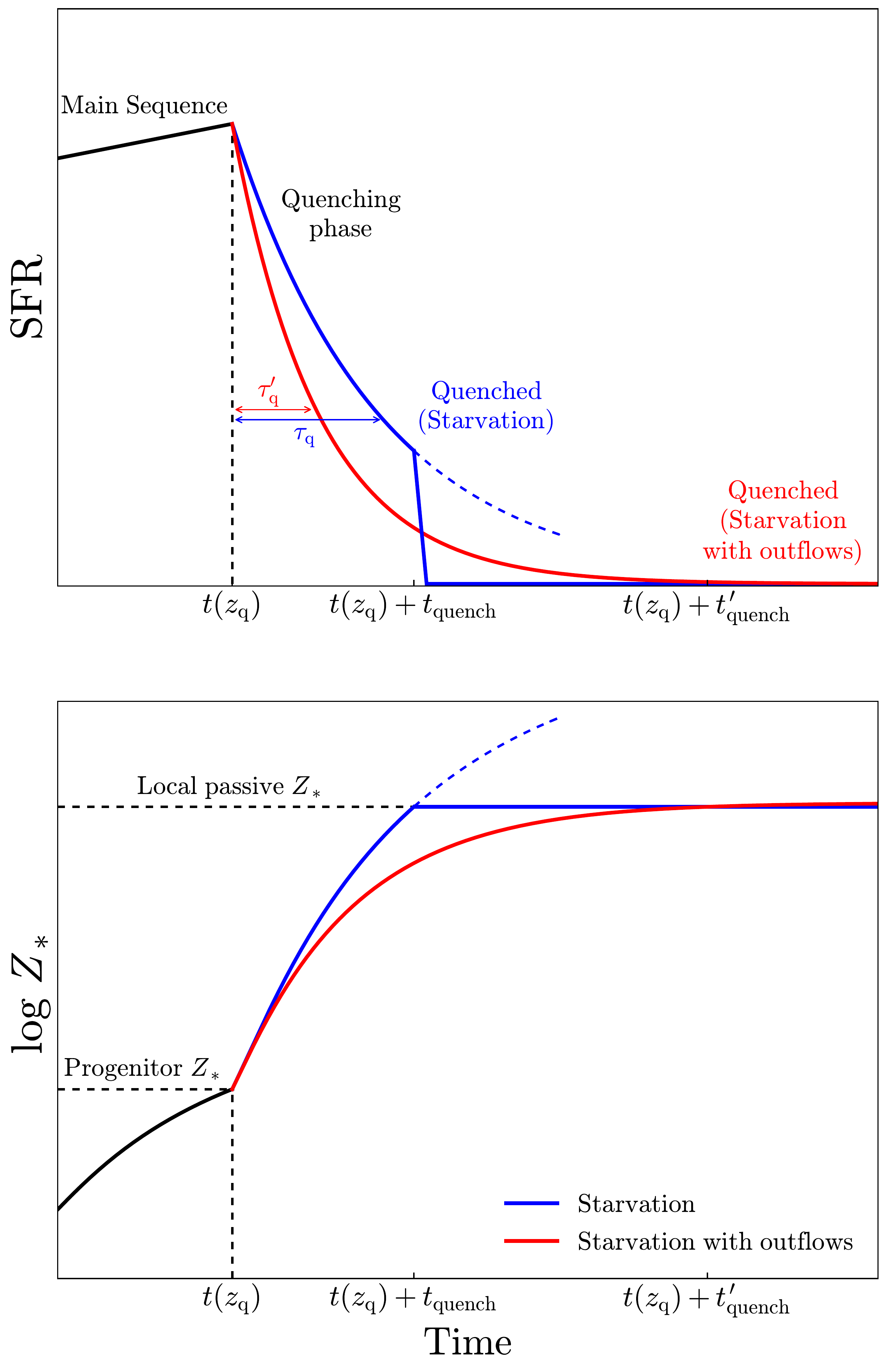}}
\caption{A schematic illustration of the evolution of the star formation rate (SFR, top panel) and the evolution of the logarithmic stellar metallicity ($\log Z_*$, bottom panel) during the quenching phase in our models. The galaxy initially evolves along the star-forming main sequence. At a time $t(z_\mathrm{q})$, the accretion of cold gas is halted and the galaxy begins quenching through starvation ($\lambda _\mathrm{eff} = 0$, blue) or through starvation with outflows ($\lambda _\mathrm{eff} > 0$, red). In the starvation scenario, the galaxy quenches for a time-scale $t_\mathrm{quench}$ before it reaches the level of chemical enrichment seen in local passive galaxies, at which point the onset of an ejective or heating mode prevents any further star formation and chemical enrichment, and the galaxy is quenched. In the starvation with outflows scenario, after a time $t_\mathrm{quench}^\prime$ has elapsed the galaxy has completed quenching, and, for our analysis in Section \ref{subsubsec:qg_sfr}, both its stellar metallicity and star formation rate are similar to that seen in local passive galaxies. $\tau _\mathrm{q}$ and $\tau _\mathrm{q}^\prime$ represent the $e$-folding time-scales in the starvation, and starvation with outflows scenarios, respectively.}
\label{fig:sfr_and_z_evolution}
\end{figure}

\subsection{Passive galaxies (quenching at high-z)} \label{subsec:qg_models}

We begin by investigating the processes responsible for quenching the star-forming progenitors
of local passive galaxies. Since these passive systems stopped forming stars early on in
cosmic history, our analysis in this subsection addresses the quenching of star formation at
high redshift. We focus on reproducing the observed stellar metallicity differences $\Delta
Z_*$ between star-forming and passive galaxies with our models. The observed $\Delta Z_*$ that
we will use in the subsequent analysis corresponds to the stellar metallicity difference
between local passive galaxies and their star-forming progenitors at high redshift, rather
than the observed $\Delta Z_*$ between local passive galaxies and local star-forming galaxies.
It should be noted that our approach is different to what was done in \citetalias{Peng2015},
who instead analysed the observed $\Delta Z_*$ between local passive galaxies and local
star-forming galaxies. As was discussed in Section \ref{subsec:ic}, the stellar metallicities
of the star-forming progenitors are estimated by evolving the stellar metallicities of local
star-forming galaxies to higher redshift using the cosmic evolution of the mass-metallicity
relation described by \citet{Maiolino2008}. The observed stellar metallicities of local
passive galaxies and our estimates for the stellar metallicities of their star-forming
progenitors are shown in Fig.\@ \ref{fig:mzr_progenitor}. Since star-forming galaxies at higher
redshift are less metal-rich than their local counterparts, the observed stellar metallicity
differences between local passive galaxies and their progenitors that are studied in this
section are even larger than the metallicity differences between local passive and local star-forming galaxies seen in Fig.\@ \ref{fig:combined_mzr_sgq}. 

\begin{figure}
\centering
\centerline{\includegraphics[width=1.0\linewidth]{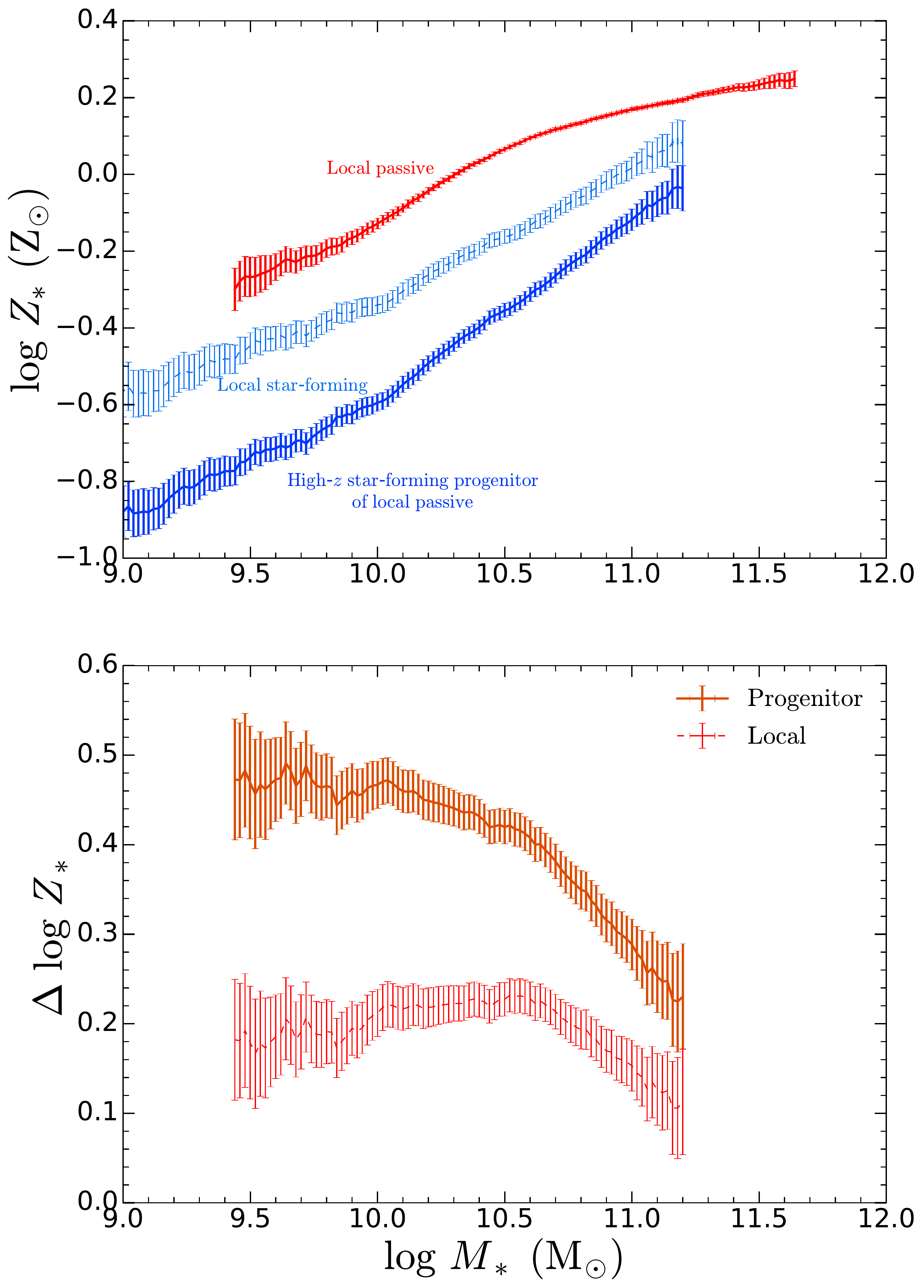}}
\caption{Top panel: The stellar mass--stellar metallicity relation observed for local passive galaxies (red) and local star-forming galaxies (light blue), as well as our estimates for the stellar metallicities for the star-forming progenitors of local passive galaxies (dark blue). Bottom panel: The observed difference in stellar metallicity between local star-forming and local passive galaxies (red) and the estimated difference in stellar metallicity between local passive galaxies and their star-forming progenitors at higher redshift (orange).}
\label{fig:mzr_progenitor}
\end{figure}

\subsubsection{Pure starvation} \label{subsubsec:qg_pure}

We first consider closed-box models for galaxy evolution, where galaxies quench purely through starvation and no outflows are included ($\lambda _\mathrm{eff} = 0$). A comparison between the observed and model-predicted stellar metallicity differences is shown in the top panel of Fig.\@ \ref{fig:qg_model_deltaZ_qt_lambda_0}. The horizontal axis refers to the initial stellar mass $M_*$ of the star-forming progenitor. The observed stellar metallicity difference between star-forming and passive galaxies of the same stellar mass is given by the dashed black curve. If galaxies formed a negligible amount of additional stellar mass $\Delta M_*$ during the quenching phase (i.e.\@ if galaxies followed a $\sim$vertical trajectory in the top panel of Fig.\@ \ref{fig:mzr_progenitor}), then this would be the amount of chemical enrichment that takes place during the quenching phase. However, star-forming galaxies at high-redshift are gas rich, and, if they quench through starvation, can convert a considerable portion of this gas into new stars, resulting in a non-negligible increase in stellar mass during quenching (i.e.\@ galaxies actually follow a diagonal trajectory in Fig. \ref{fig:mzr_progenitor}). Hence the actual increase in stellar metallicity during the quenching phase is larger than in the $\Delta M_* = 0$ case, since the stellar metallicity of passive galaxies increases with increasing stellar mass. In our models, the quenching phase ends when the metallicity of the model galaxy $Z_\mathrm{*, model}$ with stellar mass  $M_* + \Delta M_*$ is the same as that of a local passive galaxy with that same mass, i.e.\@ when $Z_\mathrm{*, model}(M_* + \Delta M_*) = Z_\mathrm{*, PG}(M_* + \Delta M_*)$. The black solid curve shows the difference in stellar metallicity between star-forming progenitors and their local passive descendants, after taking the increase in stellar mass that occurs during the quenching phase into account.

Fig.\@ \ref{fig:qg_model_deltaZ_qt_lambda_0} also shows the model-predicted stellar metallicity enhancements at various times $\Delta t$ after the onset of quenching, given by the coloured curves. The stellar metallicity enhancements grow with time as successively more metal-rich stars form out of the ISM, which increases the average mass-weighted stellar metallicity of the galaxy. Our models predict that the rate of stellar metallicity enrichment ($\Delta{\log Z_*} / \Delta t$) decreases with increasing mass, where the change in stellar metallicity in a given time step is smallest for the most massive galaxies. This is primarily because gas fractions decrease with increasing stellar mass, so the most massive galaxies have the smallest gas fractions. As a result, these massive galaxies form relatively fewer new stars compared to the population of stars already present, and so their stellar metallicity increases more slowly. 

\begin{figure}
\centering
\centerline{\includegraphics[width=1.0\linewidth]{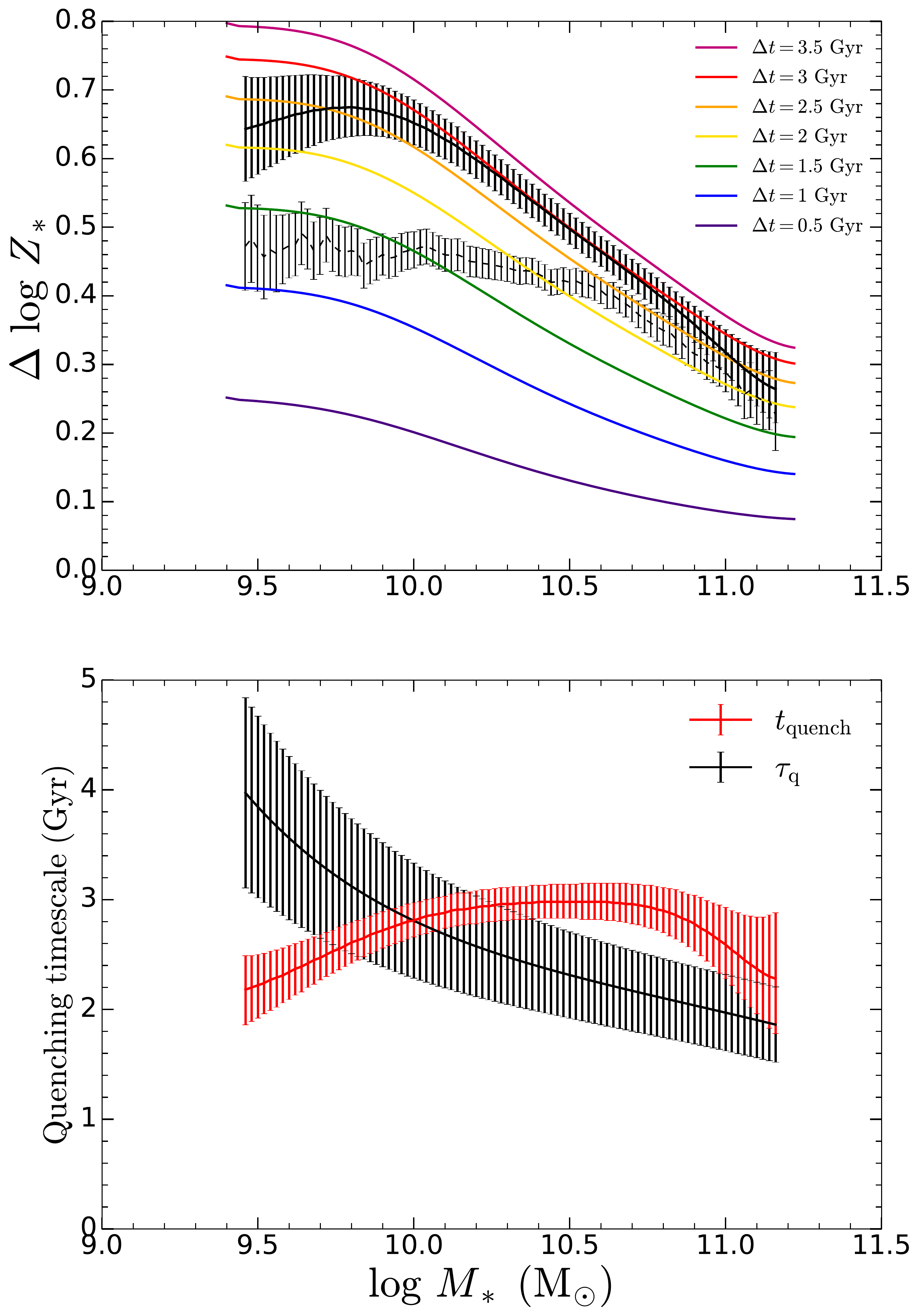}}
\caption{Top panel: The dashed black curve shows the observed stellar metallicity difference between local passive galaxies and their star-forming progenitors of the same stellar mass. The solid black curve shows the actual increase in stellar metallicity during the quenching phase (taking into account the increase in stellar mass of the galaxy during quenching). The coloured lines show the stellar metallicity difference predicted by a closed-box model ($\lambda _\mathrm{eff} = 0$) at different times after the onset of starvation. The horizontal axis refers to the initial stellar mass of the star-forming progenitor. Bottom panel: The quenching time-scales $t_\mathrm{quench}$ derived from the stellar metallicity difference between star-forming and passive galaxies are shown in red, where the error bars take into account the 1$\sigma$ errors on the stellar metallicity differences. The $e$-folding timescales for star-formation $\tau_\mathrm{q}$ during quenching are shown in black, where the error bars are due to the uncertainties on the depletion times from \citet{Tacconi2018}.}
\label{fig:qg_model_deltaZ_qt_lambda_0}
\end{figure}

From Fig.\@ \ref{fig:qg_model_deltaZ_qt_lambda_0}, we see that closed-box models are easily able
to reproduce the observed stellar metallicity differences represented by the solid black curve. The models typically take about 2--3~Gyr to reach the level of chemical enrichment seen in local passive galaxies (following a diagonal trajectory in Fig.\@ \ref{fig:mzr_progenitor}), which suggests that the progenitors of local passive galaxies up to $M_* \sim 10^{11.2}~{\rm M_\odot} $ that quenched purely through starvation did so over a 2--3~Gyr time-scale. We study this mass-dependence of the duration of quenching in more detail in the bottom panel of Fig.\@ \ref{fig:qg_model_deltaZ_qt_lambda_0}. We define the duration of quenching $t_{\mathrm{quench}}$ as the time-scale over which a galaxy quenching through starvation enriches in stellar metallicity by an amount equal to the observed stellar metallicity difference between star-forming and passive galaxies, after accounting for the increase in stellar mass during quenching. In other words, $t_{\mathrm{quench}}$ is the time when our model-predicted enrichment equals the observed stellar metallicity difference. We find that the duration of quenching (shown in red) is between 2--3 Gyr, growing for $\log (M_*/\mathrm{M_\odot}) < 10.5$ and then declining. The error bars on $t_\mathrm{quench}$ reflect the $1\sigma$ error on the stellar metallicity differences. We also show (in black) the $e$-folding timescale for star-formation $\tau_\mathrm{q}$, which decreases with increasing mass due to the increase of total star formation efficiency $\epsilon$ with mass in our models \citep[which also matches observations, e.g.\@][]{Boselli2014}. The error bars are due to the uncertainties on the depletion times from \citet{Tacconi2018}.

\citet{Thomas2010} measured the $\alpha$-enhancements in local early-type galaxies and used this information to put constraints on the typical time-scale over which the progenitors of these local passive galaxies formed the bulk of their stars. They find that the progenitors of more massive galaxies formed most of their stars at earlier epochs and over shorter time-scales than their less massive counterparts, which is qualitatively consistent with the trend shown by $\tau_\mathrm{q}$. However, at the low-mass end, we find that quenching ends even before one $e$-folding timescale has elapsed, which is rather inconsistent with the continuously declining star formation histories in \citet{Thomas2010}. Furthermore, the 2~Gyr timescale is substantially shorter than the star formation time-scales in \citet{Thomas2010}. These results potentially suggest that the progenitors of low-mass passive galaxies did not quench purely through starvation. We explore alternative scenarios, where galaxies quench through a combination of starvation and outflows later in this section.

We note that the quenching time-scales $t_{\mathrm{quench}}$ of 2--3~Gyr that we obtain are smaller than the 4~Gyr that was found by \citetalias{Peng2015}. There are a number of reasons for this. On the observational side, we use a different set of stellar metallicities to the original work. We explore this point in more detail in Section \ref{sec:caveats}. On the modelling side, we utilise different scaling relations to specify the initial conditions of the star-forming progenitors in our model, as well as an alternative method to estimate the cosmic epoch when these progenitors begin quenching. In addition, we also distinguish between quenching time-scale  $t_\mathrm{quench}$ and epoch of the quenching onset $z_\mathrm{q}$. In our model, the quenching time-scale refers to the duration of quenching i.e.\@ our findings suggest that galaxies quench purely through starvation for 2--3~Gyr, at which point quenching has completed. On the other hand, in \citetalias{Peng2015}, the quenching time-scale refers to the amount of time that has elapsed since the onset of quenching through starvation, i.e.\@ their results suggest that we are typically seeing passive galaxies 4~Gyr after the onset of quenching through starvation. 

\citetalias{Peng2015} also compared their time-scale of 4~Gyr from the stellar metallicity analysis with the difference in stellar age $\Delta \mathrm{age}$ between star-forming and passive galaxies. If one assumes that a negligible amount of stars are formed (i.e.\@ passive evolution) during the starvation phase, then this stellar age difference represents the time elapsed since the onset of quenching (which is exactly what $t_\mathrm{quench}$ measures in the \citetalias{Peng2015} model). They find this age difference to be mass-independent at  $\sim$4~Gyr, which is consistent with the time-scale that was obtained from the stellar metallicity difference analysis. We do not make this comparison between $t_\mathrm{quench}$ and $\Delta \mathrm{age}$ for passive galaxies in our work. This is because our $t_\mathrm{quench}$ measures the time-scale required to complete quenching, whereas  $\Delta \mathrm{age}$ measures the time elapsed since the onset of quenching, so these two quantities trace different time-scales and are therefore not comparable. To clarify, after a galaxy completes quenching (i.e.\@ it becomes passive), $t_\mathrm{quench}$ remains fixed, but its $\Delta \mathrm{age}$ continues to increase indefinitely as more time elapses, and so these two quantities begin to diverge. We note that we will make a comparison between $t_\mathrm{quench}$ and $\Delta \mathrm{age}$ for green valley galaxies in Section \ref{subsec:gvg}. Unlike passive galaxies, green valley galaxies are still in the process of quenching and so $t_\mathrm{quench}$ actually measures the time since the onset of quenching (rather than the time to complete quenching), so a meaningful comparison between $t_\mathrm{quench}$ and $\Delta \mathrm{age}$ can be made. 

Our closed-box model suggests that the progenitors of local passive galaxies quenched through starvation for 2--3~Gyr, as that is the time required to reproduce the observed stellar metallicity differences. However, we can see from Fig.\@ \ref{fig:qg_model_deltaZ_qt_lambda_0} that the model-predicted curves can exceed the observed stellar metallicity differences, as they continue to grow even after 2~Gyr of quenching through starvation has elapsed. There is clearly still a substantial increase in the stellar metallicity enhancement $\Delta Z_*$ during the subsequent timesteps $\Delta t$ after the observed stellar metallicity difference has been reached. If these galaxies were actually quenched, then there should only be a residual amount of chemical enrichment after 2--3~Gyr as all the gas has been exhausted, and there is only a marginal amount of star formation remaining. Evidently, the `quenched' galaxies in our closed-box model still have a relatively large gas reservoir and are still forming a considerable number of stars. This point is made clearer when comparing the duration of quenching $t_{\mathrm{quench}}$ with the $e$-folding timescale $\tau_\mathrm{q}$. At most, only $\sim$1 $e$-folding timescale has elapsed, with roughly $e^{-0.5}\approx 60\%$ and $e^{-1} \approx 37\%$ of the initial gas reservoir still present for galaxies of $\log (M_*/\mathrm{M_{\odot}}) = 9.5$ and $11.0$, respectively. The properties of the 'quenched' galaxies in our models are therefore in stark contrast with the observed properties of local passive galaxies, which have small gas reservoirs and are not actively forming stars. Furthermore, since the `quenched' galaxies in our model continue to form stars and become progressively more enriched, their stellar metallicities begin to exceed the stellar metallicities seen in local passive galaxies, as shown by the top panel in Fig.\@ \ref{fig:qg_model_deltaZ_qt_lambda_0}. \\
\indent 
In order to prevent this prolonged chemical enrichment in our closed-box model, an additional mechanism must kick in, when the observed stellar metallicity difference has been reached ($\sim$2--3~Gyr), which abruptly stops any further star formation and enrichment. This mechanism could be some form of an ejective mode or heating mode that starts to play a role near the end of the starvation phase, to prevent further chemical enrichment from taking place. There are a number of mechanisms that could eject the remaining gas in the galaxy at the end of the starvation phase. For example, an AGN that is being activated could clear or heat the gas \citep[e.g.\@][]{Ciotti2007,Ciotti2009}. Satellite galaxies plunging through the hot ICM could rapidly lose their gas through ram pressure stripping, and this may happen only when the galaxy enters deep into the cluster \citep{Muzzin2014}. However, it should be noted that our sample of SDSS galaxies consists mostly of centrals, see \citet{Yang2007}. Alternatively, the integrated feedback from many Type Ia supernovae could eject or heat the remaining gas \citep{Pipino2004, Matteucci2006, Pipino2006, Pipino2008}. Since supernovae are more effective at driving away material in low density conditions, this mechanism is especially effective at ejecting gas at the end of the starvation phase, when the gas content has diminished and the densities have fallen. Moreover, Type Ia SNe have a well defined time-scale ($\sim$1--2~Gyr) before they start contributing significantly to the energy injection into the ISM, which would explain well why the additional ejective/heating mode kicks in only around 2.5~Gyr after the last episode of major star formation, i.e.\@ after the onset of quenching. \\
\indent
To summarise, using a closed-box model we find that the progenitors of local passive galaxies quenched purely through starvation over a time-scale of 2--3~Gyr, and then additional star formation and additional chemical enrichment was abruptly halted by the onset of an ejective or heating mode.  However, as an alternative, our results may also suggest that the progenitors of local passive galaxies did not quench purely through starvation. Indeed, other models, such as a combination of starvation and outflows during the quenching, may potentially provide a better description of the quenching process as discussed in the following sections.

\subsubsection{Starvation with outflows}\label{subsubsec:qg_of}

We now study leaky-box models for galaxy evolution, where galaxies quench through a combination of starvation and outflows. In this section we assume the outflow mass loading factor of one classically assumed by many models, i.e.\@ $\lambda _\mathrm{eff}= 1$. Our results are shown in Fig.\@ \ref{fig:qg_model_deltaZ_qt_lambda_1}. The main effect of introducing outflows in our model is that the gas reservoir depletes faster than in the closed-box case, as now both star formation and galactic winds drain the gas content. This has two main consequences for our model predictions. Firstly, since the gas reservoir depletes more rapidly, the star formation rate also declines faster and less stars are formed within a given time interval. As a result, the stellar metallicity increases more slowly in the leaky-box model. Hence any quenching time-scales we derive in our analysis (i.e.\@ the time derived to match $\Delta Z_*$) will be longer than in the closed-box model. Secondly, only a fraction of the initial gas reservoir is now converted into new, metal-rich stars. The rest of the gas is expelled, so the amount of material that is actually available to increase the stellar metallicity of the galaxy has been reduced. As a result, the maximum stellar metallicity increase of the galaxy, which occurs when the gas reservoir is completely depleted, is lower in the leaky-box models. Hence it may become more difficult for our model to reproduce the stellar metallicity differences that have been observed.

\begin{figure}
\centering
\centerline{\includegraphics[width=1.0\linewidth]{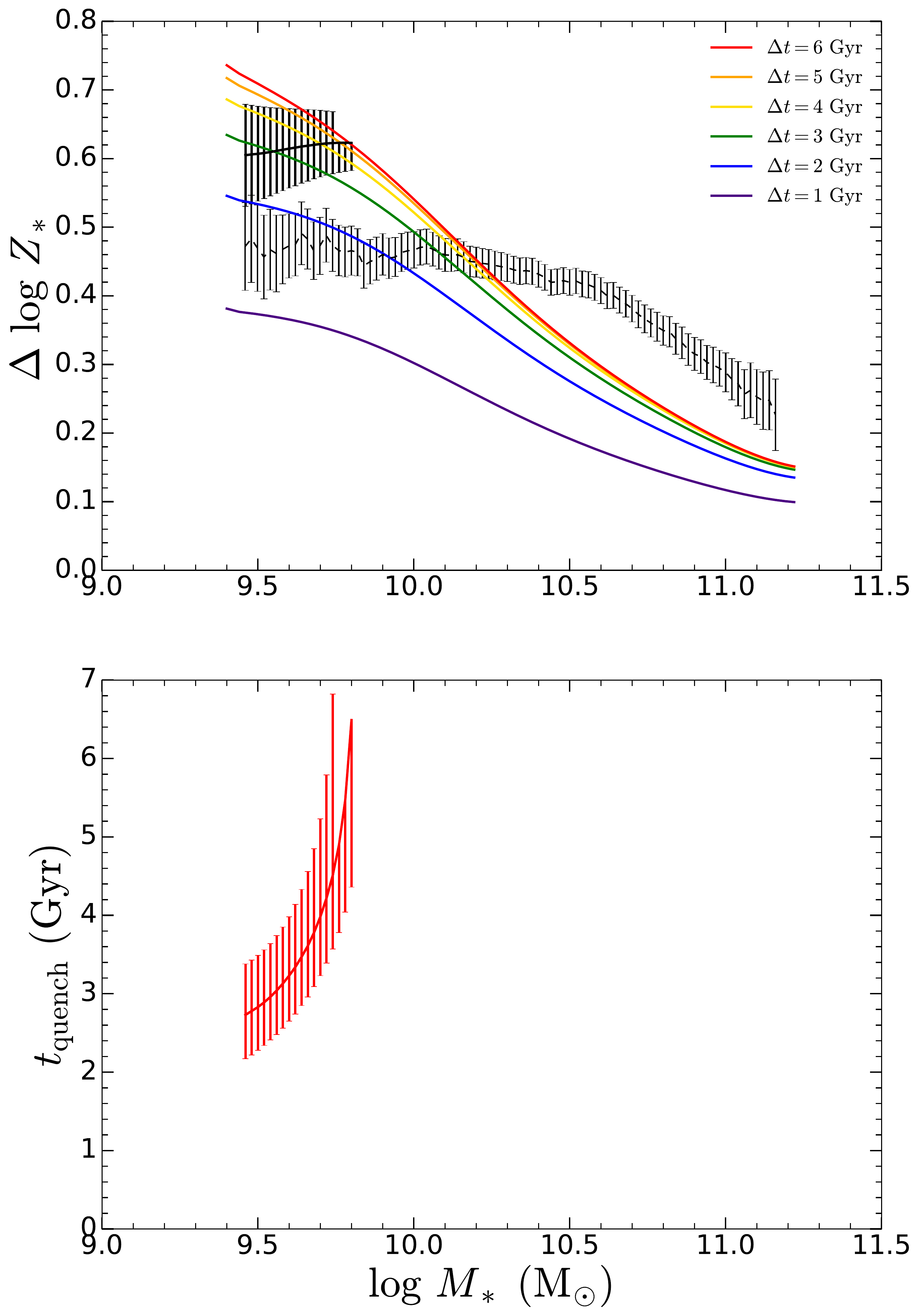}}
\caption{Similar to Fig.\@ \ref{fig:qg_model_deltaZ_qt_lambda_0}, but we now consider a leaky-box model, incorporating outflows with $\lambda _\mathrm{eff} = 1$. Also note that the timesteps in our model are now 1~Gyr (instead of the 0.5~Gyr used in Fig.\@ \ref{fig:qg_model_deltaZ_qt_lambda_0}). We omit the upper limit on the increase in stellar metallicity during quenching $\Delta Z_*$ (given by the solid black curve) when it cannot be reproduced by our model, and in that case we only show the lower limit on the duration of quenching $t_\mathrm{quench}$.} 
\label{fig:qg_model_deltaZ_qt_lambda_1}
\end{figure}

We find that models incorporating outflows are also capable of reproducing the observed stellar metallicity differences seen in low-mass galaxies, with $\log (M_*/{\rm M_\odot}) < 9.8$. These low-mass systems have sufficiently large gas fractions so that the observed stellar metallicity enhancement can be achieved, even if a substantial amount of gas is removed by outflows. Due to the reduced rate of chemical enrichment caused by the removal of gas, the typical quenching time-scales are longer than those found with the closed-box model. In the specific case of $\lambda _\mathrm{eff}= 1$, the bottom panel of Fig.\@ \ref{fig:qg_model_deltaZ_qt_lambda_1} shows that the duration of quenching is 3--6~Gyr for galaxies with $\log (M_*/{\rm M_\odot}) < 9.8$. Since both closed-box models (pure starvation) and leaky-box models (starvation with outflows) are consistent with the observations, we conclude that outflows could have potentially played an important role in shutting down the star formation in the progenitors of low-mass passive galaxies. \citet{Lian2018a, Lian2018b} also find that outflows could potentially be more important in driving evolution in low-mass galaxies, as they show that either strong, time-dependent outflows with large metal-loading factors, or a time-varying IMF must be invoked to simultaneously reproduce both the gas MZR and the stellar MZR in local low-mass star-forming galaxies. 

On the other hand, the predictions from our leaky-box model are inconsistent with the stellar metallicity differences observed in high-mass galaxies with $\log (M_*/{\rm M_\odot}) > 9.8$. Outflows with $\lambda _\mathrm{eff} = 1$ remove too much gas in these massive systems, preventing sufficient chemical enrichment during quenching. Our findings suggest that powerful outflows played a minor role in quenching the progenitors of high-mass passive galaxies. These massive galaxies either quenched purely through starvation for 2--3~Gyr before any future star formation was abruptly halted by the onset of an ejective or heating mode, or through a combination of starvation and weak outflows, with $\lambda _\mathrm{eff} < 1$. Note that this does not mean that outflows are not present in massive galaxies, but that most of the outflowing gas does not escape and is being recycled (i.e. the `classical' loading factor can be $\lambda \sim 1$ even
if $\lambda _\mathrm{eff} < 1$).

\subsubsection{Constraints on $\lambda_\mathrm{eff}$ from observed $Z_*$ and SFR} \label{subsubsec:qg_sfr}

Until now, we have only been using the observed stellar metallicity differences to put
constraints on the nature of quenching. We have investigated whether given models can
reproduce the observed stellar metallicity enhancement and we have determined the time-scale
required to achieve this. So far we have only been considering models consisting of pure
starvation ($\lambda _\mathrm{eff} = 0$) and models incorporating a combination of starvation
and outflows ($\lambda _\mathrm{eff} = 1$). As we have seen, increasing $\lambda
_\mathrm{eff}$ decreases the rate of stellar metallicity enrichment and lengthens the
time-scale required to reproduce the observed $\Delta Z_*$. At any mass, if $\lambda
_\mathrm{eff}$ is set too large, outflows become too prominent and remove a significant
fraction of the available gas, rendering it impossible to enrich by $\Delta Z_*$. Therefore,
there are a range of $\lambda _\mathrm{eff}$ values that can reproduce the observed stellar
metallicity differences, each with different quenching time-scales. Hence there is a level of degeneracy in our analysis, as the quenching time-scale that is derived depends on the value of $\lambda _\mathrm{eff}$ that is assumed.

A further, important constraint that should help break the $\lambda _\mathrm{eff}$ -- $t_{\mathrm{quench}}$
degeneracy comes from analysing the star formation rates of these galaxies that are being
quenched. At the moment when $\Delta Z_*$ enrichment is achieved, our model galaxy has a
stellar metallicity that is comparable to that of local passive galaxies. In principle this
galaxy should therefore also have a star formation rate that is consistent with those of local
passive galaxies. However, as was found in \ref{subsubsec:qg_pure}, the `quenched' galaxies
in our $\lambda _\mathrm{eff} = 0$ model still had a non-negligible star formation rate.
Rather than invoke a second, abrupt quenching phase (as was done in Section \ref{subsubsec:qg_pure}), we aim, in this subsection, to simultaneously reproduce both the observed stellar metallicity and the star formation rate in local passive galaxies, by using a non-zero time-independent mass loading factor $\lambda _\mathrm{eff}$. We require that when the galaxy's stellar metallicity enrichment equals the observed stellar metallicity difference $\Delta Z_*$,  its current SFR and $M_*$ place it in the passive region of the SFR-M$_*$ plane in Fig.\@ \ref{fig:sfr_m}. Since galaxies that quench with different $\lambda _\mathrm{eff}$ values will have different star formation rates at the moment $\Delta Z_*$ enrichment is satisfied, if $\lambda _\mathrm{eff}$ is chosen to be too small, then the star formation rate will still be too large. On the other hand, if $\lambda _\mathrm{eff}$ is too large, then too much gas is removed and it is impossible for the galaxy to ever satisfy the enrichment criterion. Hence there will only be a range of $\lambda _\mathrm{eff}$ values that can simultaneously reproduce both the observed stellar metallicity and star formation rate in local passive galaxies.
\begin{figure}
\centering
\centerline{\includegraphics[width=1.0\linewidth]{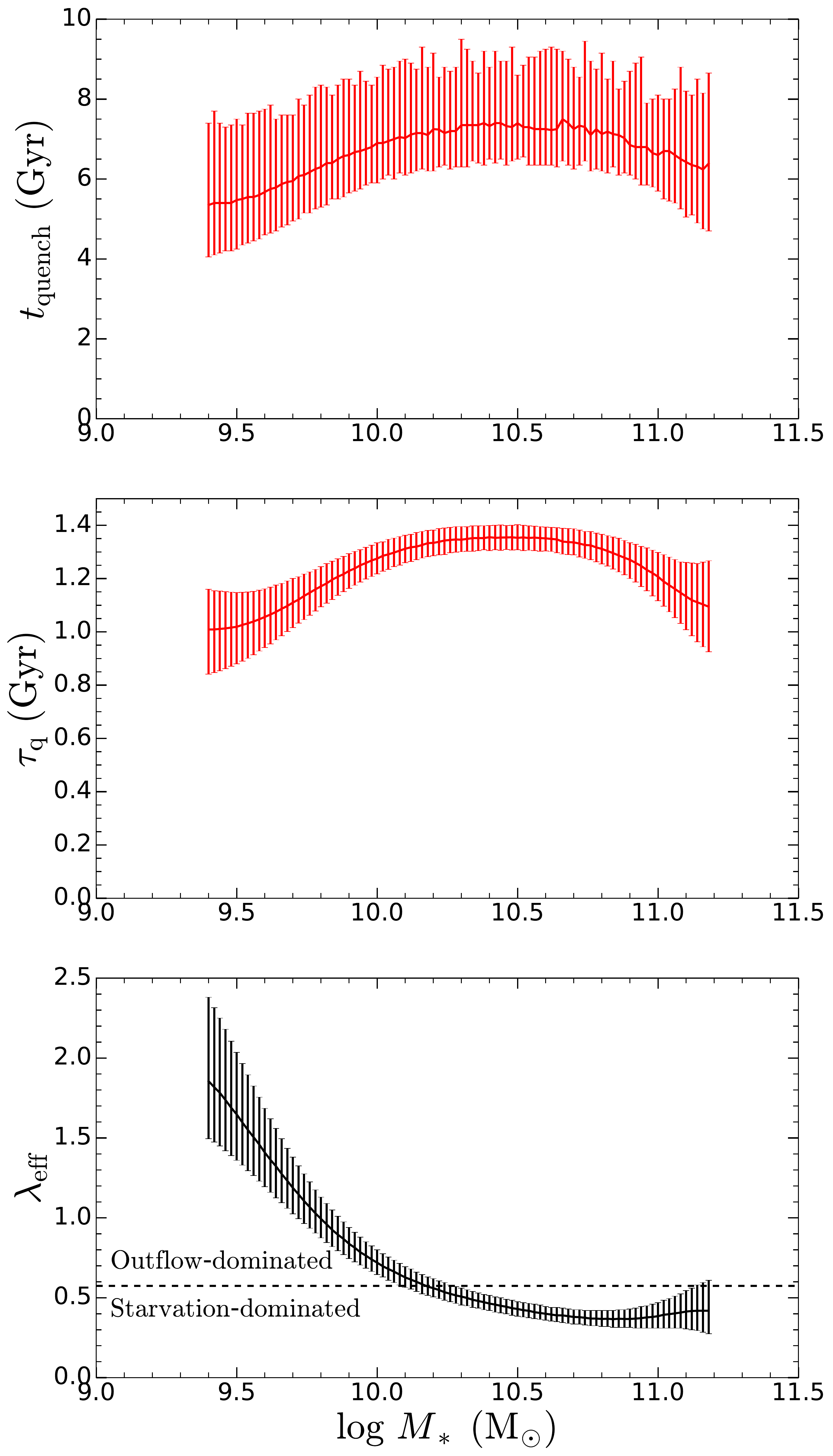}}
\caption{Top panel: Similar to the bottom panel in Fig.\@ \ref{fig:qg_model_deltaZ_qt_lambda_0}, but we now apply our joint metallicity-SFR analysis. Middle panel: The $e$-folding time-scales $\tau _\mathrm{q}$, which indicate the typical time-scale over which the star formation rate declines and the stellar metallicity enriches, as a function of stellar mass. Bottom panel: The mass-loading factors $\lambda _\mathrm{eff}$ required to simultaneously satisfy the $\Delta Z_*$ and the SFR quenching criteria. The horizontal dashed line is the $\lambda_\mathrm{eff}$ value for which the outflow rate ($\lambda_\mathrm{eff}\Psi$) is the same as the net gas consumption rate ($(1-R)\Psi = 0.575\Psi$). Galaxies with $\lambda_\mathrm{eff}$ above (below) this boundary are considered to be outflow-dominated (starvation-dominated). We show the median $e$-folding time-scale and mass-loading factor in each stellar mass bin, with the error bars representing the standard deviation.}
\label{fig:qg_model_qt_tau_variable_lambda}
\end{figure}
\indent We show the duration of quenching derived from our joint metallicity-SFR analysis in the top panel of Fig.\@ \ref{fig:qg_model_qt_tau_variable_lambda}. We find that $t_\mathrm{quench}$ is typically 5--7~Gyr, which is considerably larger than the time-scales we found in the $\lambda _\mathrm{eff} = 0$ scenario. In the middle panel we show the $e$-folding time-scales for star formation $\tau_\mathrm{q}$ during the quenching phase. We find that $\tau_\mathrm{q} \sim 1$~Gyr. While the star formation efficiency $\epsilon$ and the `effective' mass-loading factor $\lambda _\mathrm{eff}$ depend on stellar mass (see bottom panel), we find that $\tau _ \mathrm{q}$ is mostly mass-independent.  Although the duration of quenching $t_\mathrm{quench}$ is long, the typical time-scale over which the bulk of the stellar metallicity enrichment occurs is considerably shorter (2--3$\tau_\mathrm{q}$). This point is demonstrated more clearly by inspecting the model-predicted curves in the top panel of Fig.\@ \ref{fig:qg_model_deltaZ_qt_lambda_1}. We can see that the stellar metallicity initially increases quite rapidly, but quickly slows down as it asymptotically approaches its limiting value. 
\\ \indent We show the mass-loading factors required to simultaneously reproduce the stellar metallicity and SFR of passive galaxies in the bottom panel of Fig.\@ \ref{fig:qg_model_qt_tau_variable_lambda}. We find that the mass-loading factor strongly decreases with increasing stellar mass. This anti-correlation between stellar mass and $\lambda _\mathrm{eff}$ is qualitatively consistent with the mass dependence in theoretical models, which predict that $\lambda \propto M_*^{-1/3}$ \citep{Murray2005} for momentum-driven winds and $\lambda \propto M_*^{-2/3}$ for energy-driven winds \citep[e.g.\@][]{Dekel1986}, as well as other observational evidence \citep[e.g.\@][]{Heckman2015, Chisholm2017, Fluetsch2019}. Our results indicate that  `effective' outflows (which are capable of permanently removing gas from the galaxy) are, together with starvation, of increasing importance in low-mass galaxies. In particular, since the rate at which gas is locked up into long-lived stars is given by $(1-R)\Psi = 0.575\Psi$, and the rate at which gas is ejected from the galaxy is given by $\lambda_\mathrm{eff}\Psi$, then, for galaxies with $\log (M_*/\mathrm{M_\odot}) < 10.2$, the rate at which gas is lost through galactic winds is roughly 1--3 times larger than the rate at which gas is locked up into long-lived stars. Clearly outflows play an essential role in depleting the gas reservoir of low-mass galaxies during the quenching phase. Outflows are relatively weaker ($\lambda_\mathrm{eff} \leq 0.6$) in more massive galaxies ($\log (M_*/\mathrm{M_\odot}) > 10.2$), with starvation becoming the dominant quenching mechanism, as illustrated in Fig.\@ \ref{fig:qg_model_qt_tau_variable_lambda}. However, outflows still play an important role in quenching star formation, as comparable amounts of gas are removed through galactic winds and through the formation of long-lived stars. Although these massive galaxies may be ejecting large amounts of gas in the form of outflows (i.e.\@ a large $\lambda$), our results suggest that these ejection events are short-lived and/or most of the outflowing gas does not escape the galaxy and is instead recycled (i.e.\@ a relatively small $\lambda _\mathrm{eff}$).

\subsection{Green valley galaxies (quenching in the local Universe)} \label{subsec:gvg}

We now investigate the processes responsible for quenching star formation in galaxies in the
local Universe, by studying the green valley galaxy population. Our analysis is similar to Section \ref{subsec:qg_models}, but we now consider the stellar metallicity difference between star-forming galaxies and green valley galaxies.
Since green valley galaxies have only recently begun quenching (if we assume that all green valley galaxies follow a one-way evolution with little contribution from rejuvenation), the stellar metallicities of their progenitors are only slightly smaller than the stellar metallicities of local star-forming galaxies. Hence the stellar metallicity differences between green valley galaxies and their star-forming progenitors are only slightly larger than the differences seen in Fig.\@ \ref{fig:combined_mzr_sgq}.

\subsubsection{Pure starvation} \label{subsubsec:gvg_pure}

A comparison between the observed and model-predicted stellar metallicity differences is shown in the top panel of Fig.\@ \ref{fig:gvg_model_deltaZ_qt_lambda_0}. We find that our closed-box model is able to reproduce the observed stellar metallicity differences, which could potentially suggest that the majority of green valley galaxies are quenching through starvation. However, this model is not constrained to match the gas content in green valley galaxies. We will explore models that simultaneously reproduce both the chemical content and star formation rates in green valley galaxies in Section \ref{subsubsec:gvg_sfr}. The closed-box models typically require 2--5~Gyr of quenching to reach the level of enrichment seen in observations, which implies that we are typically seeing these green valley galaxies 2--5~Gyr after the onset of quenching. These quenching time-scales for local green valley galaxies are comparable to, or even longer than the quenching time-scales of $\sim$2--3~Gyr that were found for the progenitors of local passive galaxies in Section \ref{subsec:qg_models}. Coupled with the fact that passive galaxies have completed quenching, while green valley galaxies are still in the process of quenching (and so would take even longer to fully complete quenching), galaxies in the local Universe must be quenching much more slowly than their counterparts at higher redshift. This increase in quenching time-scale is mostly due to the fact that gas depletion time-scales in the local Universe are longer than at high redshift \citep[e.g.\@][]{Tacconi2013, Santini2014, Genzel2015, Schinnerer2016, Scoville2017, Tacconi2018}, so it takes more time for galaxies to exhaust their gas reservoir during starvation. 

\begin{figure}
\centering
\centerline{\includegraphics[width=1.0\linewidth]{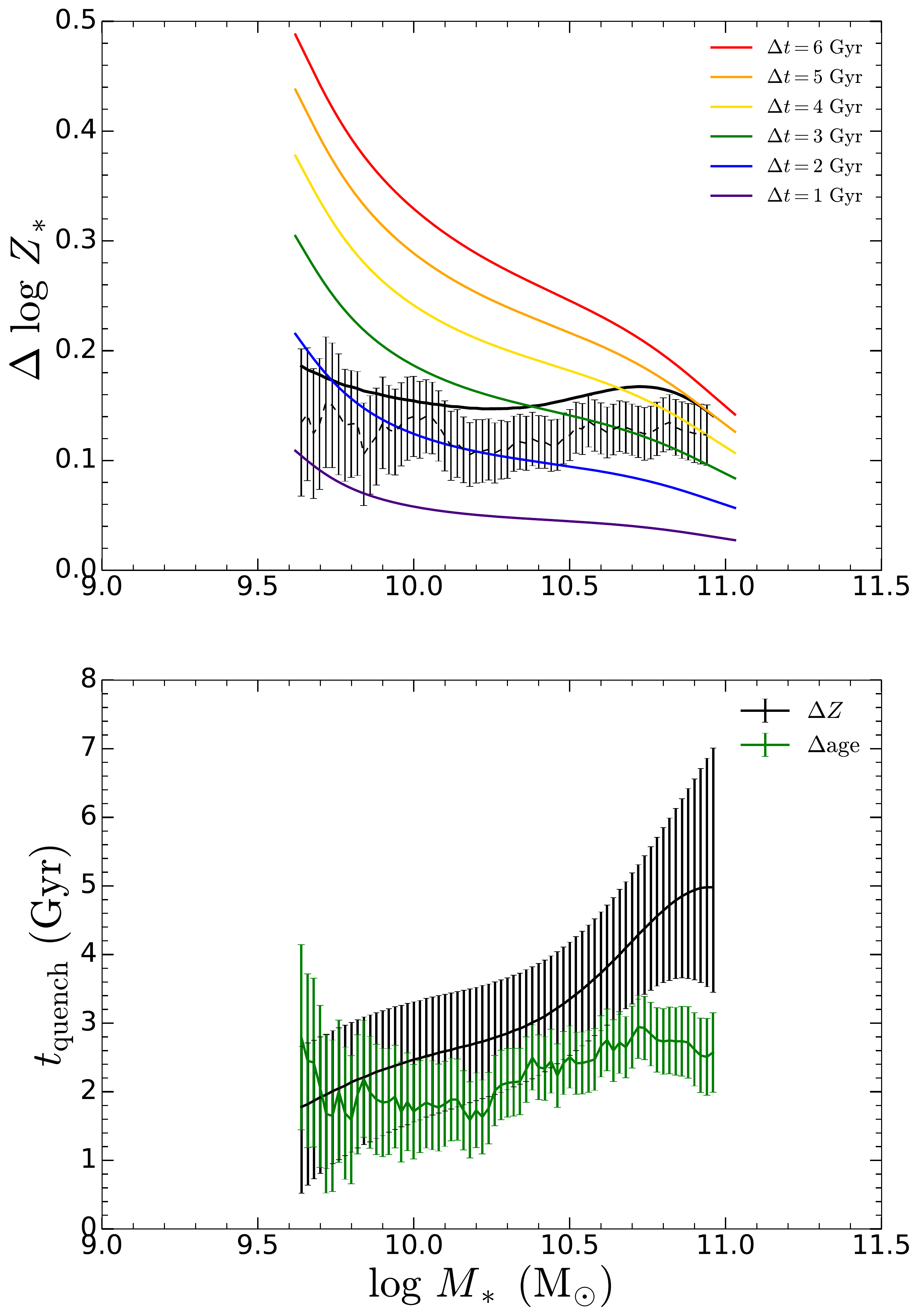}}
\caption{Top panel: The dashed black curve shows the observed stellar metallicity difference between local green valley galaxies and their star-forming progenitors of the same stellar mass. The solid black curve shows the actual increase in stellar metallicity during the quenching phase (taking into account the increase in stellar mass of the galaxy during quenching), with error bars omitted for clarity. The coloured lines show the stellar metallicity difference predicted by a closed-box model ($\lambda _\mathrm{eff} = 0$) at different times after the onset of starvation. Bottom panel: Duration of quenching derived from the stellar metallicity difference ($\Delta Z$) between star-forming galaxies and green valley galaxies (black), as well as the quenching time-scales obtained from the stellar age difference ($\Delta$age) between star-forming and green valley galaxies (green). The error bars on the $\Delta$Z-derived and $\Delta$age-derived quenching time-scales are due to the 1$\sigma$ errors on the stellar metallicity and stellar age differences, respectively.}
\label{fig:gvg_model_deltaZ_qt_lambda_0}
\end{figure}

The mass-dependence of the time elapsed since the onset of quenching $t_\mathrm{quench}$ is shown more clearly in the bottom panel of
Fig.\@ \ref{fig:gvg_model_deltaZ_qt_lambda_0}, where the quenching time-scale obtained from our stellar
metallicity difference analysis is shown by the black curve. We find that the duration of quenching increases with increasing stellar mass, i.e.\@, on average, more massive green valley galaxies began quenching at an earlier epoch than less massive green valley galaxies. This result mostly comes about because the stellar metallicity difference between star-forming and green valley galaxies is roughly mass-independent, while the rate of chemical enrichment is slower in more massive systems due to the smaller gas fractions present. 

We also estimate the time elapsed since the onset of quenching from the observed stellar ages of star-forming and green
valley galaxies. If we make the simplifying assumption that galaxies evolve passively (i.e.\@
no additional stars are formed) during the quenching phase, then from equation
(\ref{eq:passive_age_evolution}) and the discussion that followed, the time since the onset of
quenching is just given by the difference in the stellar ages of star-forming and green valley
galaxies. These stellar age differences then provide an alternative, independent estimate for
the duration of quenching. The quenching time-scales derived from the stellar age differences
($\Delta$age) are shown by the green curve in the bottom panel of Fig.\@
\ref{fig:gvg_model_deltaZ_qt_lambda_0}. The typical stellar age difference between
star-forming and green valley galaxies is within 2--3~Gyr, indicating that we are witnessing green valley galaxies roughly 2.5~Gyr after the onset of quenching. We find that, for $\log (M_*/{\rm M_\odot}) < 10.5$, the quenching time-scales derived from our stellar metallicity analysis are roughly consistent with the time-scales obtained from the stellar age differences. However, at the high-mass end, there is a significant offset between these two estimates for the time elapsed since the onset of quenching, which we shall discuss further.

We made the simplifying assumption that galaxies evolve passively during quenching. Of course, young, metal-rich stars will be formed during quenching which will increase the average stellar metallicity and also decrease the average stellar age of the galaxy. Hence the observed stellar age of real green valley galaxies, which do not evolve passively, will be smaller than the idealised, passively-evolving green valley galaxies that we were considering. As a result, the age difference between star-forming and green valley galaxies that we observe will be smaller than the actual duration of quenching. In other words, the quenching time-scales from our  $\Delta${age} analysis are likely underestimated. Furthermore, since the onset of quenching $z_\mathrm{q}$ in our models is estimated using $\Delta${age}, any uncertainties related to the onset of quenching will also affect the $\Delta Z$-derived estimates for the duration of quenching. At the high-mass end, relaxing the assumption of passive evolution will result in a larger $z_\mathrm{q}$ and therefore higher initial gas fractions and shorter depletion times, resulting in a smaller $\Delta Z$-derived estimate for $t_\mathrm{quench}$, hence further narrowing the gap between our two estimates for the duration of quenching. However, while this point does potentially help to explain the offset between the estimates for the duration of quenching at the high-mass end, we do note that it will worsen the agreement between the $\Delta Z$ ages and  $\Delta$ages at the low-mass end. We will investigate whether models that include outflows can provide better agreement between these two time-scales at the low-mass end later in this section.
\\ \indent Numerous works have also studied the typical time-scales over which galaxies in the local Universe quench. For example, \citet{Wetzel2013a} and \cite{Fossati2017} used N-body simulations/semi-analytic models to track satellite orbits, and subsequently determined the quenching time-scales that were needed to match the quenched fraction of local satellite galaxies observed in SDSS DR7. They find that the typical quenching time-scale decreases with increasing mass, with \citet{Wetzel2013a} finding the time-scale to decrease from 4~Gyr at $M_* = 10^{10}~{\rm M_\odot}$ to 2~Gyr at $M_* = 10^{11}~ {\rm M_\odot}$, while \citet{Fossati2017} find that the time-scale decreases from 8~Gyr at $M_* = 10^{10}~{\rm M_\odot}$ to 6~Gyr at $M_* = 10^{11}~{\rm M_\odot}$. \citet{Fillingham2015} argue that the quenching time-scales derived for massive satellites ($M_* \sim 10^\text{8--11}~\mathrm{M}_\odot$) are broadly consistent with being driven by starvation, with the quenching time-scale corresponding to the cold gas depletion time. \citet{Fillingham2015} also undertake a similar analysis to the aforementioned works, but instead study the much lower mass ($M_* = 10^\text{6--8}~\mathrm{M}_\odot$) dwarf satellites in the Local Group. They find substantially shorter quenching time-scales of $\sim1$--$1.5$~Gyr, which they posit may be due to ram pressure stripping \citep[see also][]{Fillingham2016}, which is likely to be more effective at removing gas in these low-mass systems. In addition, \citet{Wheeler2014} studied the quenched fractions of intermediate mass ($M_* = 10^\text{8.5--9.5}~\mathrm{M}_\odot$) satellites and find that the quenching time-scales are much longer ($>9.5$~Gyr). These satellites are likely to be too massive to quench through ram pressure stripping, so they instead quench through starvation \citep[albeit very slowly due to the long depletion times in these low-mass galaxies, e.g.\@][]{Boselli2014}.
In a similar vein, \citet{Guo2017a} also determined estimates for the time-scale upon which satellite galaxies must quench, following accretion into the group/cluster halo. Using dynamical arguments, they find that intermediate-mass ($9.5 < \log (M_*/{\rm M_\odot}) < 10.5$) galaxies typically quench over a time-scale of 7~Gyr. 

As is expected, the quenching time-scales, $t_\mathrm{passive}$, obtained from the aforementioned works (except the 2~Gyr from \citet{Wetzel2013a}) are larger than the time-scales, $t_\mathrm{green}$, we derive in our analysis. This should be the case because $t_\mathrm{passive}$ measures the total time required to complete quenching (transforming a galaxy from star-forming to passive), while our time-scales measure the time that green valley galaxies have been quenching for so far. Since green valley galaxies are still in the process of quenching and require additional time to complete the transformation into passive galaxies, $t_\mathrm{green} < t_\mathrm{passive}$. However, we do note that such a comparison between quenching time-scales is complicated by the fact that the aforementioned works investigated the quenching of satellite galaxies, while our sample of SDSS galaxies consists mostly of centrals \citep[see][]{Yang2007}. We postpone an analysis separating satellite from central galaxies to a future paper.

\subsubsection{Starvation with outflows}\label{subsubsec:gvg_of}

A comparison between the observed stellar metallicity differences and the predictions from models with the `classical' $\lambda _\mathrm{eff} = 1$ is shown in Fig.\@ \ref{fig:gvg_model_deltaZ_qt_lambda_1}. We find that models incorporating outflows struggle to reproduce the observed stellar metallicity differences at the high-mass end, suggesting that powerful outflows play a minor role in quenching star formation in massive galaxies in the local Universe. In contrast, the stellar metallicity differences are still easily reproduced at the low-mass end, indicating that even stronger outflows may be required to quench these galaxies. We will investigate this in more detail in the next section.

\begin{figure}
\centering
\centerline{\includegraphics[width=1.0\linewidth]{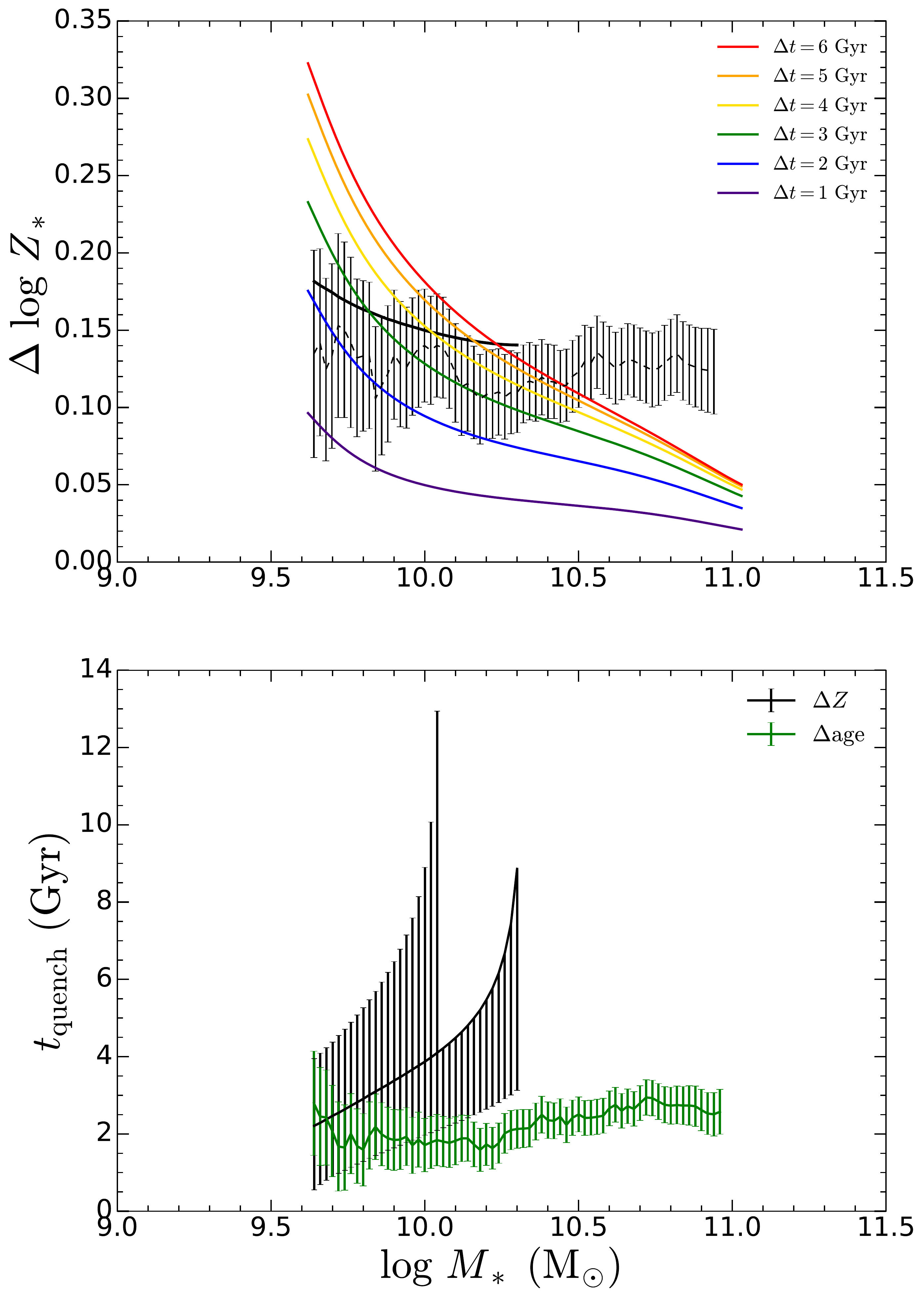}}
\caption{Similar to Fig.\@ \ref{fig:gvg_model_deltaZ_qt_lambda_0}, but we now consider a leaky-box model, incorporating outflows with $\lambda _\mathrm{eff} = 1$. We omit the upper limit on the increase in stellar metallicity during quenching $\Delta Z_*$ (solid black curve in the top panel) when it cannot be reproduced by our model, and in that case we only show the lower limit on the $\Delta Z$-derived duration of quenching $t_\mathrm{quench}$ (black curve in the bottom panel).}
\label{fig:gvg_model_deltaZ_qt_lambda_1}
\end{figure}

\subsubsection{Constraints on $\lambda_\mathrm{eff}$ from observed $Z_*$ and SFR}\label{subsubsec:gvg_sfr}

Similar to Section \ref{subsubsec:qg_sfr}, we determine the range of $\lambda _\mathrm{eff}$ such that both the $\Delta Z_*$ and the SFR quenching criteria are simultaneously satisfied. This means that when the galaxy's stellar metallicity enrichment in our model equals the observed stellar metallicity difference $\Delta Z_*$, its SFR and $M_*$ place it in the green valley galaxy locus of the SFR-M$_*$ plane in Fig.\@ \ref{fig:sfr_m}.

The quenching time-scales $t_\mathrm{quench}$ derived from our joint metallicity-SFR analysis are shown in the top panel of Fig.\@ \ref{fig:gvg_model_qt_tau_variable_lambda}. We also show the $e$-folding time-scales $\tau _\mathrm{q}$ in the middle panel, and the associated mass-loading factors in the bottom panel. The quenching duration $t_\mathrm{quench}$ increases with increasing stellar mass from 4 to 6.5~Gyr. These time-scales are not significantly different from what was seen for passive galaxies in Fig.\@ \ref{fig:qg_model_qt_tau_variable_lambda}. As mentioned earlier, since local green valley galaxies are still in the process of quenching, this means that the total time required to complete the quenching phase and fully transition to the passive sequence will be longer than the $t_\mathrm{quench}$ that has been obtained, indicating that local green valley galaxies quench more slowly than their counterparts at higher redshift. This is illustrated more clearly by the $e$-folding time-scales $\tau _\mathrm{q}$,  which increase with mass in the range from 1.5 to 2.5~Gyr, and are larger than what was obtained in our study of local passive galaxies ($\sim$1~Gyr). Furthermore, we find that the mass-loading factors $\lambda _\mathrm{eff}$ are similar to those at high-z, indicating that outflows in the local Universe are, on average, likely to be comparable in power and frequency to those at higher redshift. Similar to our study of passive galaxies, we find that the typical mass-loading factor decreases with increasing mass, indicating that `effective' outflows play an important role in quenching local low-mass galaxies, with 1--3 times more gas lost through outflows than through gas consumption for $\log (M_*/\mathrm{M}_\odot) < 10.4$. Outflows become relatively less important and sub-dominant relative to starvation at high masses (but still eject comparable amounts of gas to what is lost through gas consumption), likely indicating that the outflowing gas in local massive galaxies does not escape and is instead reaccreted and/or that these ejection events are short-lived.
\begin{figure}
\centering
\centerline{\includegraphics[width=1.0\linewidth]{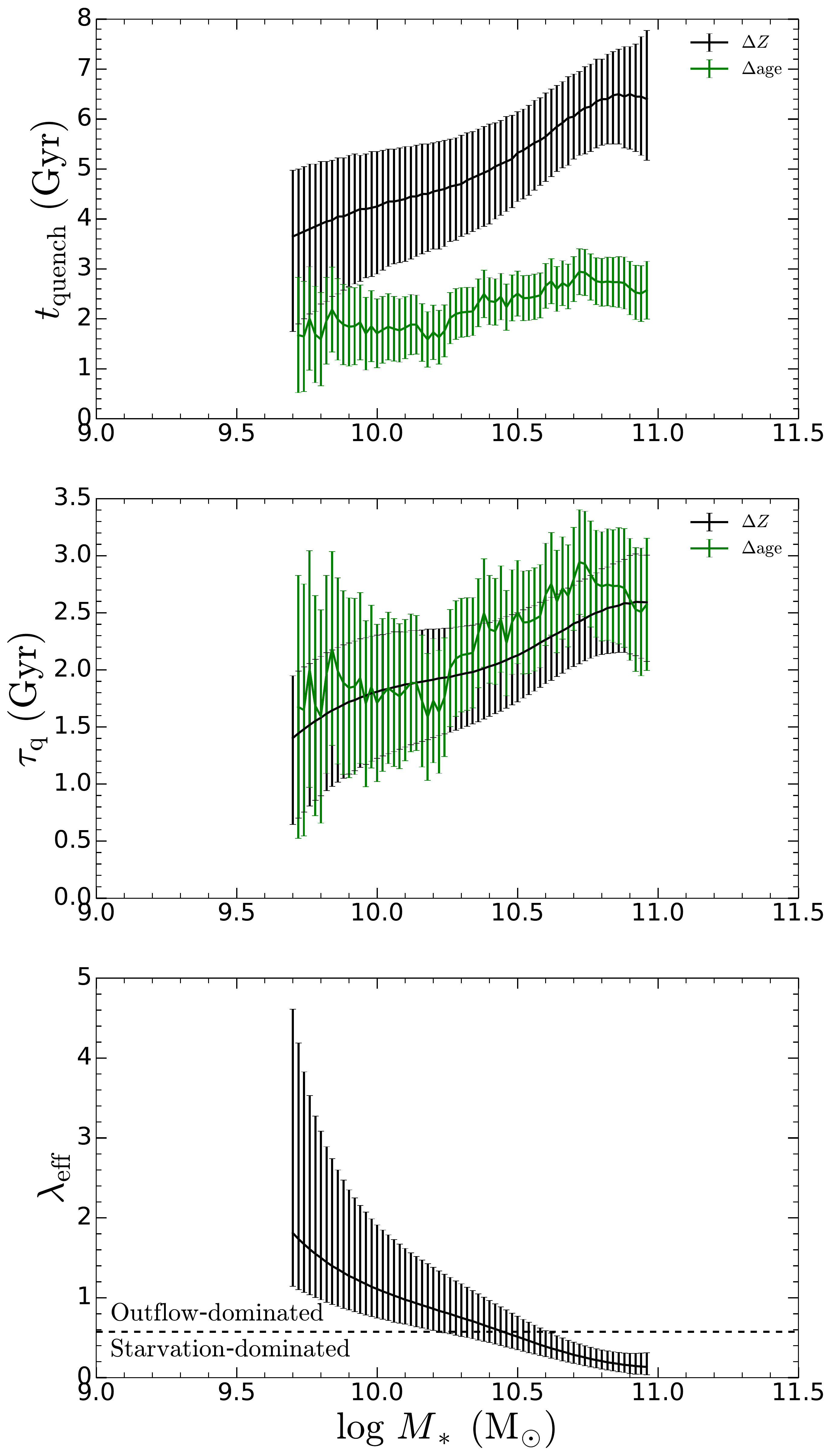}}
\caption{Top panel: Similar to the bottom panel in Fig.\@ \ref{fig:gvg_model_deltaZ_qt_lambda_0}, but we now apply our joint metallicity-SFR analysis. Middle panel: The $e$-folding time-scales $\tau _\mathrm{q}$, which indicate the typical time-scale over which the star formation rate declines and the stellar metal- licity enriches, as a function of stellar mass. Bottom panel: The mass-loading factors $\lambda _\mathrm{eff}$ required to simultaneously satisfy the $\Delta Z_*$ and the SFR quenching criteria. The horizontal dashed line is the $\lambda_\mathrm{eff}$ value for which the outflow rate ($\lambda_\mathrm{eff}\Psi$) is the same as the net gas consumption rate ($(1-R)\Psi = 0.575\Psi$). Galaxies with $\lambda_\mathrm{eff}$ above (below) this boundary are considered to be outflow-dominated (starvation-dominated). We show the median $e$-folding time-scale and mass-loading factor in each stellar mass bin, with the error bars representing the standard deviation.}
\label{fig:gvg_model_qt_tau_variable_lambda}
\end{figure}
\par
However, we do note that the long quenching time-scales $t_\mathrm{quench}$ that we have derived are substantially larger than the time-scales obtained from
the stellar age difference analysis (Fig.\@ \ref{fig:gvg_model_qt_tau_variable_lambda}), while our $e$-folding time-scales $\tau _\mathrm{q}$ are more comparable. As discussed in \ref{subsubsec:gvg_pure}, our estimates for the onset of quenching $z_\mathrm{q}$ based off of the age difference $\Delta$age may be inaccurate, which is likely to result in underestimates and overestimates for the duration of quenching from the $\Delta$age and $\Delta Z_*$ methods, respectively. It is also possible that our model which assumes continuous outflows, with a constant, large mass-loading factor does not provide an adequate description of how star formation is shut down in local green valley galaxies. This would suggest that another mechanism, which together with starvation, allows for a significant increase in stellar metallicity together with a rapid reduction in the star formation rate. 

In addition, the inconsistency between quenching time-scale inferred from the metallicities and the age differences observed in Fig.\@ \ref{fig:gvg_model_qt_tau_variable_lambda} may reflect the limits of the dataset. Indeed, the Sloan fibre only probes the central region of galaxies. While this is not a major problem for star-forming galaxies and passive galaxies \citep[which have mild metallicity and age gradients,][]{Belfiore2017b,Goddard2017a} 
it is a potential issue for green valley galaxies, which often have a quenched central region and an outer (much younger) star-forming disc \citep{Belfiore2017a,Belfiore2018}. The physical properties in green valley galaxies that vary rapidly with galactocentric radius are not properly probed by the single-fibre SDSS data, and may likely introduce some inconsistencies in our analysis. It will be possible to further investigate the quenching of green valley galaxies, overcoming the limitations of a single fibre, by analysing the spatially-resolved spectral data provided by integral field spectroscopic galaxy surveys, such as SDSS-MaNGA (although at the expense of statistics).

\section{Assumptions and caveats} \label{sec:caveats}

In this section we clarify modelling assumptions and also discuss caveats regarding our work. In particular, in Section \ref{subsec:sp} we investigate how the stellar metallicities and ages that are derived are affected by the fitting technique and stellar population synthesis model used, and the impact this may have on our results. In Section \ref{subsec:mzr_evo} we study the validity of our simplifying assumption regarding the evolution of the stellar MZR in our models. In Section \ref{subsec:prog_mass_offset} we investigate how our results are affected by introducing a mass offset between star-forming progenitors and local passive galaxies in our models. Finally, in Section \ref{subsec:mass_return} we investigate how modifications to the amount of mass return to the ISM in our models can affect our results.

\subsection{Stellar population modelling techniques}\label{subsec:sp}

In this subsection we discuss the stellar metallicities and ages derived by {\footnotesize FIREFLY}, summarising the outcomes of the various performance tests of \citet{Wilkinson2017} and \citet{Goddard2017b}. We also compare the stellar-mass--stellar metallicity and stellar-mass--stellar age relations that are obtained using mass-weighted and light-weighted quantities from {\footnotesize FIREFLY} with those obtained using the metallicities and ages derived by \citet{Gallazzi2005} which were used in \citetalias{Peng2015}.

The performance of {\footnotesize FIREFLY} has been extensively tested in \citet{Wilkinson2017}. We summarise the main findings here. Mock galaxies, with a wide range of star formation histories (short, intermediate and extended), dust content and S/N, were constructed. {\footnotesize FIREFLY} was able to recover stellar population properties such as age, metallicity and stellar mass remarkably well down to a S/N $\sim 20$ (which is the S/N cut used in this work), for moderately dusty systems. When using a lower S/N ($\sim$5), metallicity may be underestimated by roughly 0.3 dex, but this depends on the reddening of the system, which may alter the age determination and hence the metallicity. The accurate recovery of the stellar population parameters indicates that the full spectral fitting approach with no additional prior over the wide model grid of ages and metallicity is able to remove most degeneracies. However, it should be noted that there are residual degeneracies between age and dust especially, and between different star formation histories, for systems with a substantial amount of dust ($\mathrm{E(B}-\mathrm{V)}=0.75$) and an extended star formation history ($\sim$10~Gyr), but these properties should pertain only to a minority of galaxies. Still, even in these very dusty systems {\footnotesize FIREFLY} is able to distinguish between short and extended formation histories, albeit with less precision. The effect of the adopted wavelength range on the derivation of metallicity was also studied, for two mock galaxies with very different ages (55~Myr and 7~Gyr). The input solar metallicities were recovered to within $\pm$0.02~dex, with not too much dependence on the adopted wavelength range.
\\ \indent The performance of  {\footnotesize FIREFLY} at recovering the ages and metallicities of Milky Way globular clusters was also tested. \citet{Wilkinson2017} found a very good match between the ages and metallicities derived by {\footnotesize FIREFLY}, and the ages and metallicities determined via CMD fitting, and stellar spectroscopy, respectively, with no bias in the derivation of metallicity over a wide range ($-2 <  [\mathrm{Z}/\mathrm{H}] < 0$). Furthermore, the ages and metallicities derived by {\footnotesize FIREFLY} were also compared with those obtained by \citet{Koleva2008}, who used a different fitting code \citep[\begin{footnotesize}NBURSTS,\end{footnotesize}][]{Chilingarian2007} and models \citep[Pegase-HR,][]{LeBorgne2004}. Overall, the results are consistent, but \citet{Wilkinson2017} note that the {\footnotesize FIREFLY} results do not contain artificially low ages and display a smaller scatter around the resolved ages and metallicities. 
\\ \indent The loci of SDSS DR7 galaxies on the age-metallicity plane using {\footnotesize FIREFLY} was also compared with the results obtained using other fitting codes, namely {\footnotesize VESPA} \citep{Tojeiro2007, Tojeiro2009} and {\footnotesize STARLIGHT} \citep{CidFernandes2005}. The comparison with {\footnotesize VESPA} is performed using their calculations based on the \citet{Maraston2005} models. \citet{Wilkinson2017} find that the distributions of galaxies are broadly similar (most galaxies peak at old ages and high metallicity), but the exact age and metallicity solution will of course depend on what the model grid allows. In {\footnotesize FIREFLY} the models are not restricted, hence the solutions appear more spread than what is obtained with {\footnotesize VESPA}, who construct model vectors with a more restricted range of input properties. The comparison with {\footnotesize STARLIGHT} is performed using their calculations based on the \citet{Bruzual2003} models. \citet{Wilkinson2017} find that the main density of age and metallicity points cluster around 8 Gyr in age and 0.1 to 0.3 dex in [Z/H] for both fitting codes and model setup. Hence for the total sample, the codes agree very well. Differences are seen at the edges of the galaxy distribution in age and metallicity, which is attributable to the assumed model grid in the codes. \citet{Goddard2017b} further investigate the impact that different spectral fitting codes, stellar population synthesis models and stellar libraries can have on the derived ages and metallicities in the context of radial gradients. They find that the systematic uncertainties introduced can be of a similar size to the signal being measured, which may help explain the variety in the age and metallicity gradients that have been obtained from various IFU studies \citep[e.g.\@][]{GonzalezDelgado2015, Goddard2017b, Zheng2017a, Li2018}.
\\ \indent Outside of the spectral fitting code, stellar population model and stellar library used, the manner in which the average of a stellar population parameter is calculated also affects the derived stellar metallicities and ages. Using a linear average, one sums the linear quantities, e.g.\@ 
\begin{equation}
\langle Z_* \rangle = \sum_i {w_iZ_{*, i}},
\end{equation}
where $w_i$ and $Z_{*,i}$ are the weight (e.g.\@ light-weight or mass-weight) and stellar metallicity of SSP $i$, respectively. On the other hand, using a logarithmic average, one sums the logarithmic quantities, e.g.\@
\begin{equation}
\langle \log Z_* \rangle = \sum_i {w_i \log Z_{*, i}}.
\end{equation}
The inequality of the arithmetic and geometric means implies that $\langle \log Z_* \rangle \leq \log \langle Z_* \rangle$, as the more metal-poor stellar populations are weighted more strongly when using the logarithmic average \citep[see also the discussion in][]{Sanchez-Blazquez2011, GonzalezDelgado2015}. As the stellar metallicities in our simple analytical model (see Section \ref{subsec:diff_eq}) are calculated using the linear average, we have chosen to use linearly-averaged ages and metallicities from {\footnotesize FIREFLY} in our analysis to ensure a consistent treatment of metallicity across the models and the observations. However, other works advocate using the logarithmic average \citep[e.g.\@][]{Sanchez-Blazquez2011, GonzalezDelgado2015}, because, for example, this results in values for the average that are more similar to the SSP-equivalent parameters. We note that the gap in stellar metallicity between star-forming and passive galaxies is likely to widen when using a logarithmic average, especially at the low-mass end, due to the greater prevalence of metal-poor stars in star-forming galaxies.

We now compare the scaling relations derived using {\footnotesize FIREFLY} with those obtained using the \citet{Gallazzi2005} results. In Fig.\@ \ref{fig:mzr_mar_gallazzi}, we show the stellar mass--stellar metallicity relation (top panel) and stellar mass--stellar age relation (bottom panel) for star-forming, green valley and passive galaxies, using both the mass-weighted (which have been used throughout this work, solid) and light-weighted (dashed) values from {\footnotesize FIREFLY}, as well as the ages and metallicities derived by \citet[][dotted]{Gallazzi2005}. The median error bars for each galaxy type are shown in the bottom right corner of both panels. We briefly note that the relative ordering of the star-forming, green valley and passive galaxies is preserved, independent of which stellar populations values are used, i.e.\@ star-forming galaxies are always the youngest and most metal-poor, and green valley galaxies are always intermediate in metallicity and age between star-forming and passive galaxies.

First we will focus on the comparison between the mass- and light-weighted quantities from {\footnotesize FIREFLY}. As expected, the light-weighted metallicities tend to be higher than the mass-weighted metallicities (with a typical difference of 0.05--0.20~dex for star-forming galaxies that appears to decrease with mass), and the light-weighted ages tend to be younger than the mass-weighted ages. Interestingly, we find that for massive passive ($\log M_* > 10.5$) and massive green valley ($\log M_* > 10.8$) galaxies, the light-weighted metallicities are in fact slightly lower than the mass-weighted metallicities. This may perhaps be an indication that the gas in these passive galaxies has been gradually diluting \citep[e.g.\@][]{Yates2012, Yates2014}, or that some of the high-mass green valley galaxies are in fact rejuvenated systems, rather than galaxies currently in the act of quenching. However, since the offset between mass- and light-weighted metallicities is small we will not explore this point any further.

\begin{figure}
\centering
\centerline{\includegraphics[width=1.0\linewidth]{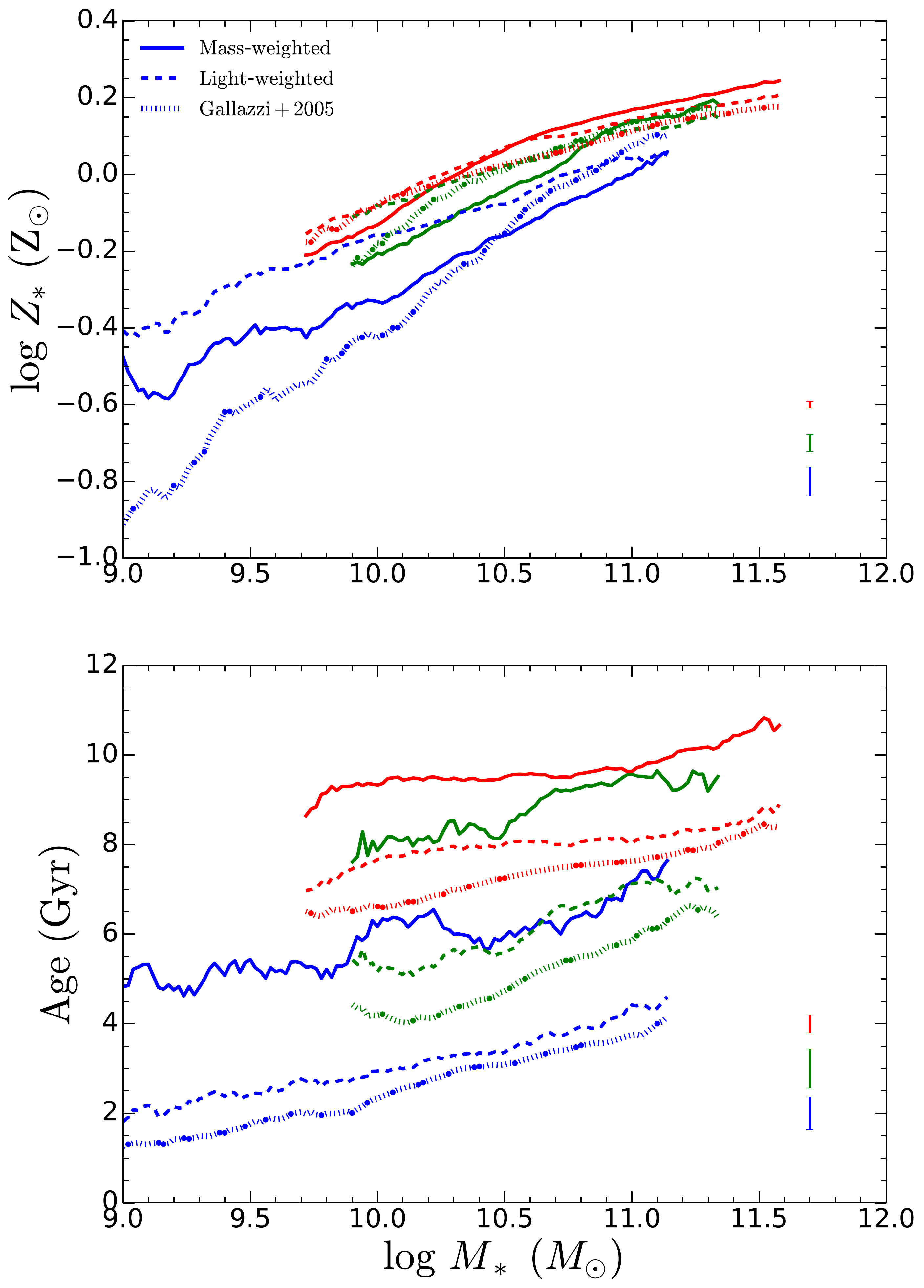}}
\caption{The stellar mass--stellar metallicity relation (top panel) and stellar mass--stellar age relation (bottom panel) for star-forming (blue), green valley (green) and passive (red) galaxies, using mass-weighted (solid) and light-weighted (dashed) metallicities/ages from FIREFLY, and metallicities/ages from \citet[][dotted]{Gallazzi2005}. The median error bars for each galaxy type are shown in the bottom right corner of both panels. Note that the mass-weighted stellar metallicities and ages from FIREFLY have been used throughout this work.}
\label{fig:mzr_mar_gallazzi}
\end{figure}

Since the ages and metallicities derived by \citet{Gallazzi2005} were obtained by simultaneously fitting five spectral absorption features in the optical SDSS spectra, their quantities are light-weighted and we will compare them with the light-weighted quantities from {\footnotesize FIREFLY}. We find that there is excellent agreement between the stellar ages derived from the two works. The \citet{Gallazzi2005} ages are consistently younger, with a typical age offset of 0.25--1~Gyr from the {\footnotesize FIREFLY} ages. In order to compensate for the younger (and therefore bluer) stellar populations in \citet{Gallazzi2005}, one might expect their metallicities to be higher (to produce a redder spectrum) than the {\footnotesize FIREFLY} metallicities, due to the age-metallicity degeneracy. However, we can see from the figure that this is not always the case, especially not for low-mass star-forming galaxies, where the \citet{Gallazzi2005} metallicities are substantially smaller than the light-weighted metallicities from {\footnotesize FIREFLY}, and, surprisingly, are also smaller than the mass-weighted metallicities as well. We suspect that the aforementioned systematics associated with the choice of stellar population model and stellar library play an important role in driving this difference between the measured metallicities, as \citet{Gallazzi2005} used the \citet{Bruzual2003} models with STELIB, while we use the \citet{Maraston2011} models with MILES. This is likely further exacerbated by the fact that the techniques used to derive ages and metallicities are different: {\footnotesize FIREFLY} uses a full spectral fit of the optical spectrum, while \citet{Gallazzi2005} derive stellar populations by simultaneously fitting multiple optical spectral indices. \citet{Maraston2011} compared the metallicities obtained via full spectral fitting and via indices with those from resolved spectroscopy. They found that metallicities derived via indices depend strongly on which exact indices are used, whereas the full spectral fitting method, by using the full spectrum, provides a better match to the actual metallicity. We further note that, while the metallicities in \citet{Gallazzi2005} are averaged logarithmically \citep[see discussion in][]{GonzalezDelgado2015}, we instead use their median stellar metallicities, which are not affected by the averaging scheme used. Hence the large offset between the {\footnotesize FIREFLY} and \citet{Gallazzi2005} metallicities that we see are not due to this effect.\\
\indent We acknowledge that the quantitative aspects of our results are affected by the set of stellar population parameters that are adopted in the analysis. On the one hand, the stellar age differences are not greatly affected by the choice of stellar population parameter set, so our estimations for the onset of quenching in the models will be largely unchanged. However, at least from Fig.\@ \ref{fig:mzr_mar_gallazzi}, the stellar metallicities appear to be rather sensitive to the parameter set that is adopted. We note however that the difference in stellar metallicity between star-forming and passive galaxies is rather substantial, especially when comparing against the high-$z$ star-forming progenitors, even when using the mass-weighted metallicities from {\footnotesize FIREFLY}. Our results suggest that the stellar metallicity of a galaxy typically increases by a factor of 2--3 during the quenching phase (see Fig.\@ \ref{fig:mzr_progenitor}). The gap in metallicity widens even further when using the \citet{Gallazzi2005} data, which will result in longer quenching time-scales in our closed-box models, and smaller $\lambda_\mathrm{eff}$ values in our joint SFR-$Z_*$ analysis in Section \ref{subsubsec:qg_sfr}, further strengthening the case for starvation as a necessary component of quenching.

Finally, we reiterate that the qualitative aspects of our results seem to be less strongly affected by the set of stellar population parameters used. As noted earlier, we find that star-forming galaxies are more metal-poor and younger than passive galaxies of the same stellar mass, independent of the three stellar population parameter sets that were investigated in Fig.\@ \ref{fig:mzr_mar_gallazzi}. We further note that, in addition to this work and \citetalias{Peng2015}, a wide gap in stellar metallicity between star-forming and passive galaxies has also been seen in other works (using different spectral fitting codes, stellar population models and libraries), albeit indirectly. \citet{GonzalezDelgado2015, Scott2017, Li2018} studied the global and, with the exception of \citet{Scott2017}, also the spatially-resolved stellar population properties for galaxies in the CALIFA \citep{Sanchez2012}, SAMI \citep{Bryant2015}, and MaNGA \citep{Bundy2015} surveys, respectively. Using visual morphology, or, in some cases, also Sércic indices \citep{Sersic1963}, each work finds that early-type/elliptical galaxies have larger stellar metallicities than late-type/spiral galaxies of the same stellar mass. Given the correlation between galaxy morphology and star formation rate \citep[e.g.\@][]{Wuyts2011}, early- and late-type galaxies roughly correspond to passive, and star-forming galaxies, respectively, so the finding of a stellar metallicity difference between different galaxy populations in those works is qualitatively consistent with what has been found in this work.

\subsection{Evolution of the stellar mass--stellar metallicity relation}\label{subsec:mzr_evo}

In our model, the initial stellar metallicities of the high-$z$ progenitors of local passive galaxies were estimated by assuming that the stellar MZR evolves in the same way as the gas MZR in \citet{Maiolino2008}. This method was adopted because observations of the stellar metallicities of star-forming galaxies beyond $z=0$ currently lack sufficient statistics and coverage in stellar mass and redshift to adequately constrain the evolution of the stellar MZR. In this subsection we compare our predicted redshift evolution with determinations of stellar metallicity and stellar mass from data available in the literature. Given that those determinations depend on a range of assumptions regarding stellar population models and fitting techniques it is quite hard to quantitatively interpret any agreement or disagreement. We note upfront however that the overall picture seems to be consistent and we provide some general statements to aid the interpretation.

The redshift-evolution of the stellar mass--stellar metallicity relation is shown in Fig.\@ \ref{fig:mzr_evolution}. The solid curves represent the stellar MZR that is used in our model, evaluated at various redshifts. The $z=0$ (indigo) relation corresponds to the mass-weighted stellar metallicities of star-forming galaxies shown in Fig.\@ \ref{fig:combined_mzr_sgq}, while the higher redshift relations were obtained by evolving the local relation in redshift according to the evolution of the gas MZR in \citet{Maiolino2008}. Since the gas-phase metallicities of high-redshift star-forming galaxies tend to be lower than local galaxies of the same stellar mass, the normalisation of the stellar MZR decreases with redshift in our model. Furthermore, since the evolution of the gas MZR is stronger for lower-mass galaxies in \citet{Maiolino2008}, the stellar MZR steepens with redshift, which is qualitatively consistent with the findings of \citet{Lian2018c}, who used a chemical evolution model to investigate the evolution of the gas and stellar MZR from z$\sim$3.5 to 0.

On Fig.\@ \ref{fig:mzr_evolution}, we also show the stellar metallicities of star-forming galaxies modelled from data at high-redshift, which follow the same colour coding as our model estimates for the stellar MZR. We show the $z=0.7$ stellar MZR measured from 40 galaxies by \citet[blue dashed curve]{Gallazzi2014}, the stellar metallicity of a stack of 75 galaxies at $z=2$ from \citet[green triangle]{Halliday2008}, the stellar metallicities of $z=3$ galaxies from \citet[orange circles]{Sommariva2012} and the $z \sim 3.5$ stacks of galaxies from \citet[red diamonds]{Cullen2019}. Note that the stellar metallicities from the high-redshift observations are all light-weighted, while our model estimates of the stellar MZR are mass-weighted. Mass-weighted stellar metallicities tend to be lower than light-weighted metallicities, with the offset between them depending on the star formation history (but is typically 0.1--0.2 dex \citep[e.g.\@][]{Zahid2017}). This offset needs to be taken into account when comparing the model estimates with the observations.
\\ \indent From inspection of Fig.\@ \ref{fig:mzr_evolution}, we find that there is good agreement between our model estimates and the results from \citet{Gallazzi2014} at $z=0.7$, with at most a 0.09 dex disagreement at the high-mass end. However, this disagreement becomes worse when accounting for the mass-weight--light-weight offset. We note that \citet{Gallazzi2014} adopted \citet{Bruzual2003} models and obtained an estimate of stellar metallicity from solar-scaled Lick indices. This modelling approach is substantially different from the one we adopt here, where the \citet{Maraston2011} models and full spectral fitting were used to obtain metallicities and ages. We also find reasonable agreement (0.15 dex offset) between our model and the $z=2$ \citet{Halliday2008} results, which is further improved when accounting for the mass-weight--light-weight offset. There is significant scatter in the stellar metallicities for the $z=3$ galaxies obtained by \citet{Sommariva2012}. While our model estimates provide a reasonable agreement with the two high-mass measurements, there is considerable disagreement with the three low-mass measurements, even when accounting for the mass-weight--light-weight offset. We also find some disagreement between our model estimates and the $z \sim 3.5$ results from \citet{Cullen2019}, with our models consistently underestimating the stellar metallicities of these galaxies. Indeed, \citet{Cullen2019} find that there is little evolution in the stellar MZR for $2.5 < z < 5.0$. These results suggest that our assumed evolution of the stellar MZR may become inaccurate at $z=3$. However, this inaccuracy is not a significant issue, since the maximum redshift that we use for our quenching epochs $z_\mathrm{q}$ is $z\sim2$ (see Fig.\@ \ref{fig:quenching_epochs}), and up to that redshift the stellar MZR evolution in our models provides reasonable agreement with the observations. Moreover, for what concerns the residual discrepancies, one should also take into account that the metallicity measurements based on the rest-frame UV spectrum of distant galaxies mostly probe the iron abundance, while the optical spectra of local galaxies probe a mixture of iron and alpha-elements, while gas phase mostly traces the oxygen (or alpha-element) abundance, so it should not be surprising that there is some residual disagreement.
\\ \indent Finally, we note that the shape and normalisation of the gas MZR is affected by the calibration used \citep[e.g.\@][]{Kewley2008}. Hence different calibrations may predict stronger or weaker evolution of the gas MZR with redshift. The strong evolution of the gas MZR in \citet{Maiolino2008} agrees well with the predictions of several hydrodynamical simulations \citep[e.g.\@][]{Ma2016, Dave2017}. However, other simulations predict a weaker evolution of the gas MZR \citep[e.g.\@][]{Dave2011a, DeRossi2015, DeRossi2017b,  Torrey2019}, which agrees better with the weaker evolution of the gas MZR observed by \citet{Hunt2016a}, who used a different metallicity calibration to \citet{Maiolino2008} in their analysis. We note that assuming a weaker evolution of the gas MZR in our analysis will result in narrowing the metallicity gap between local passive galaxies and their high-$z$ star-forming progenitors. This will result in shorter quenching timescales in our closed-box analysis in Section \ref{subsubsec:qg_pure}. For our analysis in Section \ref{subsubsec:qg_sfr}, the narrower metallicity gap means that galaxies can afford to lose more gas in the quenching process and therefore our  $\lambda_\mathrm{eff}$ values will increase.

\begin{figure}
\centering
\centerline{\includegraphics[width=1.0\linewidth]{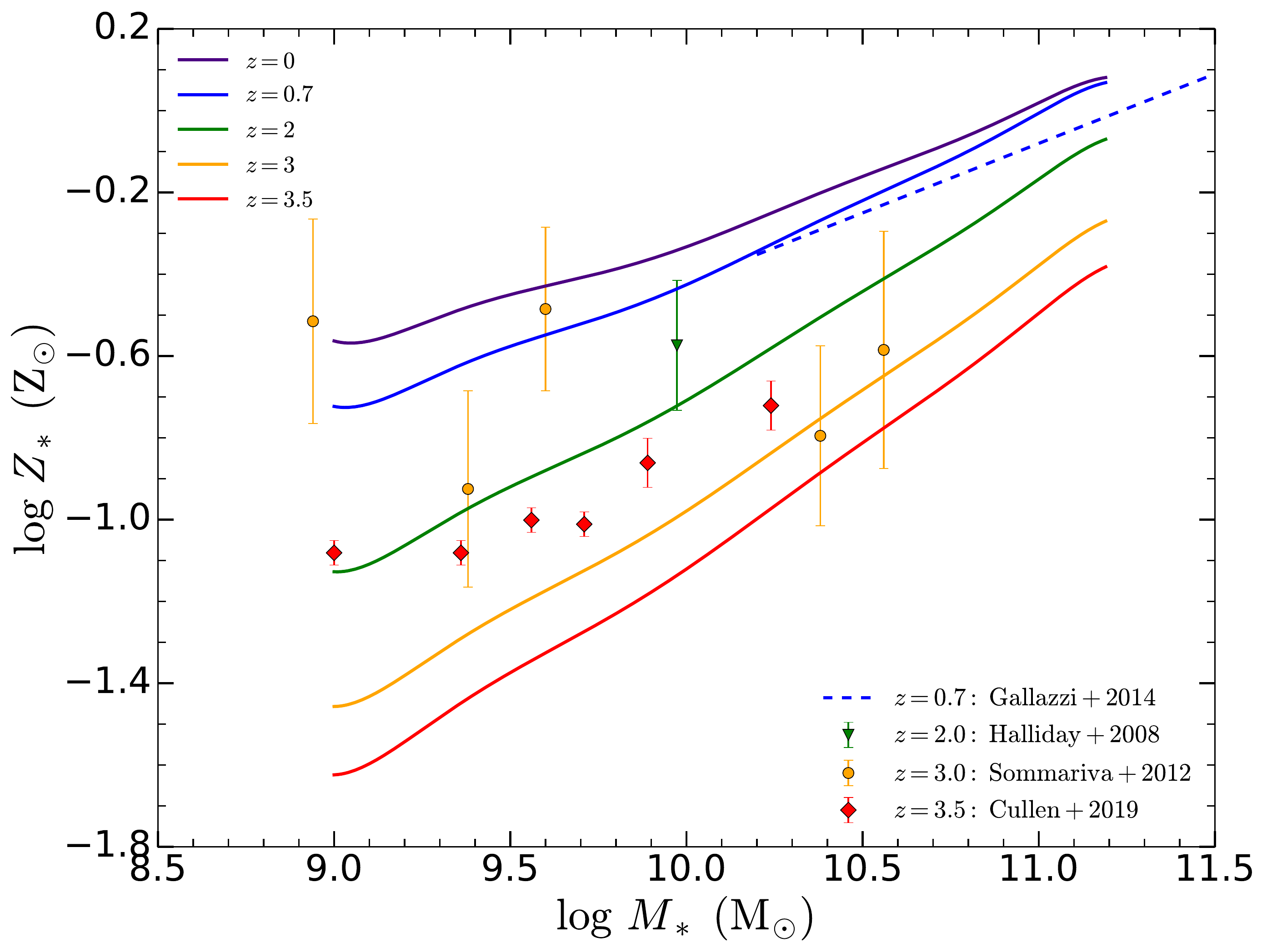}}
\caption{The redshift-evolution of the stellar mass--stellar metallicity relation in our model is shown by the solid curves, evaluated at $z=0$ (indigo), $z=0.7$ (blue), $z=2$ (green), $z=3$ (orange) and $z=3.5$ (red). The $z=0$ curve corresponds to the mass-weighted stellar metallicities of star-forming galaxies shown in Figure \ref{fig:combined_mzr_sgq}. The $z>0$ curves were obtained by evolving the $z=0$ curve in redshift according to the evolution of the gas MZR in \citet{Maiolino2008}. We also show the stellar metallicities of star-forming galaxies observed at $z=0.7$ \citep[blue dashed curve]{Gallazzi2014}, $z=2$ \citep[green triangle]{Halliday2008}, $z=3$ \citep[orange circles]{Sommariva2012, Cullen2019} and $z=3.5$ \citep[red diamonds]{Cullen2019}.}
\label{fig:mzr_evolution}
\end{figure}

\subsection{Progenitor--descendant stellar mass offset}\label{subsec:prog_mass_offset}

In our model we assume that high-$z$ progenitors have the same stellar mass as their local passive descendants. In the absence of stellar mass loss \citep[e.g.\@ through tidal stripping,][]{Chang2013, Tollet2017, Wang2017}, this implies that galaxies form a minor amount of additional stellar mass during the quenching phase. However, high-$z$ star-forming galaxies have large gas fractions \citep[e.g.\@][]{Daddi2010, Genzel2015, Tacconi2018}, and, if they quench through starvation, are able to convert a significant portion of this gas into new stars. This results in a non-negligible increase in stellar mass during quenching, which is inconsistent with the assumption of progenitors and descendants having the same stellar mass in our model. In this subsection, we investigate how introducing a stellar mass offset between progenitors and descendants affects the timescales and mass-loading factors predicted by our model.

In principle, the stellar mass increase increment during quenching could be mass-dependent, because, for instance, gas fractions are mass-dependent \citep[e.g.\@][]{Genzel2015, Tacconi2018}. Indeed, in our closed-box models from Section \ref{subsubsec:qg_pure}, we find that galaxies typically increase in mass by 0.2--0.6 dex (at the high-mass and low-mass ends, respectively), while for our models in Section \ref{subsubsec:qg_sfr} the mass increase range is 0.20--0.45 dex. Hence the stellar mass offset, defined as 
\begin{equation}
\Delta \log M_* = \log M_{*,\mathrm{progenitor}} - \log M_{*,\mathrm{descendant}}
\label{eq:mass_offset}
\end{equation}
is mass-dependent. However, rather than assume a mass-dependent stellar mass offset between progenitors and descendants, we instead study the effect of using various fixed stellar mass offsets (that span the range of mass increases seen in our models) on our model predictions. 

The effect of introducing a progenitor--descendant stellar mass offset is shown in Fig.\@ \ref{fig:progenitor_mass_offsets} (which follows the format of Fig.\@ \ref{fig:qg_model_qt_tau_variable_lambda}). The different curves correspond to different stellar mass offsets, $\Delta \log M_*$, ranging from 0 to -0.6, in 0.2 dex increments. Error bars are omitted for clarity. We find that introducing a stellar mass offset between progenitors and descendants has a very minor effect on the quenching timescales ($t_\mathrm{quench}$ and $\tau_\mathrm{q}$) and mass-loading factors $\lambda_\mathrm{eff}$ obtained by our model, which indicates that our simplifying assumption that progenitors and descendants have the same stellar mass does not significantly affect our results. Note that the horizontal axis refers to the stellar mass of the progenitors, rather than the stellar mass of the descendants. For a given progenitor stellar mass, a more negative stellar mass offset corresponds to a descendant of higher stellar mass, which will tend to be older (see Fig.\@ \ref{fig:combined_mar_sgq}) and therefore have an earlier quenching epoch $z_\mathrm{q}$ in our model (see Fig.\@ \ref{fig:quenching_epochs}). However, since $z_\mathrm{q}$ is only weakly dependent on stellar mass, using a different stellar mass offset will only result in a small change in $z_\mathrm{q}$ and will therefore only introduce a small change in the initial gas mass, depletion time and stellar metallicity of the progenitor. Hence the results from our model are not strongly affected by the stellar mass offset that is assumed. 

\begin{figure}
\centering
\centerline{\includegraphics[width=1.0\linewidth]{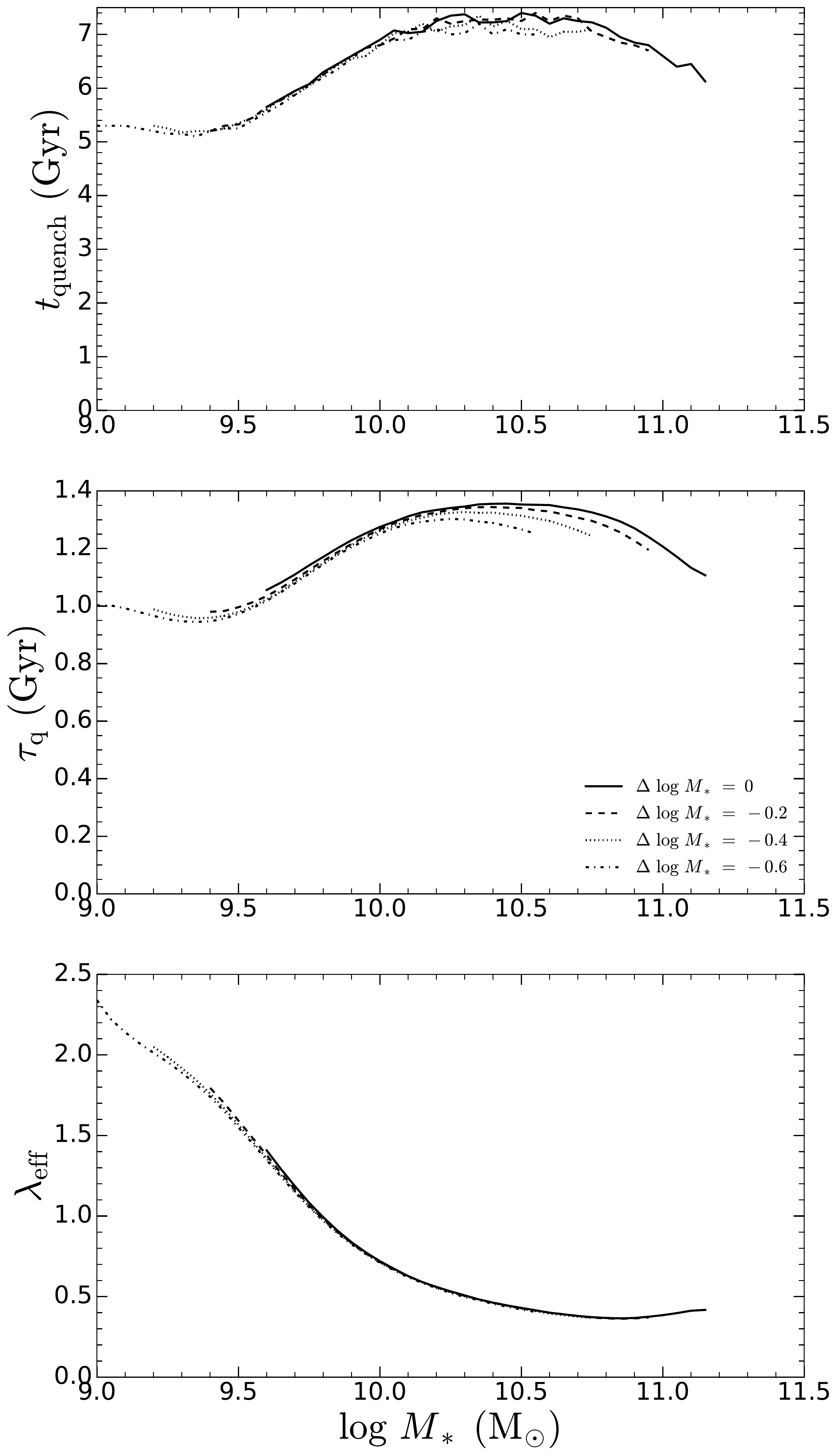}}
\caption{Similar in format to Fig.\@ \ref{fig:qg_model_qt_tau_variable_lambda}. The horizontal axis represents the stellar mass of the star-forming progenitors, which are assumed to be offset from the stellar mass of their local passive descendants by $\Delta \log M_*$ (defined in equation (\ref{eq:mass_offset})). The model predictions for $\Delta \log M_*$ = 0 (solid), -0.2 (dashed), -0.4 (dotted), -0.6 (dot-dashed) are shown.}
\label{fig:progenitor_mass_offsets}
\end{figure}

\subsection{Mass return to the ISM}\label{subsec:mass_return}

In our model we assume the instantaneous recycling approximation (IRA), where sufficiently massive stars instantly die upon formation, returning some of their stellar mass to the ISM. Massive stars, which recycle their gas quickly, significantly contribute to the return of stellar mass to the ISM. However, AGB stars (and to a much lesser extent Type Ia SNe) also return a considerable amount of gas to the ISM \citep[e.g.\@][]{Segers2016a}, albeit over much longer timescales than massive stars. The majority of this gas is returned over the first few Gyr, but there is still some mass return even after 10~Gyr \citep[e.g.\@][]{Maraston2005, Leitner2011, Segers2016a}. This delayed recycling of gas will affect the evolution of the gas reservoir in the starvation scenario. In this subsection, we investigate how modifications to the return fraction $R$ affect the timescales and mass-loading factors predicted by our model.

The return fraction $R$ is weakly dependent on stellar metallicity ($\Delta R \sim 0.02$), but is sensitive to the IMF adopted \citep[$\Delta R \sim 0.15$,][]{Vincenzo2016a}, with $R\sim0.3$ for a more bottom-heavy IMF like \citet{Salpeter1955} and $R\sim0.45$ for a more bottom-light IMF like \citet{Kroupa2001} and \citet{Chabrier2003}. Modifying the mass threshold used in the IRA to, for example, take into account the delayed recycling of gas from AGB stars, will also affect $R$. In this work, we assumed $R=0.425$ \citep{Vincenzo2016a}. We now consider the impact of assuming different values of $R$, using values that are larger/smaller ($\pm$25\%) and substantially larger/smaller ($\pm$50\%) than our adopted value. Thus, the maximum variations in $\Delta R$ that we test are much greater than what is to be expected from adopting different IMFs or changing the mass threshold in the IRA.

The effect of modifying $R$ is shown in Fig.\@ \ref{fig:return_fraction}. The various curves correspond to different values of $R$. Error bars are omitted for clarity. We find that the value of $R$ that is adopted does not significantly affect the shape of the quenching timescale and mass-loading factor curves. However, the normalisation of these curves changes quite strongly ($\sim$50\%) when substantial ($\pm50\%$) variations in $R$ are considered, while the effects are more minor ($\sim$25\%) when smaller variations in $R$ are used. 

We find that increasing $R$ results in longer quenching timescales and smaller mass-loading factors. The reason for this is that, for a given mass-loading factor $\lambda _\mathrm{eff}$, increasing $R$ increases the amount of gas that is being lost through winds, $\lambda_\mathrm{eff}\Psi$, relative to the amount of gas that is being locked up in stars, $(1-R)\Psi$. In order to still reproduce the observed $\Delta Z_*$ between star-forming and passive galaxies, $\lambda_\mathrm{eff}$ therefore has to decrease when $R$ increases. Furthermore, by definition (see equation (\ref{eq:tau_q})), $\tau_\mathrm{q}$ therefore increases (and therefore also $t_\mathrm{quench}$ increases) with increasing $R$.

\begin{figure}
\centering
\centerline{\includegraphics[width=1.0\linewidth]{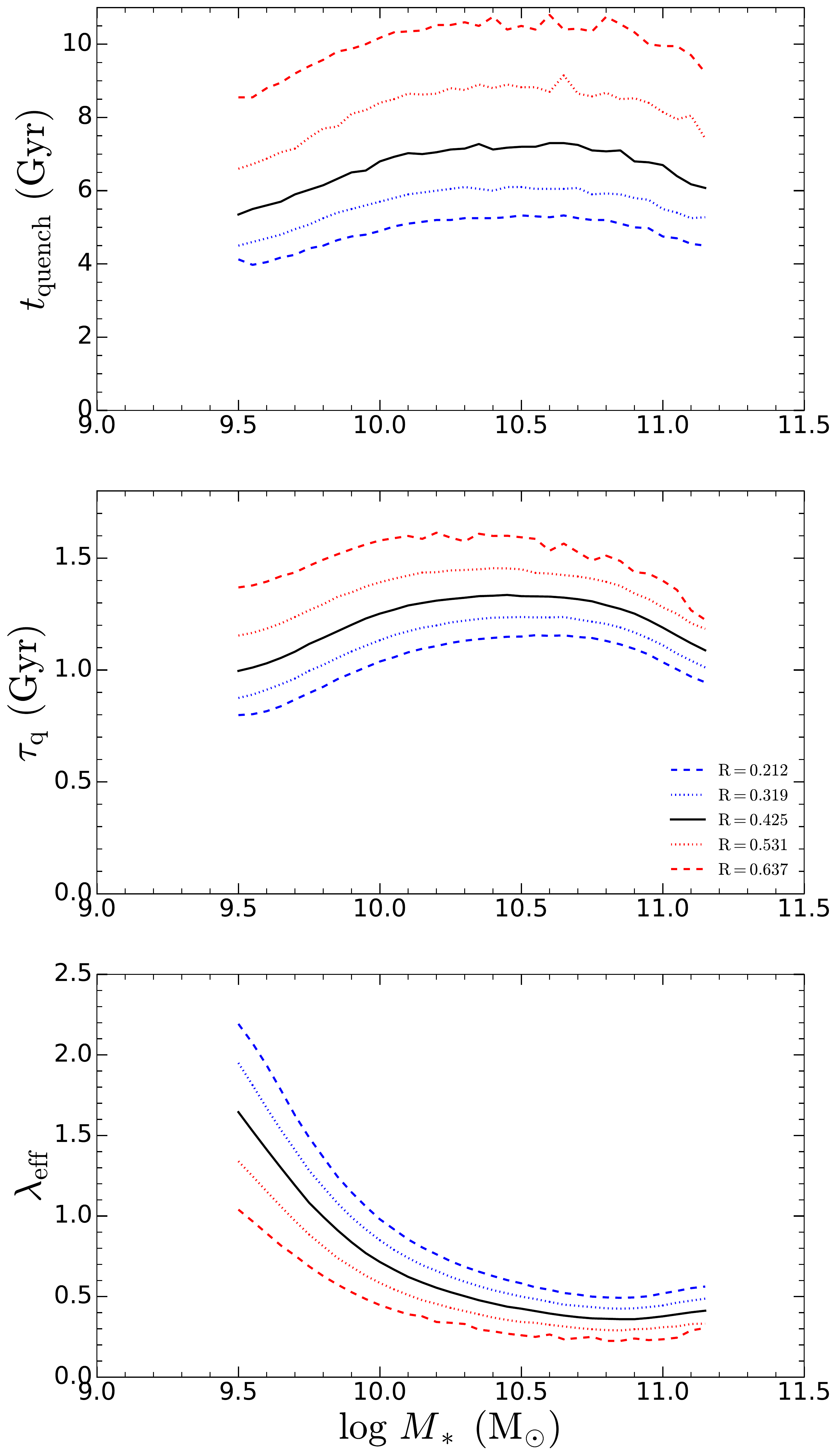}}
\caption{Similar in format to Fig.\@ \ref{fig:qg_model_qt_tau_variable_lambda}.  Here we investigate how the adopted value of $R$ affects our model predictions, for $R=0.212$ (dashed blue), $R=0.319$ (dotted blue), $R=0.425$ (solid black), $R=0.531$ (dotted red) and $R=0.637$ (dashed red). Note that $R=0.425$ was assumed throughout this work.}
\label{fig:return_fraction}
\end{figure}

\section{Summary and conclusions} \label{sec:summary}

We have investigated the mechanisms responsible for
quenching star formation in galaxies across the cosmic
epochs, i.e.\@ the processes responsible for transforming
star-forming galaxies into passive systems.
We leverage on the method initially developed by \citetalias{Peng2015},
who suggest that the stellar metallicity difference between local
passive galaxies and their star-forming progenitors is a powerful
tracer of the quenching process.
We expand and improve upon the \citetalias{Peng2015} method in the following
ways. We undertake a more extensive comparison between models and observations, deriving stronger constraints on the mass-dependent role of outflows in quenching star formation, by simultaneously reproducing the stellar metallicities and the star formation rates observed in local passive galaxies in our models. 
We use the much larger spectroscopic sample of galaxies available in SDSS DR7. Leveraging on the tripling in statistics that this new dataset has enabled, we now also study the quenching of star formation in local green valley galaxies.
We have introduced a more sophisticated treatment of the progenitor--descendant comparison. We now compare the stellar metallicity difference between local passive galaxies and their high-$z$ star-forming progenitors. This results in a widening of the gap in stellar metallicity between star-forming and passive galaxies, which qualitatively strengthens the case for starvation. In addition, we now also adopt the scaling relations, in terms of gas fractions and star formation efficiency, that have been observed for star-forming galaxies at high-redshift.
We use the mass-weighted stellar ages of local galaxies to provide a more empirically-motivated estimation of the onset of quenching $z_\mathrm{q}$ in the models.
We adopt a consistent treatment of stellar metallicity in this work, comparing mass-weighted metallicity predictions from models with mass-weighted stellar metallicities from observations.

We obtain the following observational results for local passive galaxies:
\begin{itemize}

\item The (mass-weighted) stellar metallicity of passive galaxies is always
higher than the stellar metallicity of local star-forming galaxies of the same stellar mass.

\item Under the empirically- and theoretically-motivated assumption that the normalisation of the stellar mass--stellar metallicity relation for star-forming galaxies decreases with increasing redshift, then the metallicity difference becomes even larger if passive galaxies are compared with their star-forming progenitors at high-redshift.

\item The metallicity difference is a function of stellar mass,
being highest at low masses and decreasing at high masses, but still
remaining significant in the most massive galaxies analysed
($M_* \sim 10^{11}~\mathrm{M}_{\odot}$).

\item The stellar metallicities of green valley galaxies (which probe the quenching process in the local Universe) are intermediate between those of star-forming and passive galaxies.

\end{itemize}

We investigated galaxy quenching using both closed-box models (where galaxies quench purely through starvation, with `effective' outflow mass-loading factor $\lambda_\mathrm{eff} = 0$) and leaky-box models (where galaxies quench through a combination of starvation and outflows, with $\lambda_\mathrm{eff} > 0$). We explored two different leaky-box models in this work. Firstly, we investigated the case when $\lambda_\mathrm{eff} = 1$ for galaxies of all stellar masses. Secondly, we did not fix $\lambda_\mathrm{eff}$, but instead found the values of $\lambda_\mathrm{eff}$ for which our model simultaneously reproduced both the $Z_*$ and SFR seen in local passive (and green valley) galaxies. By comparing the observed stellar metallicity difference between passive galaxies and their star-forming progenitors at high-redshift with the predictions from these simple analytical models of chemical evolution, we infer the following:

\begin{itemize}

\item The prominent metallicity difference between passive galaxies and their star-forming progenitors implies that for galaxies at \textit{all} masses, quenching involved an extended phase of starvation, i.e. halting (or very substantial decrease) of gas accretion from the circumgalactic/intergalactic medium.

\item However, we find that some form of gas ejection has to be introduced into the models to best match the observed properties (i.e.\@ stellar metallicities and SFRs) of local passive galaxies, indicating that outflows also play an important role in quenching star formation in galaxies. Thus we find that the combination of starvation together with outflows is responsible for quenching the majority of galaxies. 

\item In our closed-box models we find that the duration of the quenching phase is 2--3~Gyr, with an $e$-folding time of 2--4~Gyr, after which a sudden ejection or heating of gas must have occurred to prevent further star formation and chemical enrichment of the galaxy. This delayed ejection/heating phase may have resulted from e.g.\@ the cumulative energy of Type Ia SNe, AGN energy injection or the onset of ram pressure stripping.

\item In our leaky-box models which instead incorporate continuous (rather than invoke sudden) outflows, we find that the duration of quenching is longer at 5--7~Gyr, with an $e$-folding time-scale for star formation of $\sim$1~Gyr.

\item Furthermore, we find that the `effective' mass loading factor decreases with increasing stellar mass, indicating that `effective' outflows (which are capable of permanently removing gas from the galaxy) are, together with starvation, of increasing importance in low-mass galaxies. In particular, for galaxies with $\log(M_*/\mathrm{M}_\odot) < 10.2$, we find that the rate at which gas is lost through galactic winds is roughly 1--3 times larger than the rate at which gas is locked up into long-lived stars. Outflows become relatively less important and sub-dominant relative to starvation at high masses (but still eject comparable amounts of gas to what is lost through gas consumption), likely indicating that a significant fraction of the outflowing gas in these massive galaxies does not escape and is instead reaccreted and/or that these ejection events are short lived.
\end{itemize}

While passive galaxies have enabled us to explore the role of different quenching mechanisms at high redshift, when most of the star formation and quenching process took place, the analysis of green valley galaxies has offered us the opportunity of studying the quenching process in the local Universe. By comparing the observed stellar metallicity with simple models we infer the following:

\begin{itemize}

\item The stellar metallicity difference between green valley galaxies and local star-forming galaxies indicates that also locally the quenching process is likely to involve an extensive period of starvation.

\item The duration of quenching tends to increase with increasing stellar mass, indicating that, on average, more massive green valley galaxies began quenching at an earlier epoch than less massive green valley galaxies. However, this time-scale is dependent on the contribution from outflows, with our closed-box model predicting 2--5~Gyr, while our leaky-box model that matches the $Z_*$ and SFR in local green valley galaxies predicts 4--6.5~Gyr.

\item We find that the duration of quenching is comparable to, or larger than the time-scales obtained for passive galaxies. Since green valley galaxies are still in the transition phase and have yet to quench completely, the time-scales derived are in fact lower limits, suggesting that galaxies in the local Universe quench more slowly than their counterparts at higher redshift. 

\item We also find that the $e$-folding time-scales for star formation are longer in green valley galaxies (1.5--2.5~Gyr) than for passive galaxies ($\sim$1~Gyr), which further supports the notion that, on average, the quenching of star formation occurs more slowly in the local Universe.

\item We find that the effective mass-loading factors $\lambda_\mathrm{eff}$ decrease with increasing stellar mass, with a similar normalisation to the results obtained for our analysis of passive galaxies. This result suggests that the power and prevalence of outflows in galaxies in the local Universe are comparable to those in galaxies at higher redshift ($z\sim$1--2).

\end{itemize}

We note however some inconsistencies between the quenching time-scale inferred
from the metallicity difference between green valley and star-forming galaxies
and the age difference between the two galaxy populations. There are various
possible explanations for this inconsistency. One of the most important effects
is that green valley galaxies are characterised by steep radial gradients in their
properties (in particular age and star formation), while the SDSS DR7 single
fibre captures only the central properties of these galaxies. Therefore, it is
important to repeat this analysis by exploiting integral field spectroscopic galaxy surveys.

\section*{Acknowledgements}

We thank the anonymous referee for their comments and suggestions which contributed to greatly improving this article. JT thanks S.\@ Wuyts for helpful discussions. 
JT and RM acknowledge support from the ERC Advanced Grant 695671 `QUENCH'. RM acknowledges support by the Science and Technology Facilities Council (STFC). 
YP acknowledges support from the National Key Program for Science and Technology Research and Development under grant number 2016YFA0400702, and the NSFC grant no.\@ 11773001.
Funding for the SDSS and SDSS-II has been provided by the Alfred P. Sloan Foundation, the Participating Institutions, the National Science Foundation, the U.S. Department of Energy, the National Aeronautics and Space Administration, the Japanese Monbukagakusho, the Max Planck Society, and the Higher Education Funding Council for England. The SDSS Web Site is http://www.sdss.org/. 
The SDSS is managed by the Astrophysical Research Consortium for the Participating Institutions. The Participating Institutions are the American Museum of Natural History, Astrophysical Institute Potsdam, University of Basel, University of Cambridge, Case Western Reserve University, University of Chicago, Drexel University, Fermilab, the Institute for Advanced Study, the Japan Participation Group, Johns Hopkins University, the Joint Institute for Nuclear Astrophysics, the Kavli Institute for Particle Astrophysics and Cosmology, the Korean Scientist Group, the Chinese Academy of Sciences (LAMOST), Los Alamos National Laboratory, the Max-Planck-Institute for Astronomy (MPIA), the Max-Planck-Institute for Astrophysics (MPA), New Mexico State University, Ohio State University, University of Pittsburgh, University of Portsmouth, Princeton University, the United States Naval Observatory, and the University of Washington.




\bibliographystyle{mnras}
\bibliography{pDR7.bib,Supplementary.bib} 





\appendix

\section{Molecular gas relations} \label{app:molec_gas}

We used the integrated Schmidt-Kennicutt (SK) law, given by equation (\ref{eq:sk_law}), to parametrize star formation in our models. The total gas mass, which includes both the atomic and molecular components, as well as the total gas depletion time were used in the SK law in the main body of the paper. In this section, we investigate how our model predictions and conclusions change when only the molecular gas component is considered. In this case, the law for star formation becomes

\begin{equation}
\Psi = \epsilon _m m,
\end{equation}

\noindent where $\epsilon _m$ is the star formation efficiency for molecular gas and $m$ is the molecular gas mass. $\epsilon _m$ is related to $t_{\mathrm{depl,m}}$, the molecular gas depletion time, through $\epsilon _m = 1/t_{\mathrm{depl,m}}$. 

There are two main differences between models that use the total gas mass and models that only use the molecular component. Firstly, molecular gas reservoirs are smaller than the total gas reservoirs. Hence outflows with a given outflow rate $\Lambda$ will be more effective at quenching star formation in models that only contain molecular gas. This means that leaky-box models that only contain molecular gas will have more difficulty in reproducing the large stellar metallicity differences that have been observed. As a result, these models will tend to disfavour quenching through outflows and $\lambda _\mathrm{eff}$ will be smaller. Secondly, molecular gas depletion time-scales are shorter than total gas depletion time-scales. Models that only contain molecular gas will process and enrich the gas in the ISM more rapidly. As a result, the stellar metallicity increases more quickly and so the quenching time-scales derived from our analysis of stellar metallicity differences tend to be shorter. These differences between the total gas and molecular gas models are most apparent for low-mass galaxies, as these tend to have relatively larger atomic-to-molecular gas mass ratios than the more massive galaxies. The model predictions for low-mass galaxies are therefore more strongly affected by the removal of the atomic gas component.

\subsection{Passive galaxies}

We compare the observed stellar metallicity differences between star-forming and passive galaxies with the predictions from gas regulator models that only contain molecular gas. Our results using a closed-box model with $\lambda _\mathrm{eff} = 0$ is shown in Fig.\@ \ref{fig:qg_model_molec_deltaZ_qt_lambda_0}. We find that our models are able to reproduce the observations, and the derived quenching time-scales are shorter than what was seen in Fig.\@ \ref{fig:qg_model_deltaZ_qt_lambda_0}. We also show the quenching time-scales and constraints on $\lambda_\mathrm{eff}$ derived using our leaky-box model in Fig.\@ \ref{fig:qg_model_molec_qt_tau_variable_lambda}. The derived quenching time-scales are shorter, and the mass-loading factors are smaller than what was seen in Fig.\@ \ref{fig:qg_model_qt_tau_variable_lambda}.

\begin{figure}
\centering
\centerline{\includegraphics[width=1.0\linewidth]{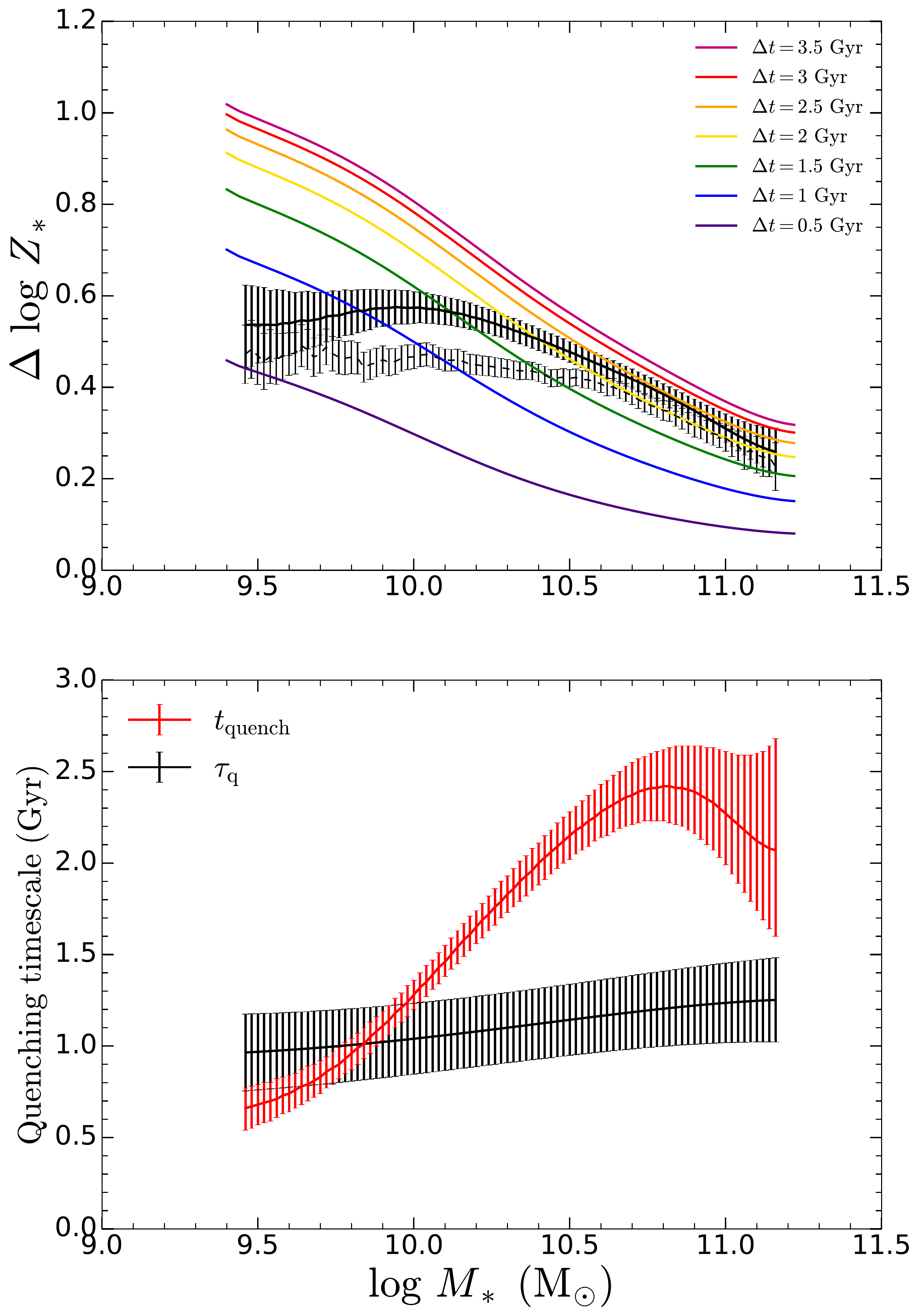}}
\caption{Similar to Fig.\@ \ref{fig:qg_model_deltaZ_qt_lambda_0}, where we study the stellar metallicity differences between star-forming and passive galaxies using a closed-box model with $\lambda _\mathrm{eff} = 0$, but now we only use the molecular gas component in our models.}
\label{fig:qg_model_molec_deltaZ_qt_lambda_0}
\end{figure}

\begin{figure}
\centering
\centerline{\includegraphics[width=1.0\linewidth]{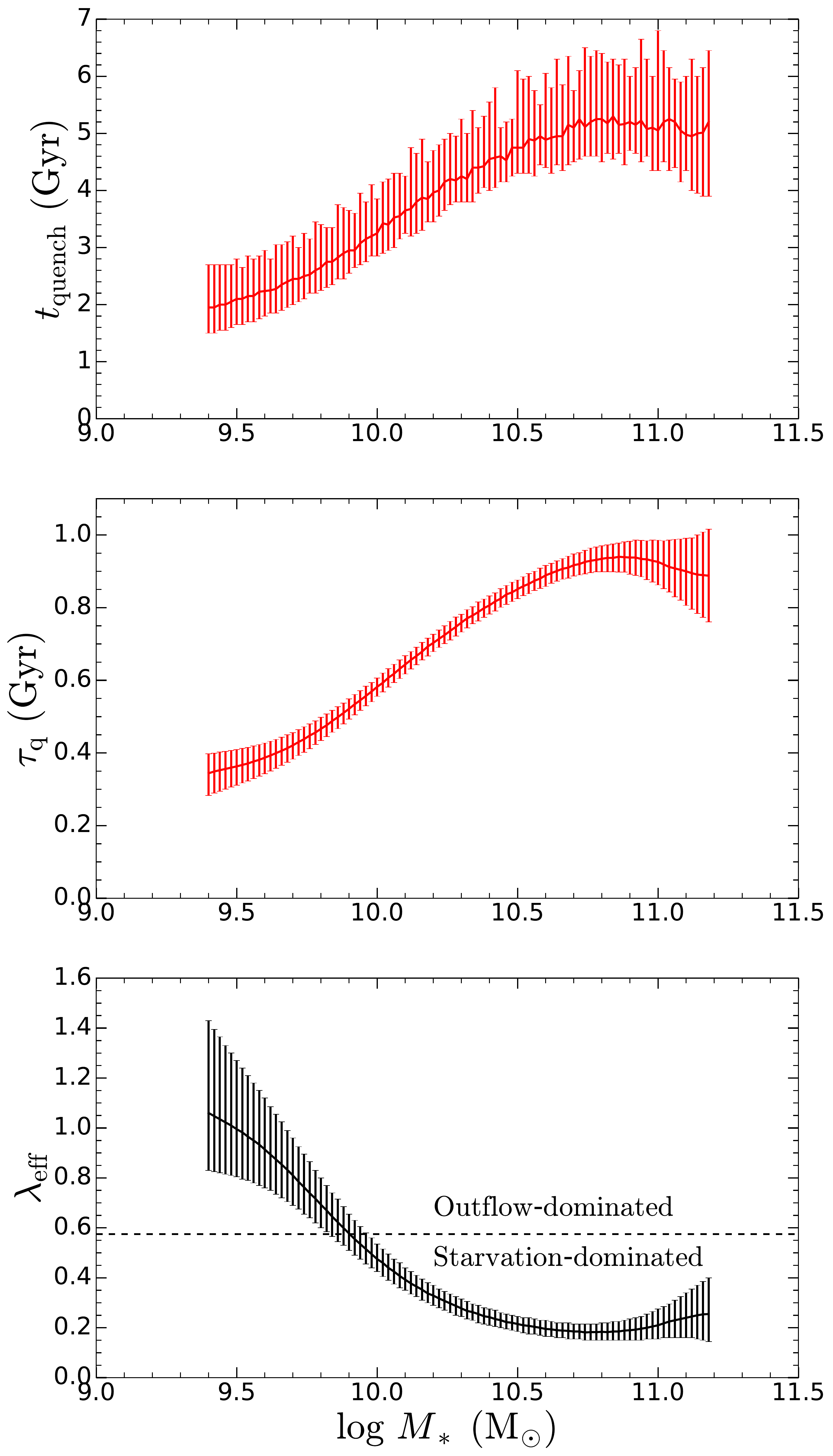}}
\caption{Similar to Fig.\@ \ref{fig:qg_model_qt_tau_variable_lambda}, but now we only use the molecular gas component in our models.}
\label{fig:qg_model_molec_qt_tau_variable_lambda}
\end{figure}

\subsection{Green valley galaxies}

\begin{figure}
\centering
\centerline{\includegraphics[width=1.0\linewidth]{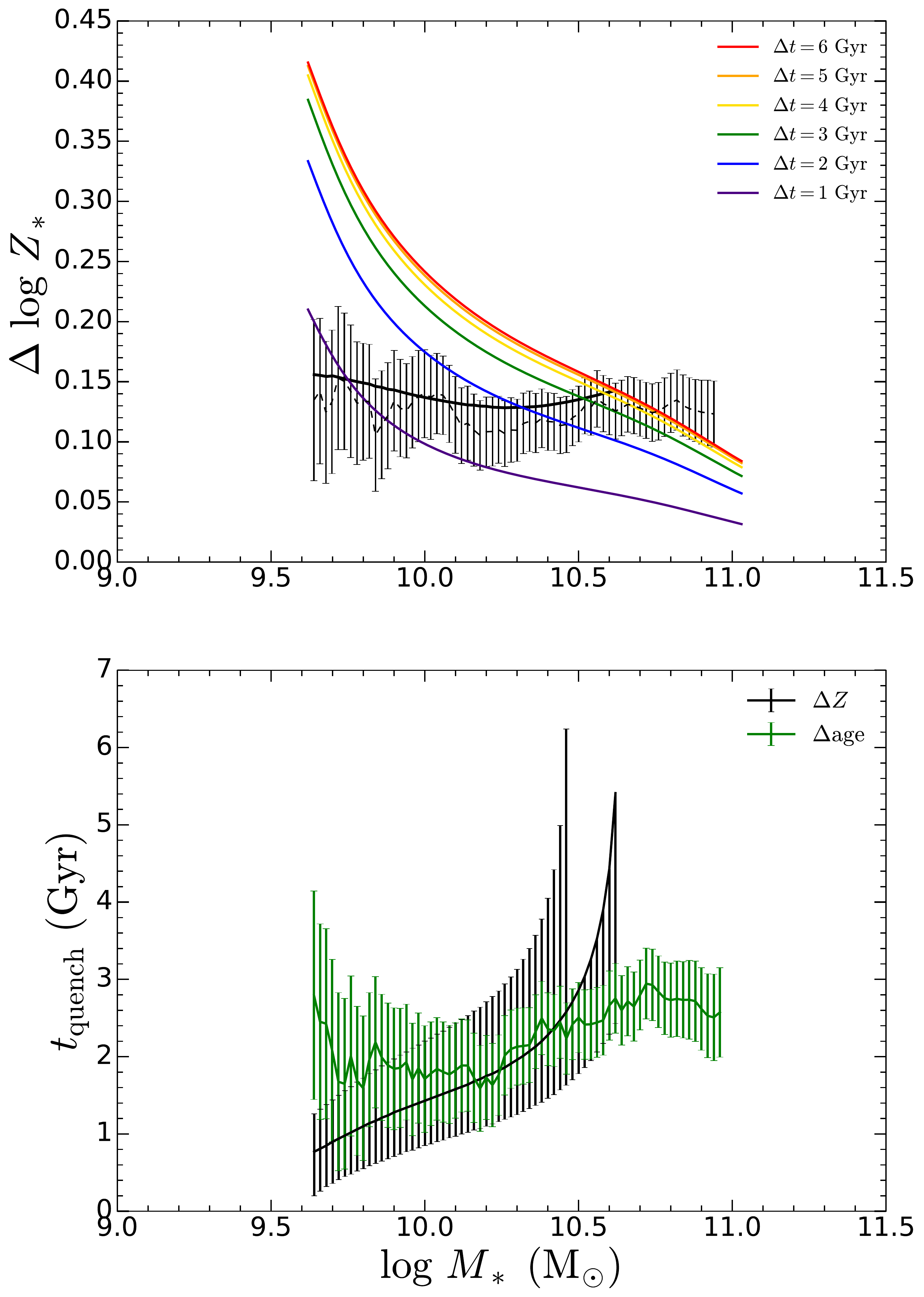}}
\caption{Similar to Fig.\@ \ref{fig:gvg_model_deltaZ_qt_lambda_0}, where we study the stellar metallicity differences between star-forming and green valley galaxies using a closed-box model with $\lambda _\mathrm{eff} = 0$, but now we only use the molecular gas component in our models. In the cases when the upper limit on the stellar metallicity difference $\Delta Z_*$ cannot be reproduced by our model, only the lower limit on the quenching time-scale $t_\mathrm{quench}$ is shown.}
\label{fig:gvg_model_molec_deltaZ_qt_lambda_0}
\end{figure}

\begin{figure}
\centering
\centerline{\includegraphics[width=1.0\linewidth]{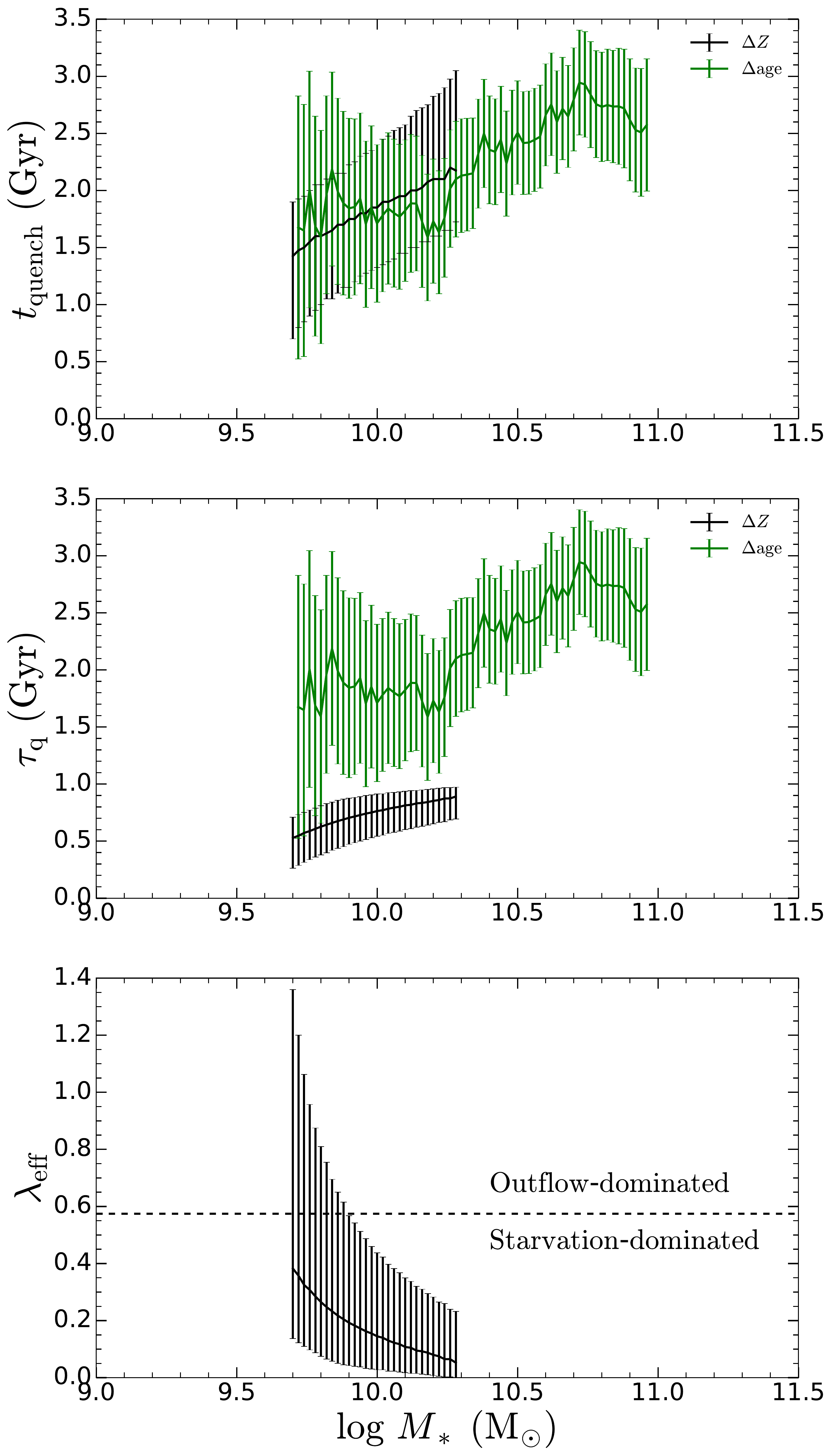}}
\caption{Similar to Fig.\@ \ref{fig:gvg_model_qt_tau_variable_lambda}, but now we only use the molecular gas component in our models.}
\label{fig:gvg_model_molec_qt_tau_variable_lambda}
\end{figure}

We now study the stellar metallicity differences between star-forming and green valley galaxies. Our results for the closed-box model are shown in Fig.\@ \ref{fig:gvg_model_molec_deltaZ_qt_lambda_0}. The quenching time-scales are typically shorter than what was seen in Fig.\@ \ref{fig:gvg_model_deltaZ_qt_lambda_0}. The predictions from the leaky-box model are shown in Fig.\@ \ref{fig:gvg_model_qt_tau_variable_lambda}. The derived quenching time-scales are shorter, and the mass-loading factors are smaller than what was seen in Fig.\@ \ref{fig:gvg_model_qt_tau_variable_lambda}. When using only molecular gas masses, both our closed-box and leaky-box models struggle to reproduce the observed stellar metallicity differences at the high-mass end.

\bsp	
\label{lastpage}
\end{document}